\documentclass{aa}  

\usepackage{graphicx}
\usepackage{rotating}
\usepackage{siunitx}
\usepackage{xcolor}
\usepackage{txfonts}

\usepackage{hyperref}
\hypersetup{
    colorlinks=true,
    citecolor=blue,
    }

\usepackage[capitalise]{cleveref}
\def\s{\rule[0mm]{0mm}{4.0mm}}

\newcommand{\Msun}{\ensuremath{\rm M_\odot}}
\newcommand{\mdot}{\ensuremath{\rm M_\odot\,yr^{-1}}}

\newcommand{\Rsun}{\ensuremath{\rm R_\odot}}

\newcommand{\fuse}{\emph{FUSE}}
\newcommand{\stis}{\emph{STIS}}
\newcommand{\cosi}{\emph{COS}}
\newcommand{\hst}{\emph{HST}}
\newcommand{\eso}{\emph{ESO}}
\newcommand{\xshooter}{X-shooter}
\newcommand{\giraffe}{GIRAFFE}
\newcommand{\heiiwr}{He\,{\sc ii}\,$\lambda4686$}

\newcommand{\zsol}{Z$_{\odot}$}
\newcommand{\msol}{M$_{\odot}$}

\newcommand{\ergs}{erg\,s$^{-1}$}
\newcommand{\teff}{$T_\text{eff}$}
\newcommand{\logg}{$\log g$}

\newcommand{\mspec}{$M_\mathrm{spec}$}
\newcommand{\mev}{$M_\mathrm{ev}$}
\newcommand{\mini}{$M_\mathrm{ini}$}

\newcommand{\lmec}{$L_\mathrm{mec}$}
\newcommand{\loglmec}{$\log L_\mathrm{mec}$}
\newcommand{\lx}{$L_\mathrm{X}$}
\newcommand{\lbol}{$L_\mathrm{bol}$}
\newcommand{\lxlbol}{$\log (L_\mathrm{X}/L_\mathrm{bol}$)}
\newcommand{\Dmom}{$D_\mathrm{mom}$}
\newcommand{\nh}{$N_\mathrm{H}$}
\newcommand{\ff}{$x_\mathrm{fill}$}
\newcommand{\rmin}{$r_\mathrm{min}$}
\newcommand{\tx}{$T_\mathrm{X}$}
\newcommand{\vini}{$\varv_\mathrm{ini}$}

\newcommand{\vmac}{$\varv_\mathrm{mac}$}
\newcommand{\vmic}{$\varv_\mathrm{mic}$}

\newcommand{\vrad}{$\varv_\mathrm{rad}$}
\newcommand{\vinf}{$\varv_\mathrm{\infty}$}
\newcommand{\vrot}{$\varv\sin\,i$}
\newcommand{\kms}{km\,s$^{-1}$}
\newcommand{\cm}{cm\,s$^{-2}$}

\newcommand{\rstar}{$R_\star$}
\newcommand{\mstar}{$M_\star$}
\newcommand{\tstar}{$T_\star$}
\newcommand{\loglstar}{$\log (L_{\star}/L_\odot)$}
\newcommand{\lstar}{$L_{\star}$}
\newcommand{\zsun}{${\rm Z}_{\odot}$}
\newcommand{\lsun}{${\rm L}_{\odot}$}
\newcommand{\ebv}{$\rm E(B-V)$}

\newcommand{\Mv}{$M_{\rm v}$}

\newcommand{\QH}{$Q_\mathrm{H}$}
\newcommand{\QHeI}{$Q_\mathrm{He\,{\small{\sc I}}}$}
\newcommand{\QHeII}{$Q_\mathrm{He\,{\small{\sc II}}}$}
\newcommand{\logQH}{$\log\,Q_\mathrm{H}$}
\newcommand{\logQHeI}{$\log\,Q_\mathrm{He\,{\small{\sc I}}}$}
\newcommand{\logQHeII}{$\log\,Q_\mathrm{He\,{\small{\sc II}}}$}

\newcommand{\hii}{H\,{\small{\sc ii}}}
\newcommand{\lya}{Ly$\alpha$}
\newcommand{\lyb}{Ly$\beta$}
\newcommand{\ha}{H$\alpha$}

\newcommand{\hb}{H$\beta$}
\newcommand{\hg}{H$\gamma$}
\newcommand{\hd}{H$\delta$}
\newcommand{\hep}{H$\epsilon$}

\newcommand{\hei}{He\,{\sc i}}
\newcommand{\heii}{He\,{\sc ii}}

\newcommand{\oiiic}{O\,{\sc iii}\,$\lambda5592.3$}

\newcommand{\ovi}{O\,{\sc vi}\,$\lambda\lambda1031.9,1037.6$}
\newcommand{\oiii}{O\,{\sc iii}}

\newcommand{\nv}{N\,{\sc v}}
\newcommand{\nva}{N\,{\sc v}\,$\lambda4603.8$}
\newcommand{\nvb}{N\,{\sc v}\,$\lambda4619.9$}
\newcommand{\nvuv}{N\,{\sc v}\,$\lambda\lambda1238.8,1242.8$}
\newcommand{\ciii}{C\,{\sc iii}}
\newcommand{\civ}{C\,{\sc iv}}
\newcommand{\civa}{C\,{\sc iv}\,$\lambda5801$}
\newcommand{\civb}{C\,{\sc iv}\,$\lambda5811$}
\newcommand{\pv}{P\,{\sc v}}
\newcommand{\pva}{P\,{\sc v}\,$\lambda1118$}
\newcommand{\pvb}{P\,{\sc v}\,$\lambda1128$}

\newcommand{\oiiin}{[O\,{\sc iii}]}
\newcommand{\niii}{N\,{\sc iii}}
\newcommand{\niiia}{N\,{\sc iii}\,$\lambda4515$}
\newcommand{\niiib}{N\,{\sc iii}\,$\lambda4518$}

\newcommand{\siiv}{Si\,{\sc iv}}
\newcommand{\siiva}{Si\,{\sc iv}\,$\lambda4088.9$}
\newcommand{\siivb}{Si\,{\sc iv}\,$\lambda4116.1$}

\begin{document} 

\title{X-Shooting ULLYSES: Massive stars at low metallicity X. Physical Parameters and Feedback of Massive Stars in the LMC N11\,B Star-Forming Region}
%\subtitle{\thanks{Based on observations collected at the European Southern Observatory under \eso\ program 106.211Z.001.}}

\titlerunning{X-Shooting ULLYSES -- X. Physical Parameters and Feedback of Massive Stars in N11 B}

\author{
      {V.~M.~A. G\'omez-Gonz\'alez\inst{\ref{UP}}}
    \and
      {L.~M. Oskinova\inst{\ref{UP}}}
    \and
      {W.-R. Hamann\inst{\ref{UP}}}
    \and
      {H. Todt\inst{\ref{UP}}}
    \and
      {D. Pauli\inst{\ref{UP}}}
    \and
      {S. Reyero Serantes\inst{\ref{UP}}}
    \and
      {M. Bernini-Peron\inst{\ref{ARI}}}
    \and 
      {A.~A.~C. Sander\inst{\ref{ARI}}}  
    \and
      {V. Ramachandran\inst{\ref{ARI}}}    
    \and 
      {J.~S. Vink\inst{\ref{AOP}}}
    \and
      {P.~A. Crowther\inst{\ref{UoS}}}
     \and
      {S. R. Berlanas\inst{\ref{Laguna},\ref{IAC}}}
    \and 
      {A. ud-Doula\inst{\ref{PSU}}}
    \and
      {A. C. Gormaz-Matamala\inst{\ref{ASU},\ref{UAIVina},\ref{PUC}}}
    \and 
      {C. Kehrig\inst{\ref{IAA}}}
    \and 
      {R. Kuiper\inst{\ref{DE}}}
    \and 
      {C. Leitherer\inst{\ref{stsi}}}
    \and 
      {L. Mahy\inst{\ref{ROB}}}
    \and
      {A.~F.~McLeod\inst{\ref{durham1},\ref{durham2}}} 
    \and 
      {A. Mehner\inst{\ref{ESO}}}
    \and 
      {N. Morrell\inst{\ref{CA}}}
    \and 
      {T. Shenar\inst{\ref{ISR}}}
    \and
      {O. G. Telford\inst{\ref{Princeton}, \ref{Carnegie}}}
    \and 
      {J. Th. van Loon\inst{\ref{Keele}}}
    \and 
      {F. Tramper\inst{\ref{CSIC}}}
     \and 
      {A. Wofford\inst{\ref{IA}}}
}

\institute{
    {Institut f{\"u}r Physik und Astronomie, Universit{\"a}t Potsdam, Karl-Liebknecht-Str. 24/25, 14476 Potsdam, Germany\label{UP}}
    \and
    {Zentrum für Astronomie der Universität Heidelberg, Astronomisches Rechen-Institut, Mönchhofstr.\ 12-14, 69120 Heidelberg, Germany\label{ARI}}
    \and
    {Armagh Observatory and Planetarium, College Hill, BT61 9DG Armagh, Northern Ireland \label{AOP}}
    \and
    {Department of Physics \& Astronomy, University of Sheffield, Hicks Building, Hounsfield Road, Sheffield S3 7RH, UK\label{UoS}}
    \and
    {Dpto. de Astrof{\'i}sica, Universidad de La Laguna, 38205 La Laguna, Tenerife, Spain\label{Laguna}}
    \and
    {Instituto de Astrof{\'i}sica de Canarias, 38200 La Laguna, Tenerife, Spain \label{IAC}}
    \and 
    {Penn State Scranton, 120 Ridge View Drive, Dunmore, PA 18512, USA\label{PSU}}
    \and
    {Astronomický ústav, Akademie věd České republiky, Fričova 298, 251 65 Ondřejov, Czech Republic\label{ASU}}
    \and
    {Departamento de Ciencias, Facultad de Artes Liberales, Universidad Adolfo Ib\'a\~nez, Vi\~na del Mar, Chile\label{UAIVina}}
    \and
    {Instituto de Astrof\'isica, Facultad de F\'isica, Pontificia Universidad Cat\'olica de Chile, 782-0436 Santiago, Chile\label{PUC}}
    \and 
    {Instituto de Astrof\'isica de Andalucía, Glorieta de la Astronom\'ia s/n, E-18008 Granada, Spain\label{IAA}}
    \and 
    {Faculty of Physics, University of Duisburg-Essen, Lotharstra{\ss}e 1, D-47057 Duisburg, Germany\label{DE}}
    \and
    {Space Telescope Science Institute, 3700 San Martin Drive, Baltimore, MD 21218, USA\label{stsi}}
    \and
    {Royal Observatory of Belgium, Avenue Circulaire/Ringlaan 3, 1180 Brussels, Belgium\label{ROB}}
    \and
    {Centre for Extragalactic Astronomy, Department of Physics, Durham University, South Road, Durham DH1 3LE, UK\label{durham1}}
    \and
    {Institute for Computational Cosmology, Department of Physics, University of Durham, South Road, Durham DH1 3LE, UK\label{durham2}}
    \and
    {European Organisation for Astronomical Research in the Southern Hemisphere, Alonso de Cordova 3107, Vitacura, Santiago de Chile, Chile\label{ESO}}
    \and
    {The School of Physics and Astronomy, Tel Aviv University, Tel Aviv 6997801, Israel\label{ISR}}
    \and
    {Carnegie Observatories, Las Campanas Observatory, Casilla 601, La Serena, Chile\label{CA}}
    \and
    {Department of Astrophysical Sciences, Princeton University, 4 Ivy Lane, Princeton, NJ 08544, USA \label{Princeton}} 
    \and
    {The Observatories of the Carnegie Institution for Science, 813 Santa Barbara Street, Pasadena, CA 91101, USA \label{Carnegie}}
    \and
    {Lennard-Jones Laboratories, Keele University, ST5 5BG, UK\label{Keele}}
    \and
    {Centro de Astrobiolog{\'i}a, CSIC-INTA, Carretera de Ajalvir km 4, E-28850 Torrej{\'o}n de Ardoz, Madrid, Spain\label{CSIC}}
    \and
    {Instituto de Astronom{\'i}a, UNAM, Unidad Acad{\'e}mica en Ensenada, Km 103 Carr. Tijuana–Ensenada, Ensenada, BC 22860, Mexico\label{IA}}
}
   \date{Received XXX; accepted YYY}
 
  \abstract
  {
   Massive stars lead the ionization and mechanical feedback within young star-forming regions.
   The Large Magellanic Cloud (LMC) is an ideal galaxy for studying individual massive stars and quantifying their feedback contribution to the environment.
   We analyze eight exemplary targets in LMC N11\,B from the Hubble UV Legacy Library of Young Stars as Essential Standards (ULLYSES) program, using novel spectra from \hst\ (\cosi\ and \stis) in the UV, and from VLT (\xshooter) in the optical.
   We model the spectra of early to late O-type stars by using state-of-the-art PoWR atmosphere models.
   We determine the stellar and wind parameters (e.g., \tstar, \logg, $L_{\star}$, $\dot{M}$, ${\varv}_{\infty}$) of the analyzed objects, chemical abundances (C, N, O), ionizing and mechanical feedback (\QH, \QHeI, \QHeII, \lmec) and X-rays.
   We report ages of $2-4.5$~Myr and masses of $30-60$~\Msun\ for the analyzed stars in N11\,B, consistent with a scenario of sequential star formation.
   We note that the observed wind-momentum luminosity relation is consistent with theoretical predictions.
   We detect nitrogen enrichment in most of the stars, up to a factor of seven.
   However, we do not find a correlation between nitrogen enrichment and projected rotational velocity.
   Finally, based on their spectral type, we estimate the total ionizing photons injected from the O-type stars in N11\,B into its environment.
   We report $\log$ ($\sum$\,\QH)$=50.5$~ph\,s$^{-1}$, $\log$ ($\sum$\,\QHeI)$=49.6$~ph\,s$^{-1}$ and $\log$ ($\sum$\,\QHeII)$=44.4$~ph\,s$^{-1}$, consistent with the total ionizing budget in N11.
   }


   \maketitle

%-------------------------------------------------------------
\section{Introduction}
\label{Introduction}

Massive stars, defined as those with initial masses ($M_\mathrm{i}$)
on the main sequence $\geq8$~M$_{\odot}$, are key to understanding multiple astrophysical phenomena: from fundamental nucleosynthesis processes (e.g., CNO cycle and He-burning) to interstellar medium (ISM) feedback; such as the injection of mechanical energy, ionizing photons, and fresh elements into their local environments.
Thus, massive stars are expected to have a significant impact on the evolution of their host galaxies \citep[see][and references therein]{Massey2003}.
However, understanding the individual contribution of massive stars to the total feedback of their local environments is still ongoing, and proper quantification is still needed.
In order to address this issue, a comprehensive analysis of the stellar and wind parameters, as well as chemical abundances of these objects, must include not only the optical wavelengths, but specially the ultra-violet (UV) range of the spectrum.
It is known that most of the luminosity and crucial wind features of massive stars are precisely in the UV wavelengths.
The analysis of multi-wavelength spectral observations, complemented with precise photometry, together with state-of-the-art stellar atmosphere models, is critical for establishing the feedback of individual massive stars and investigating the ecology of their environments, from local to larger scales.

Stars even more massive ($M_\mathrm{i}\geq20-25$~M$_{\odot}$), therefore hotter, more luminous, and younger on the main sequence -- the O-type stars -- are particularly important \citep[see e.g.,][]{Langer2012} as they lead to some of the most exotic phenomena in their last stages of evolution, such as:
the classical Wolf-Rayet (WR) stars \citep[][]{Crowther2007}, core-collapse supernovae (SNe) of type Ibc \citep[][]{Woosley2005}, long-duration gamma-ray bursts (GRB) \citep[][]{Woosley2006}, compact objects such as neutron stars and black holes \citep[][]{Heger2003}, and gravitational-wave sources \citep[][]{Abbott2017}.
In order to understand the role that massive stars played in the first galaxies, now being observed at higher redshifts with present-generation facilities such as the James Webb Space Telescope (JWST) \citep[e.g.,][]{Arellano2022}, we can use lower-than-solar metallicity environments as local proxy scenarios to study the astrophysical processes and feedback mechanisms occurring in earlier stages of the Universe, after the reionization epoch to the present \citep[][]{Wofford2021, Eldridge2022}.

The Large Magellanic Cloud (LMC) is known to have an average metallicity of around half-solar \citep[$\sim$0.5~\zsol;][]{Larsen2000,Hunter2007}, with no apparent metallicity gradient \citep[e.g.,][]{Dominguez2022}.
Although its metal deficiency is modest with respect to extreme metal-poor galaxies \citep[e.g., I Zw\,18; 12$+\log$(O/H)$\leq7.2$;][at 13.4~Mpc]{Izotov2019}, its nearby distance \citep[DM=18.5~mag (50~kpc);][]{Pietrzynski2013} and its relatively low reddening \citep[\ebv$=0.05$~mag;][]{Larsen2000} make it an ideal place to study \textit{individual} massive stars in great detail.
Certainly, the Small Magellanic Cloud (SMC) has an even lower metallicity ($\sim$0.2~\zsun),
and individual stars are being subject of recent studies \citep[e.g.,][]{Pauli2023}.
However, a comprehensive understanding of massive stars at any metallicity must include the O-type stars in the LMC as a reference frame for comparative analysis. Additionally, there are not many "nearby-enough" systems containing massive stars with the adequate observations (e.g., dedicated spectra with the proper signal-to-noise ratios (SNR), spatial resolution and resolving power, wavelength coverage and precise photometry) for the proper characterization of their main physical parameters with modern stellar atmosphere models.

\begin{figure*}
\begin{centering}
\includegraphics[width=0.53\linewidth]{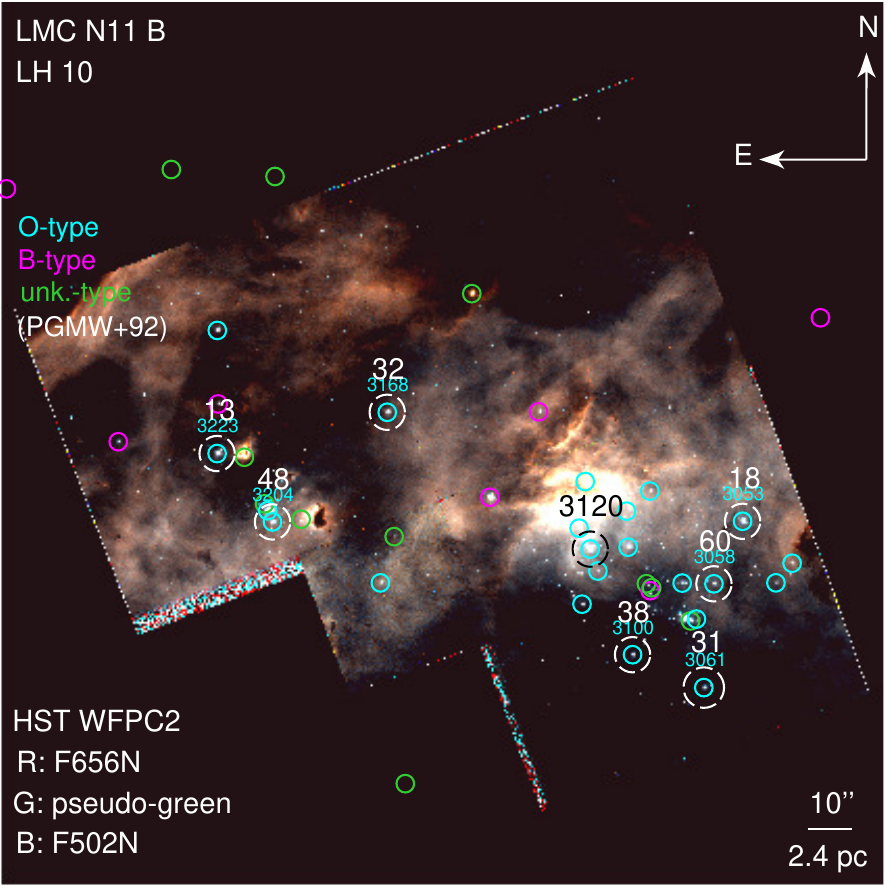}~
\par\end{centering}
\caption{
Massive stars in the N11\,B star forming region in the LMC.
The colour-composite image covers most of the stellar complex, and is formed using \hst/WFPC2 filters in F656N (\ha), pseudo-green, and F502N (\oiiin) as red, green, and blue components, respectively.
We indicate the location of the complete sample of stars
from \citet[][]{Parker1992}, identified as O-type stars (cyan), B-type (magenta),
and unclassified objects (green).
The eight ULLYSES targets studied here are identified by dashed circles with their
respective ID numbers from \citet[][]{Parker1992} and \citet[][]{Evans2006}.
Scale and orientation are indicated.
As a reference, LH\,9 \citep[][]{Lucke1970} at the center of N11, is located to the South.
}
\label{fig:n11b-hst}
\end{figure*}

In the LMC, N11 is the second-brightest star-forming region, just after the well-known 30 Doradus (30 Dor) \citep[see][]{Pellegrini2012}.
30 Dor contains a rich population of massive stars and has been the subject of numerous studies \citep[see e.g.,][]{Evans2011,Sana2022,Crowther2024}.
Here we focus our study in N11\,B \citep[][]{Henize1956}, a.k.a. LH\,10, the brightest \hii\ region of N11. N11\,B is located in the periphery of LH\,9 \citep[][]{Lucke1970}, a gas-depleted cavity of $\sim$100~pc in diameter at the center of N11.
N11\,B contains a rich population of massive stars, with at least 25 known O-type stars, and 9 B-type stars, from a catalog complete at V$=16$~mag \citep[][]{Parker1992}.
General parameters of the region are provided in Tab.~\ref{tab:parameters}.
In Fig.~\ref{fig:n11b-hst}, we show the known blue and bright stars located in N11\,B.
Additionally, there is only one SN remnant observed by XMM Newton close to N11. However, given its projected distance from N11\,B, it is unlikely to be associated with this star-forming region, where no SNe have been reported so far, neither any resolved (point-like) source of X-ray emission.
Given the brightness of this region, N11 has been the subject of multiple studies. \citet[][]{Evans2006} for example, used optical spectra (3850--6700~\AA) to determine spectral types and velocities for a sample of 124 objects, out of which 44 were classified as O-type stars.
Based on its spectral type and metallicity, \citet[][in their Fig. 12]{Evans2006} adopted a temperature for the stars, and the luminosities were determined based on their colours and distances.
Later, \citet[][]{Mokiem2007} analyzed 6~O-type stars in N11\,B from \citet[][]{Evans2006} (IDs: N11\,31, 32, 38, 48 and 60), excluding the known binaries (see Fig.~\ref{fig:n11b-hst}).
These studies were done in the optical range.

\begin{table}
\begin{center}
  \caption{Parameters of the star-forming region N11\,B in the LMC.}
  \label{tab:parameters}
\begin{tabular}{ccc}
\hline\hline
\rule{0cm}{2.2ex} Parameter & Value          & Ref.\\
\hline
Cross-identifications  & LH\,10, NGC\,1763   & (a, b) \\
R.A (J2000)            & 04h56m49.2s         & (b) \\
Dec. (J2000)           & $-$66d24m32.7s      & (b) \\
Distance modulus       & 18.5~mag (50~kpc)   & (c) \\
Metallicity            & 0.5~Z$_{\odot}$     & (d) \\
12$+\log$(O/H)         & 8.39$\pm$0.02~dex   & (e) \\
\ebv                   & 0.04~mag            & (f) \\
N$_\text{OB}$$^{\dagger}$ ($V<16$~mag) & 25 O-type; 9 B-type & (f) \\
\hline
\end{tabular}
\end{center}
(a) \citet[][]{Lucke1970};
(b) NASA/IPAC Extragalactic Database (NED);
(c) \citet[][]{Pietrzynski2013};
(d) \citet[][]{Hunter2007};
(e) \citet[][]{Dominguez2022};
(f) \citet[][]{Parker1992}.
(${\dagger}$) Number of OB stars: O-type, and B-type stars, complete at $V<16$~mag in N11\,B \citep[][]{Parker1992}.
\end{table}

UV spectra of massive stars are \emph{essential} for studying their wind parameters.
These include not only the wind mass-loss rate and terminal wind velocity, but also chemical abundances and the presence (or absence) of X-rays.
It is for this reason that the Hubble Space Telescope's (\hst) Ultraviolet Legacy Library of Young Stars as Essential Standards (ULLYSES) program \citep[][]{Roman-Duval2020} dedicated 1000~\hst\ orbits in order to construct a UV spectroscopic library of young high- and low-mass stars in the local universe.
Eight O-type stars of N11\,B are ULLYSES targets \citep[see][]{Vink2023}. In addition, these objects also include optical spectra, obtained from Very Large Telescope (VLT) \xshooter\ and \giraffe\ instruments.
On this basis, by using dedicated multi-wavelength spectra, as well as photometry in the UV, optical (including recent Gaia DR3), and near-infrared (NIR), we perform a detailed study of a sample of massive stars in the star-forming complex N11\,B.

We use novel observations of eight O-type stars in N11\,B to model their UV and optical spectra by using state-of-the-art atmosphere models.
Potsdam Wolf-Rayet (PoWR) atmosphere models \citep[][]{Grafener2002,Hamann2003,Sander2015} have proven to be ideal for analyzing massive stars \citep[e.g.,][]{Ramachandran2018,Ramachandran2019,Ramachandran2021,Pauli2023}.
We use these models to determine their main stellar parameters, such as the temperature of the star (\tstar), surface gravity (\logg), and luminosity (\lstar).
We also determine wind parameters like the mass-loss rate ($\dot{M}$), terminal wind velocity (\vinf) and wind-momentum luminosity (\Dmom), as well as their chemical abundances (C, N, O), mechanical luminosity (\lmec), ionizing photons (\QH, \QHeI, and \QHeII), and X-rays.
With additional standard tools, such as {\it iacob-broad} \citep[][]{SimonDiaz2014}, we also determine projected rotational velocities (\vrot) as well as non-rotational broadening.
With BONNSAI \citep[][]{Schneider2014} we report the predicted ages and evolutionary masses of the stars using stellar models from \citet[][]{Brott2011,Kohler2015}.
A comprehensive study of the massive stars in N11\,B, using the above-mentioned multi-wavelength observations and state-of-the-art analysis tools, is crucial for determining their physical parameters, quantifying their feedback contribution, and investigating the ecology of this environment.

This article is structured as follows: in Sec.~\ref{Spectroscopic data} we describe the spectroscopic observations of our stars;
in Sec.~\ref{Analysis} we model the observed multi-wavelength spectra with PoWR models and determine physical parameters;
our results are discussed in Sec.~\ref{discussion}. Finally, a summary and our conclusions are given in Sec.~\ref{conclusions}.

\section{Spectroscopic data}
\label{Spectroscopic data}

\begin{table*}
\begin{center}
  \caption{Sample of ULLYSES targets. O-type stars in the N11\,B star-forming region in the LMC.}
  \label{tab:sample_n11b}
\begin{tabular}{cccccccccccc}
\hline\hline
\rule{0cm}{2.2ex}$\#$ & ID & Cross ID   & \multicolumn{2}{c}{Coordinates (J2000)$^{a}$} & \Mv\     & \ebv\ & SpC  & $\varv_{r}$  & Binary \s \\
     &          &      & R.\,A.      & Dec.          & [mag]  & [mag]&                  & [\kms]$^{c}$  &  status    \\
(1)  & (2)      & (3)  & (4)         & (5)           & (6)    & (7)  & (8)              & (9) & (10)                  \\
\hline
1& PGMW\,3053& N11\,18 & 04:56:41.05 & --66:24:40.54 & $-5.8$ & 0.15 & O6.5\,II(f)$^{c}$   & 301      & single$^{\dagger}$\\
2& PGMW\,3058& N11\,60 & 04:56:42.15 & --66:24:54.61 & $-5.0$ & 0.19 & O3\,V((f*))$^{c}$  & 314      & single$^{\dagger}$ \\
3& PGMW\,3061& N11\,31 & 04:56:42.51 & --66:25:18.22 & $-5.7$ & 0.22 & ON2\,III(f*)$^{d}$ & 322      & single$^{\dagger}$\\
4& PGMW\,3100& N11\,38 & 04:56:45.20 & --66:25:10.78 & $-5.7$ & 0.26 & O5\,III(f)$^{c}$  & 318      & SB1 \\
5&PGMW\,3120a&         & 04:56:46.80 & --66:24:46.86 & $-6.7$ & 0.21 & O5.5\,V((f))$^{c}$ & 300      & cluster$^{\dagger\dagger}$ \\
6& PGMW\,3168& N11\,32 & 04:56:54.46 & --66:24:15.87 & $-5.3$ & 0.16 & O7.5\,III(f)$^{c}$  & 305      & single$^{\dagger}$\\
7& PGMW\,3204& N11\,48 &04:56:58.79$^{b}$&--66:24:40.71$^{b}$&$-5.0$&0.16& O6\,Vz((f))$^{c}$ & 299  & single$^{\dagger}$ \\
8& PGMW\,3223& N11\,13 & 04:57:00.88 & --66:24:25.21 & $-6.2$ & 0.19 & O8\,Vz$^{c}$        & \dots    & SB1\\
\hline
\end{tabular}
\end{center}
(1) Object $\#$; (2) PGMW\,$\#$ identification (ID) from \citet[][]{Parker1992}; (3) N11\,$\#$ ID from \citet[][]{Evans2006}; (4) right ascension (J2000); (5) declination (J2000); (6) absolute visual magnitude (this work); (7) reddening (this work); (8) spectral type classification; (9) radial velocity \citep[][]{Evans2006};
(10) binarity status (this work).
$^{(a)}$\citet[][]{Gaia2020};
$^{(b)}$\citet[][]{Bonanos2009};
$^{(c)}$ PAC priv. comm.;
$^{(d)}$\citet[][]{Evans2006};
%$^{(d)}$\citet[][]{Parker1992}.
$^{(\dagger)}$ Single here means: \emph{without evidence} of binarity.
SB1 status was determined here based on \giraffe\ multi-epoch spectra (see text for details).
$^{(\dagger\dagger)}$ Note: PGMW\,3120a is one of the three stars of around the same brightness clustering (with PGMW\,3120b,c) in the source known as PGMW\,3120 (see Fig.~\ref{fig:hst_zoom} and text for details).
\end{table*}

Our sample of O-type stars in N11\,B are ULLYSES targets, and their spectra are publicly available to the scientific community\footnote{\url{https://ullyses.stsci.edu/ullyses-targets-lmc.html}}.
These targets are: 1) PGMW\,3053, 2) PGMW\,3058, 3) PGMW\,3061, 4) PGMW\,3100, 5) PGMW\,3120, 6) PGMW\,3168, 7) PGMW\,3204, and 8) PGMW\,3223.
In Table~\ref{tab:sample_n11b}, we list the general information for each star, including magnitudes and extinction from this study, their known spectral types, radial velocities, and binarity status.
These objects have been observed with \hst, either with the Space Telescope Imaging Spectrograph (\stis) \citep[][]{Woodgate1998} or the Cosmic Origins Spectrograph (\cosi) \citep[][]{Green2012}.
In the Appendix (see Table~\ref{tab:spec_obs}), we list the spectroscopic observations used in this work, obtained from different instruments, including relevant information on the covered spectral range, spectral resolution, observation dates, exposure times, and SNR.

The ULLYSES targets have complementary observations in the optical range with the \xshooter\ instrument \citep[][]{Vernet2011} on the VLT, which are key for our analysis.
\xshooter\ covers the UV-Blue (UVB: $3000-5595$~\AA), visible (VIS: $5595-10240$~\AA) and Near-IR (NIR: $10240-24800$~\AA).
NIR spectra are not used in this work.
We use \xshooter\ final products, which are flux- and wavelength-calibrated spectra, provided by the data reduction group of the XShootU collaboration \citep[see][for details]{Sana2024}.
Except for PGMW\,3061 and PGMW\,3204, which final products are not yet delivered.
For these two targets, we retrieved the spectra from the ESO Science Archive Facility\footnote{\url{http://archive.eso.org/cms.html}}.
Three objects, PGMW\,3053, PGMW\,3120, and PGMW\,3223, have spectra from the Far Ultraviolet Spectroscopic Explorer (\fuse) satellite \citep[][]{Moos2000,Sahnow2000} with medium resolution (MDRS) in the UV range of $905-1187$~\AA.
We use this information to complement our analysis.
Additionally, seven targets have \giraffe\ multi-epoch spectra, except for PGMW\,3120.
\giraffe\ is a medium-high resolution (R$=5500-65000$) spectrograph in the optical range of $3700-9000$~\AA\ \citep[][]{Pasquini2002}.
We used these observations to check for binarity in Sec.~\ref{binarity}.

In Fig.~\ref{fig:n11b-hst}, covering most of N11\,B, we indicate the brightest objects, complete at $V<16$~mag from \citet[][]{Parker1992}, which are identified as O-type stars, B-type stars, and unclassified objects. The ULLYSES targets studied here are indicated.

\section{Analysis}
\label{Analysis}

In order to model the observed spectra of the stars, one needs to follow a sequence of steps: 1) determine the luminosity and extinction of the object, prioritizing recent Gaia DR3 photometry; 2) check for binarity, using available multi-epoch spectra; and 3) determine rotational and "macroturbulent" velocities, before fitting the spectral line widths. Once these steps are completed, step 4) is to model the spectra to obtain the physical parameters of the stars. This process is done iteratively. Final stages include the determination of chemical abundances, X-ray luminosities, and refining the entire set of parameters \emph{comprehensively}.
Next we describe these different parts of our analysis.

\subsection{Bolometric luminosity and extinction}
\label{Photometry}

\begin{figure*}
\begin{centering}
\includegraphics[width=0.23\linewidth]{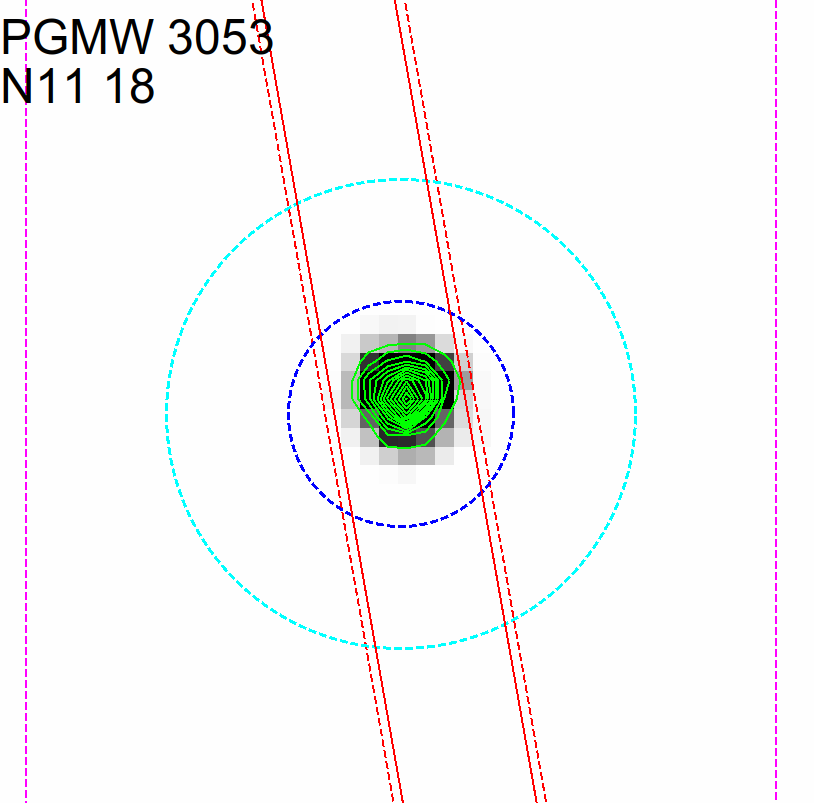}~
\includegraphics[width=0.23\linewidth]{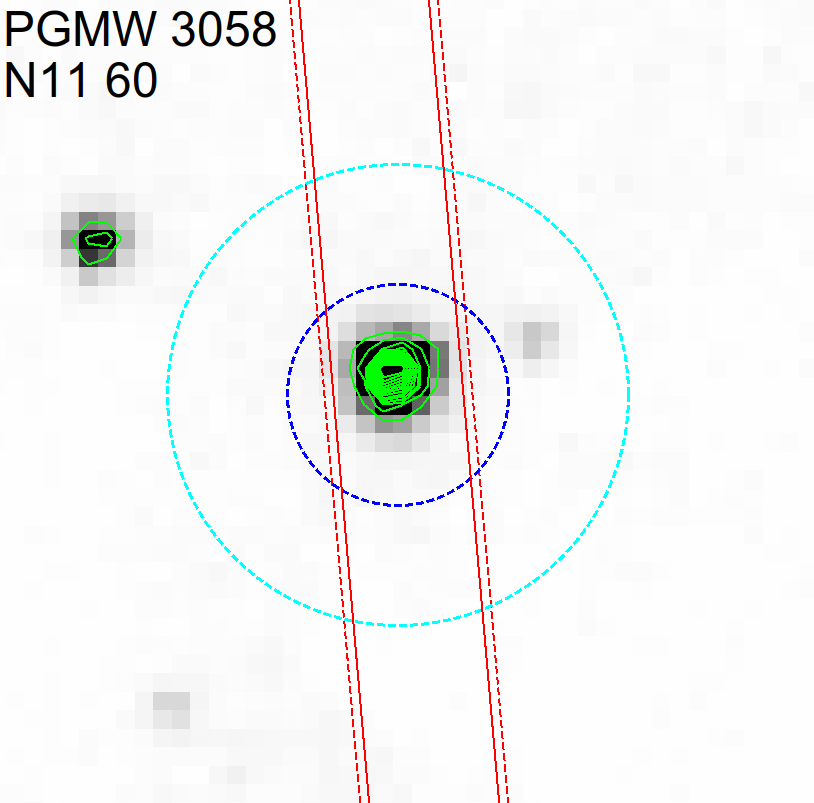}~
\includegraphics[width=0.23\linewidth]{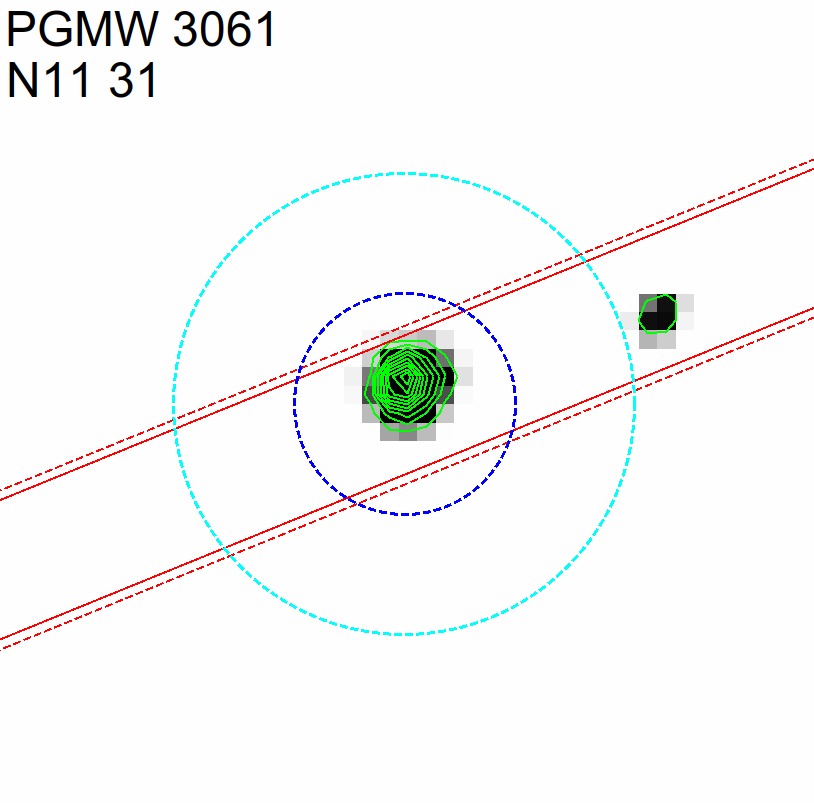}~
\includegraphics[width=0.23\linewidth]{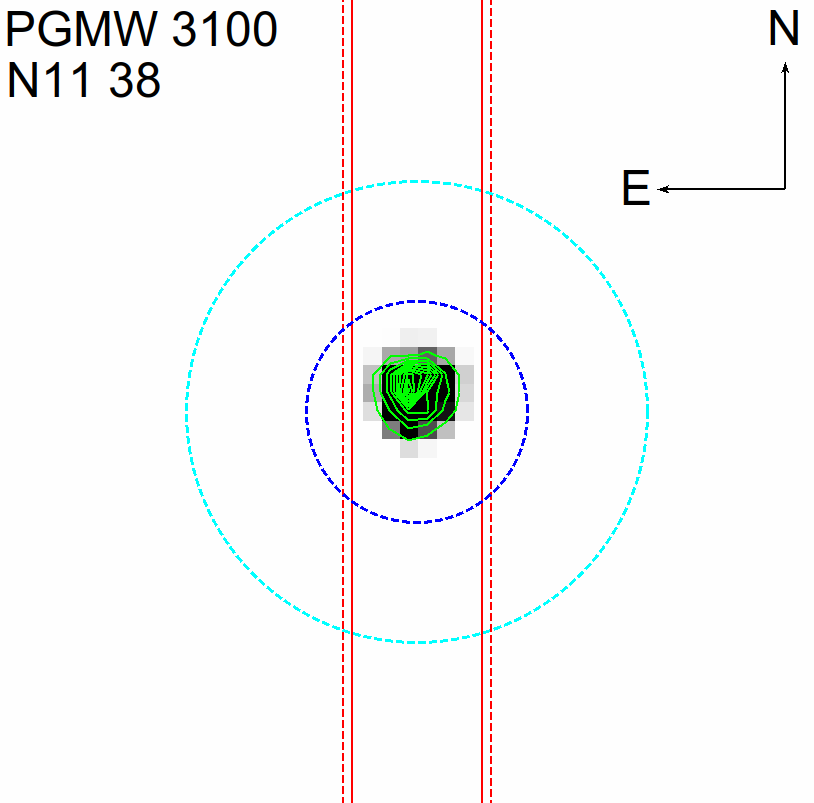}\\
\includegraphics[width=0.23\linewidth]{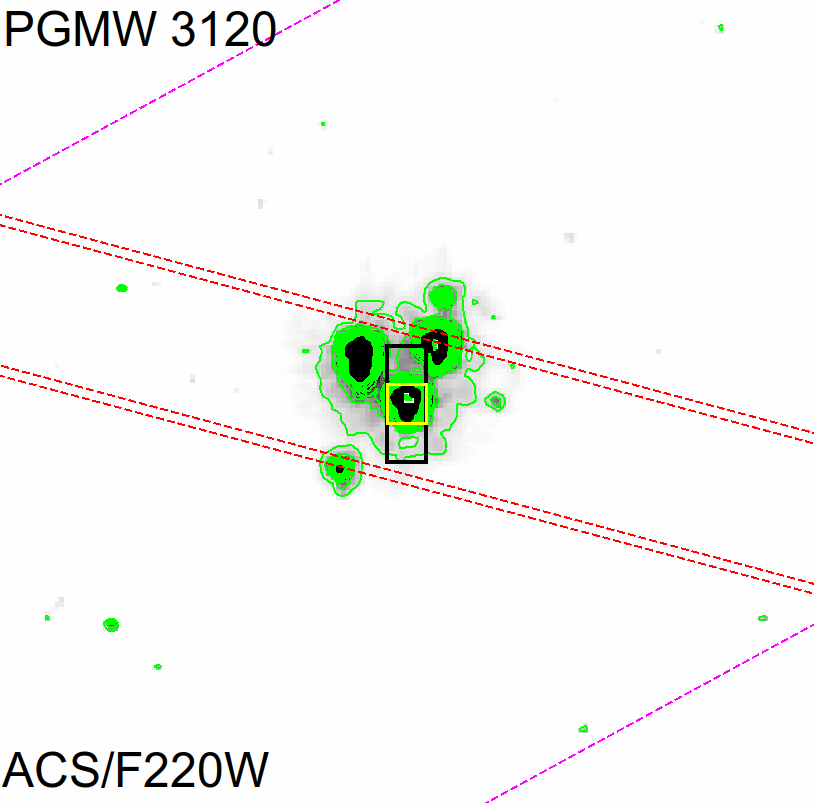}~
\includegraphics[width=0.23\linewidth]{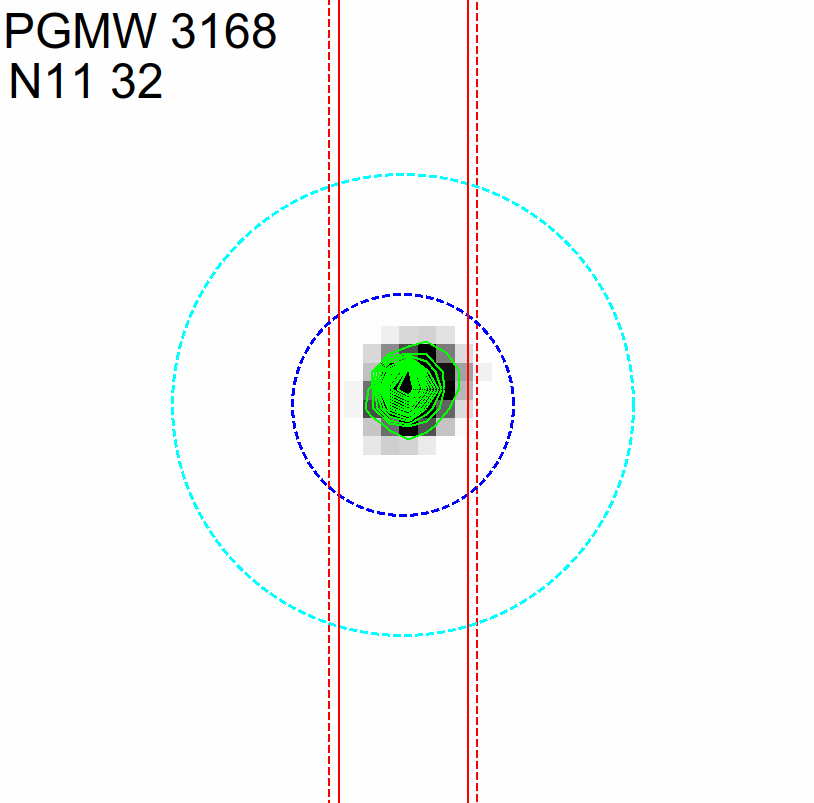}~
\includegraphics[width=0.23\linewidth]{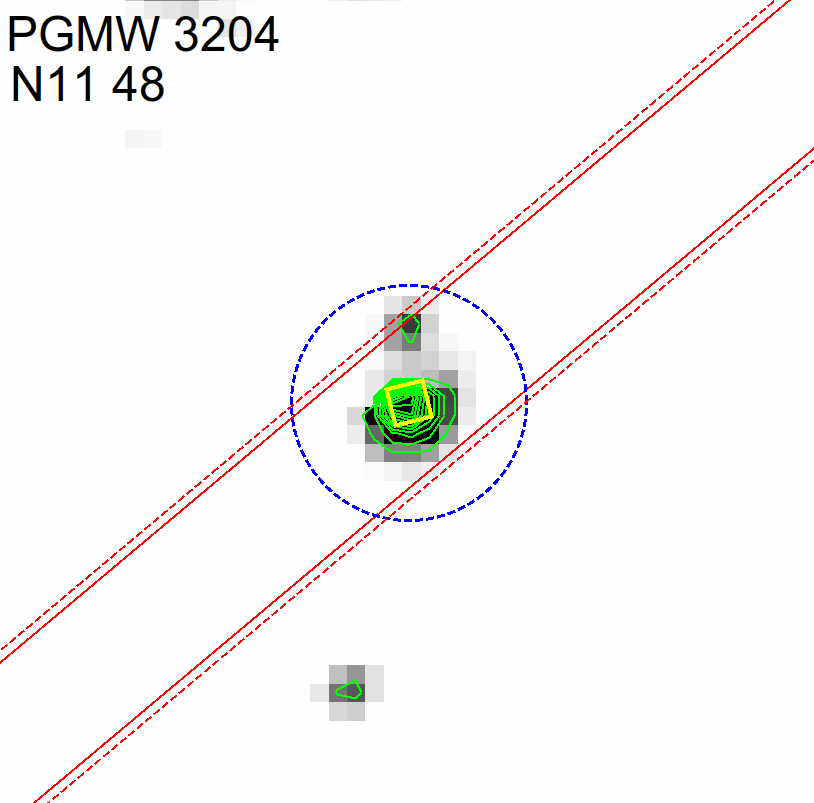}~
\includegraphics[width=0.23\linewidth]{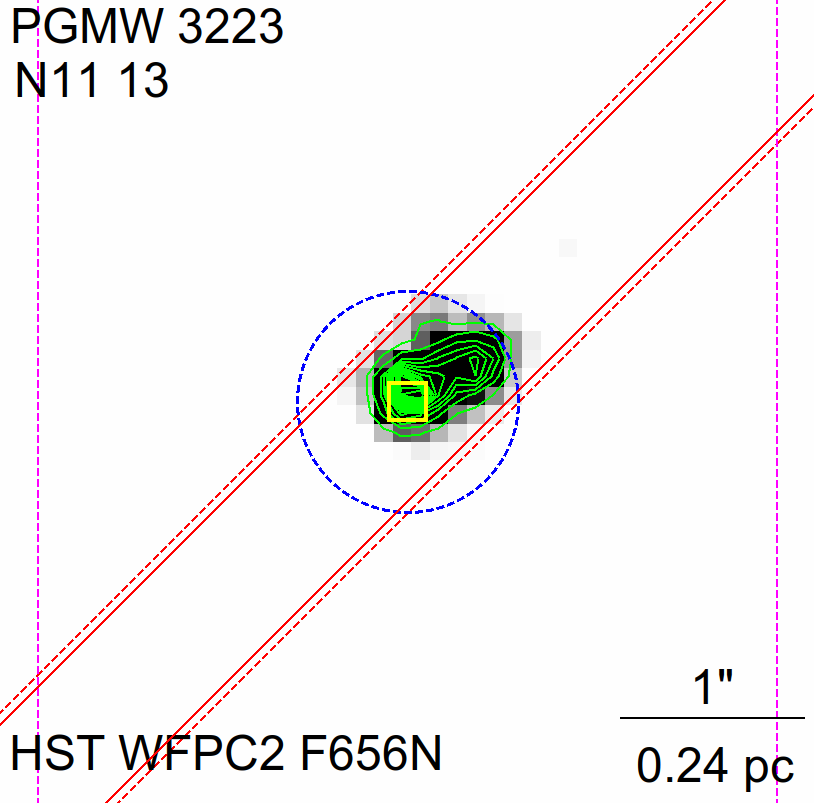}\\
\par\end{centering}
\caption{
Images of the LMC N11\,B O-type stars, from \hst/WFPC2 in the F656N filter (\ha), except for PGMW\,3120 (\hst/ACS/F220W).
The ULLYSES targets are identified with their
respective IDs from \citet[][PGMW\,$\#$]{Parker1992} and \citet[][N11\,$\#$]{Evans2006}.
The slits from different instruments are indicated, including their shapes, sizes and position angles: \fuse\ FUV/MRDS (dashed magenta; $4\times20$~arcsec); \hst\ \cosi\ (dashed cyan; D2.5~arcsec); \hst\ \stis/E140M (black rectangle; $0.2\times0.6$~arcsec); \stis/E230M (yellow square; $0.2\times0.2$~arcsec); \giraffe\ (dashed blue; D1.2~arcsec); \xshooter\ UVB (dashed red; $0.8\times11$~arcsec); and \xshooter\ VIS (red; $0.7\times11$~arcsec).
Isocontours of brightness are displayed (green) to check for multiple sources inside the slits.
Scale and orientation are indicated.
}
\label{fig:hst_zoom}
\end{figure*}

To determine the luminosity of our stars,
we need to match the continuum of the synthetic spectrum of a selected model with the available photometry on the spectral energy distribution (SED).
Reliable photometric values in the UV and optical ranges were taken from multiple references in the literature, from the U, B, V and R filters. Moreover, Gaia DR2 and DR3 values were included. In the NIR we use the J, H, and K bands. In Table~\ref{tab:photometry} we list the photometric values from different references used in this work to construct the SEDs of the analyzed stars in N11\,B.
Although we make use of most of the values available from the literature, from UV up to the NIR, those values coming from Gaia \citep[][]{Gaia2018,Gaia2023} are prioritized in the optical range.
In the UV range, the calibrated fluxes from \hst\ \cosi\ and \stis\ can help as a photometric reference.
Additionally, a careful inspection of each object was conducted using \hst/WFPC2 images in filters F656N and F502N, to ensure that we are considering the flux of a single object at the spatial resolution of \hst\ (with a pixel scale of 0.05~arcsec/pixel).
However, we did not use these filters to obtain extra photometric values, since these bands correspond to \ha\ (F656N) and \oiiin\ (F502N), which are impacted by the nebular environment of the stars.

We note that in the case of the spectra obtained with \xshooter,
using slits of 0.7 and 0.8~arcsec $\times~11$~arcsec
in the UVB and VIS ranges, respectively, we are observing single objects for PGMW\,3053, PGMW\,3061, PGMW\,3168, PGMW\,3100, PGMW\,3204, and PGMW\,3058.
This is clearly not the case for PGMW\,3223, where at least a second object as bright as our target star is inside the apertures of \xshooter, and a third one (although dimmer) is inside the aperture of \fuse. Thus, care must be taken when interpreting these values in the SED. The same case for PGMW\,3120, which has three objects inside the \xshooter\ aperture, and two more (though dimmer) inside FUSE. 
On the other hand, \hst\ \cosi\ and \stis\ apertures only cover the star of interest, and for this reason their fluxes can be considered reliable references to construct their SEDs. A zoom view of 1~arcsec (0.24~pc at the distance of N11) in \hst/WFPC2 images for the sample stars with their respective apertures from different instruments is shown in Fig.~\ref{fig:hst_zoom}. Isocontours of brightness are used to inspect for probable contamination, like other spurious sources inside the slits covering our targets.

Once these aspects were taken into consideration, the construction process of the SED is as follows: first, the flux-continuum of the model is scaled to the distance of the object. We assumed the same distance for all the objects, which is that of the LMC.
Next, the reddening is determined by matching the slope of the SED, which is particularly sensitive towards the bluer wavelengths. We take into account the extinction attributed to the Galactic foreground (\ebv$=0.04$ mag) and the LMC-law. For this purpose, we used the reddening laws by \citet[][]{Seaton1979} and \citet[][for the LMC]{Trundle2007}. The values we obtain are listed in Table~\ref{tab:sample_n11b} (columns 6 and 7).
These values are better constrained than previous results in the literature, as we also used the flux levels of novel spectra in the UV range to determine the extinction and the luminosity of the stars.
Detailed information of these differences is given in Sec.~\ref{comments} for each star.

\subsection{Checking binarity status}
\label{binarity}

When studying massive stars, one has to consider two possible scenarios for their evolution: the single \citep[][]{Conti1983} and the binary pathway \citep[e.g.,][]{Vanbeveren1997}.
Since most of the massive stars are expected to be in a binary or multiple star system \citep[][]{Sana2012}, it is important to verify any indication of a binary companion before determining the stellar parameters as if we were dealing with a presumably single object. In order to check for binarity, multi-epoch spectra are required.

Our targets have multi-epoch observations from \giraffe.
Although planned for other scientific cases, they allow us to check whether there is evidence of binarity, at least within the observed time frames, with the same resolution, exposure time, and SNR.
The observed time periods span from 1~day to 1~month.
However, the available observations are complex, and they need careful treatment for proper interpretation.
For instance, the multi-epoch dates are different for each spectral range; e.g., it is of 45~days for the spectral range of $3850-4050$~\AA; 24~h for $4030-4200$~\AA; the same day for $4180-4400$~\AA; and same day for $4340-4340$~\AA, however different day from previous spectral range; 1~day for $4540-4760$~\AA\ and 31~days for $6300-6690$~\AA.
Details are provided in Table~\ref{tab:spec_obs} at the Appendix.
With this information, we checked for binary features with orbital periods matching the duration of the available observations.

Here we check for any evidence of binarity in our sample. However, it is known that a star cannot be proven to be single. Even a lack of evidence of binarity does not necessarily mean it is not a binary. What we mean here by evidence of binarity includes radial velocity shifts for SB1 types, and variations in the line profile, e.g., double lines, for SB2 types. Also, the presence of a companion does not necessarily dominate the key features of the spectrum of the star, this would depend on its physical parameters.
According to our analysis, PGMW\,3053 displays a variable \heiiwr\ profile within a 1-day time interval,
as previously noted by \citet[][]{Evans2006}. We attribute these features to wind variability, not to binarity.
Other \heii\ lines do not show variability.
PGMW\,3223 was found to be a binary. Its double line profile of \heiiwr\ could be interpreted as SB2 feature. However, no other \heii\ lines display this characteristic feature. We classified it as SB1. The variability in radial velocity in the features of PGMW\,3100 indicate a binary SB1.
The rest of the stars can be considered single, with the precaution mentioned above.
Table~\ref{tab:sample_n11b} (column 11) lists the binary status of our sample.

\subsection{Rotation}
\label{Rotation}

\begin{figure*}
\begin{centering}
\includegraphics[trim={0 9.7cm 0 1cm},clip,width=0.80\linewidth]{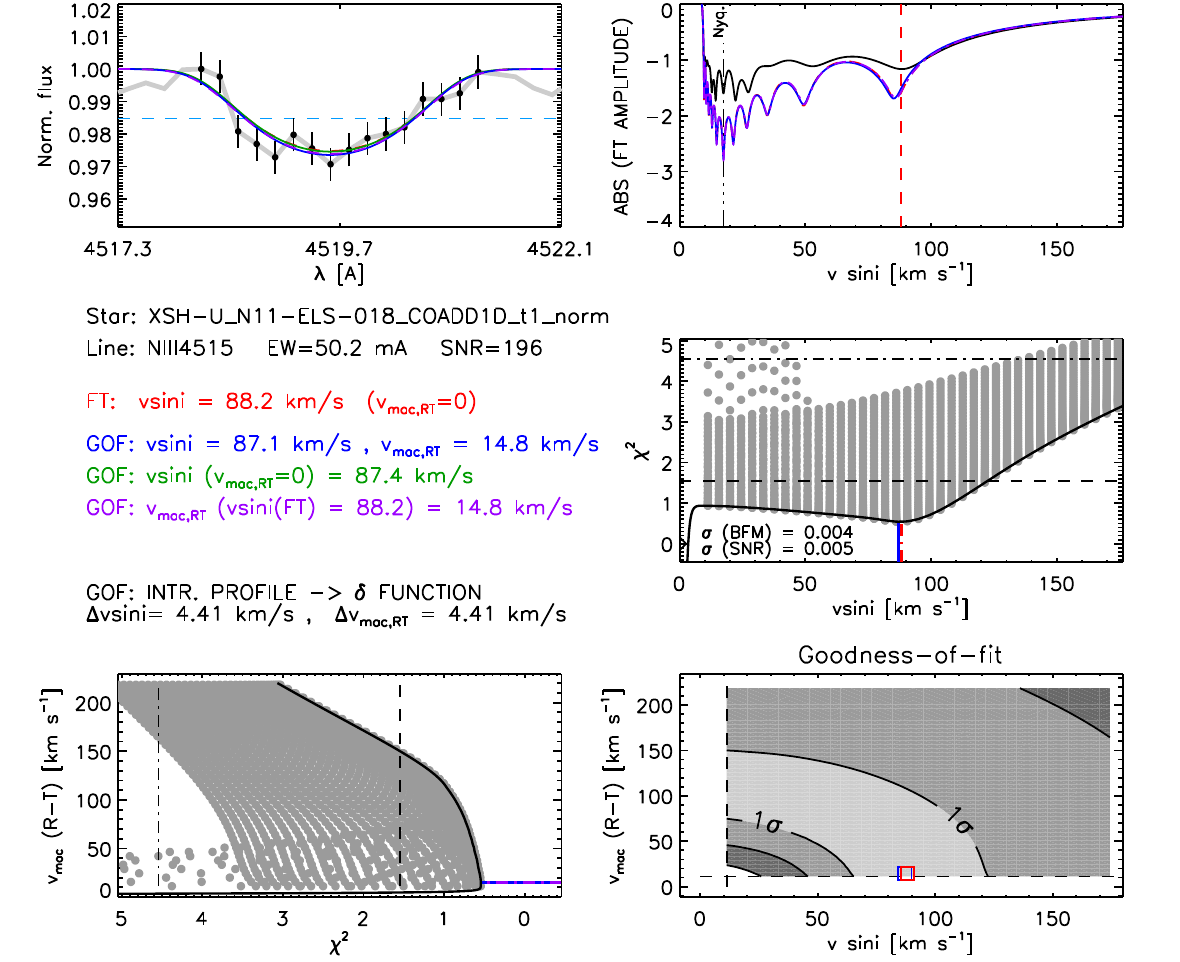}~
\par\end{centering}
\caption{
Graphical output from the {\it iacob-broad} semi-automatized tool to determine projected rotational velocity (\vrot) and non-rotational broadening. In this particular case for the star PGMW\,3053, using the
metal line of \niiia\ in absorption. Five panels are displayed:
(upper left) the line profile of the line;
(upper right) the Fourier transform (FT) of the line;
(lower right) 2D $\chi$-distributions resulting from Goodness-of-fit (GOF) analysis, and their projections (middle and lower left).
The fitting with colour red indicates the \vrot\ corresponding to the first zero of the FT;
(blue) the result for \vrot\ and "macroturbulent" velocity (\vmac) from GOF;
(green) the result for \vrot\ from the GOF, if \vmac\ is assumed to be zero;
(purple) the \vmac\ from the GOF assuming \vrot\ is the first zero of the FT.
The equivalent width (EW) and signal-to-noise ratio (SNR) measured in the neighboring continuum of the line are also indicated. In this case EW$=50$~m\AA\ and SNR$=196$.
Here we take the \vrot\ value determined with the FT, and the \vmac\ from the GOF, which fitting is shown in purple. Thus, for this star we report \vrot$=88\pm4$~\kms, and \vmac$=15\pm4$~\kms.
}
\label{fig:vrot_n11-18}
\end{figure*}

In order to determine the projected rotational velocity (\vrot) of the stars, one ought to measure the broadening of their photospheric lines in absorption.
Unlike the Balmer and He lines, the broadening of metal lines is normally attributed to macroturbulence rather than pressure.
However, metal lines are generally weaker than H and He lines, or even absent, when the SNR is not sufficient. Thus, dedicated observations should be analyzed when determining this parameter.
Here we use {\it iacob-broad} \citep[][]{SimonDiaz2014}, a standard semi-automatized tool to measure \vrot\ in OB stars \citep[e.g.,][]{Holgado2018,Berlanas2020}.
This tool employs the Fourier transform (FT) and goodness-of-fit (GOF) methods.
Basically, the model spectra are convolved to the instrumental resolution of the analyzed spectra, and the rotational velocities are obtained.
The values we report are the current rotational velocities at the equator, times the sine of the inclination towards the observer. The inclination is an unknown parameter with the available data.

As an example, in Fig.~\ref{fig:vrot_n11-18} we show the graphical output obtained from the {\it iacob-broad} tool for a particular metal line in absorption of the PGMW\,3053 star.
The plot is rich in information.
Basically, the line profile of the selected metal, in this case \niiia\ (in the upper left), is shown with different fittings from  the two different methods: 1) FT, and 2) GOF with their projections, for "macroturbulent" velocity (\vmac) to the left, and for the rotational component (\vrot) to the top of the plot.
There are four different fittings indicated with different colours: 1) red indicates the \vrot\ corresponding to the first zero of the FT; 2) blue for \vrot\ and \vmac\ obtained from GOF; 3) green indicates the \vrot\ from the GOF, if \vmac\ is assumed to be zero;
and 4), purple shows the fitting of the \vmac\ from the GOF, assuming \vrot\ is the first zero of the FT.
We select the values from the last assumption, given it provides the best fit of the observed line, considering both \vmac\ and \vrot\ contributions simultaneously.

We use exclusively metal lines in absorption to determine their \vrot\ broadening component.
The metal lines we use are \niiia\ (for PGMW\,3053 and PGMW\,3058).
When \niii\ lines are too weak or absent, then we check for more intense, and resolved lines, present in the spectrum of a particular star under analysis.
We use higher ionizing state N lines, though weaker, like \nva\ and \nvb, when \niii\ is not detected (e.g., in PGMW\,3061).
Other metal lines are also used as well.
We used \siiva\ (for PGMW\,3168) and \civa\ (for PGMW\,3100, PGMW\,3120 and PGMW\,3204).
When using \niii\ to determine rotation, we prioritize \niiia\ over \niiib, given the latter is usually weaker and more difficult to fit.
The same case applies for \siiv; \siiva\ is usually stronger than \siivb, and sometimes \siivb\ is even in emission. We do not use lines in emission to determine \vrot.
Carbon lines also help, in the case of \civ, \civa\ is prioritized over \civb, which is usually weaker.
Oxygen lines like \oiii\,$\lambda$5592.3 also helps (e.g., for PGMW\,3223). We report our results for \vrot\ in Table~\ref{tab:results}.
The {\it iacob-broad} tool provides computational errors for \vrot\ and \vmac\ of around $5\%$.
However, we assumed a value of $20\%$ to account for errors associated with the methodology.
This represents the lower error, and the upper error is consistent with the difference between \vrot\ (GOF) and \vrot\ (FT).
Detailed information, like EW and SNR, for the selected lines of each star is given in Sec.~\ref{comments}.

\begin{table}
\begin{center}
\caption[]{Broadening parameters of the O-type stars in N11\,B.}
\setlength{\tabcolsep}{0.9\tabcolsep}
\begin{tabular}{lcccc}
\hline\hline
\rule{0cm}{2.2ex} ID & \vrot\    & \vrot\ & \vmac\ & metal line \\
PGMW            & [\kms]         & [\kms] & [\kms] & used    \\ 
                & (previous)     & (here) & (here) & (here)  \\ 
(1)             & (2)            & (3)    & (4)    & (5) \\
\hline
3053 & \dots  & 88$^{+9}_{-18}$ &  15$\pm$4  & \niiia\ \\ 
3058 & 106/68 & 66$^{+25}_{-13}$ &  80$\pm$16 & \nva\   \\ 
3061 & 116    &120$^{+20}_{-24}$ &  80$\pm$16 & \nv\,$\lambda\lambda4604/20$ \\ 
3100 & 145    & 99$^{+30}_{-20}$ &  92$\pm$18 & \civa\  \\ 
3120a & \dots &143$^{+70}_{-29}$ & 150$\pm$30 & \civa\  \\
3168 & 96     & 55$^{+40}_{-11}$ &  89$\pm$18 & \siiva\ \\ 
3204 & 130    & 58$^{+45}_{-12}$ &  97$\pm$19 & \civa\  \\ 
3223 & 147    &108$^{+35}_{-22}$ &  93$\pm$19 &   \oiiic\ \\ 
\hline
\end{tabular}
\label{tab:results}
\end{center}
(1) PGMW\,$\#$ ID \citep[][]{Parker1992};
(2) projected rotational velocity (\vrot) reported by \citet[][]{Evans2006};
broadening parameters we report in this work:
(3) \vrot\ obtained with {\it iacob-broad} \citep[][]{SimonDiaz2014} using the Fourier transform (FT); (4) non-rotational parameter, the so-called "macroturbulence" (\vmac), considering the \vrot\ obtained with FT methodology;
(5) metal line used to determine the broadening parameters.
\end{table}

We remark that the {\it iacob-broad} tool determines the broadening of a given photosperic line, separating the contribution from two different effects: rotational broadening and non-rotational broadening.
Rotational broadening is the \vrot\ parameter.
On the other hand, non-rotational broadening, generally assumed to be macroturbulent broadening, is less understood.
In fact, the given name is not necessarily related with the physical meaning of term "macroturbulence", which \citet[][]{SimonDiaz2017} explain how it was originally introduced in the framework of cool stars, and it is defined as "large-turbulent motions of material in the line-forming region" in that context. Here we aim to measure \vrot\ in order to study an invoked correlation between this parameter and the chemical enrichment of the stars (see Sec.~\ref{abundances}). However, we also report the non-rotational broadening, refereed here as \vmac\ for simplicity, because this broadening effect is important.
Its contribution may even exceed the rotational velocity term \citep[see e.g.,][]{SimonDiaz2017, Holgado2018}, and therefore it ought to be considered in the modeling of the spectra, independently of its physical meaning. Otherwise, if assuming \vmac$=0$~\kms, one could be overestimating \vrot.

Discussing the physical origin of the non-rotational broadening in our stars is beyond the scope of this paper. However, we refer the reader to the detailed study by \citet[][]{SimonDiaz2017}. Basically, they conclude that the so-called macroturbulence can be attributed to "pulsational modes linked to a heat-driven mechanism" and/or "cyclic surface motions caused by turbulent pressure instabilities in subsurface convection zones".
Most of our stars have an important "macroturbulent" contribution which has to be considered in our modeling.
Reporting this value is also helpful for reproducibility purposes and future reference.
The \vmac/\vrot\ ratios we find here are consistent with the findings by \citet[][]{SimonDiaz2014}. 

Finally, we also determine the \vrot\ parameter using the \hei\ and \heii\ lines.
However, even for the same star, we obtain different values from those obtained with metal lines.
This is not surprising, because as previously mentioned, other broadening (no-rotational) mechanisms start to play a role.
Thus, we avoid using these lines to determine \vrot.
Only in cases where no metal lines are present, H and He could be considered for a rough estimation of rotational broadening.

\subsection{PoWR models}
\label{PoWR models}

Once we determined key parameters like the \lstar, \ebv, \vrot, and checked for binarity in each of our objects,
then we can proceed with the modeling of the observed spectra.
We started by using synthetic spectra from the OB model grids by \citet[][]{Hainich2019}, generated with the PoWR model atmosphere program\footnote{\url{https://www.astro.physik.uni-potsdam.de/PoWR/}}.
PoWR models are basically the simulated emergent spectra of hot stars with \emph{specified} stellar and wind parameters, which consider non-local thermodynamic equilibrium (Non-LTE) radiative transfer, spherical symmetry, stationary outflow, metal line blanketing and wind inhomogeneities.
The rate equations for statistical equilibrium and radiative transfer are solved simultaneously by the PoWR code in the co-moving frame and secure energy conservation \citep[see][for more details]{Grafener2002,Hamann2003,Sander2015}.

Next, we calculated new synthetic spectra to model the stellar and wind parameters of our targets.
The basic input parameters of a specific stellar model are the bolometric luminosity \lstar\ and the stellar temperature \tstar. These are related to the stellar radius \rstar\ via the Stefan-Boltzmann equation.
The stellar radius is defined at a Rosseland-mean optical depth of 20.
While \tstar\ is the effective temperature related to that radius, one can also define an effective temperature \teff\ or T2/3, which is related to the radius where the Rosseland-mean optical depth reaches 2/3.
For the type of atmospheres considered in this paper, the difference between \tstar\ and \teff\ is negligibly small. Further fundamental parameters of a model include the surface gravity \logg, the abundances, and the wind parameters (mass-loss rate $\dot{M}$ and terminal wind velocity $v_\infty$).
The modeling procedure is described in Sec.~\ref{Fitting}.

For establishing the non-LTE population numbers, the radiative transfer is calculated assuming the same Gaussian profiles in the absorption coefficient, with the width set equivalent to a Doppler velocity of 30~\kms. This is a usual procedure and should account for turbulence, pressure broadening, and multiplet splitting of lines, which are not accounted for at this stage.
For the hydrostatic equation in the quasi-static part of the atmosphere, we adopt a turbulence pressure (\vmic) corresponding to 20~\kms\ \citep{Shenar2016}.
When the model stratification has been established, the emergent spectrum is calculated in the observer's frame of reference. Here, detailed line broadening is taken into account, and multiplets are split into their components.
A microturbulence velocity ($\xi$) of 14~\kms\ is found to give adequate results \citep[e.g.,][]{Hainich2019}. In the wind, we tentatively adopt an additional contribution to $\xi$ as $10\%$ of the local wind speed.
We used a clumping factor, or density contrast (D) of 10 (filling factor $f_{V}=0.1$), as described in \citet[][]{Hamann1998}. We assumed a wind velocity field following the $\beta$-law from \citet[][]{Castor1975}, with the exponent $\beta=0.8$, which is a typical value for O-type stars \citep[e.g.,][]{Kudritzki1989, Repolust2004}.

We included the following atoms in our analysis: H, He, C, N, O, Mg, Si, P, S, Fe, and Ni.
The mass fractions for these elements are adopted from \citet[][for the LMC]{Trundle2007}.
The H ($X_{\rm H}$) and He ($X_{\rm He}$) mass fractions are set to 0.738 and 0.258, respectively.
The total CNO value is conserved, unless otherwise specified. When certain spectra need a different abundance (e.g., N enrichment), then the relative values of these elements are changed accordingly.
We start with typical mass fractions for CNO elements of:
$X_{\rm C}=4.7\times10^{-4}$, $X_{\rm N}=7.8\times10^{-5}$, and $X_{\rm O}=2.6\times10^{-3}$.
The values for magnesium and silicon are
$X_{\rm Mg}=2.1\times10^{-4}$ and
$X_{\rm Si}=3.2\times10^{-4}$, respectively.
Since there are not direct abundance measurements of phosphorus in the LMC \citep[see e.g.,][]{Massa2003}, P is set to 0.5~\zsun, assuming a solar abundance from \citet[][]{Asplund2009}.
The same case for Sulfur.
The values for P and S are set to: 
$X_{\rm P}=2.9\times10^{-6}$ and
$X_{\rm S}=1.5\times10^{-4}$, respectively.
The iron group elements (from Sc to Ni) are treated with the super-level approach from \citet[][]{Grafener2002}.
For Fe, the adopted value was $X_{\rm Fe}=7.0\times10^{-4}$.
Dielectronic recombination and autoionization mechanisms were taken into account for CNO ions.
PoWR can also consider embedded X-rays, and they were included to model UV features.

\subsection{Modeling the observed spectra with PoWR}
\label{Fitting}

\begin{figure*}
\begin{centering}
\includegraphics[trim={0 2cm 0 2.7cm},clip,width=0.97\linewidth]{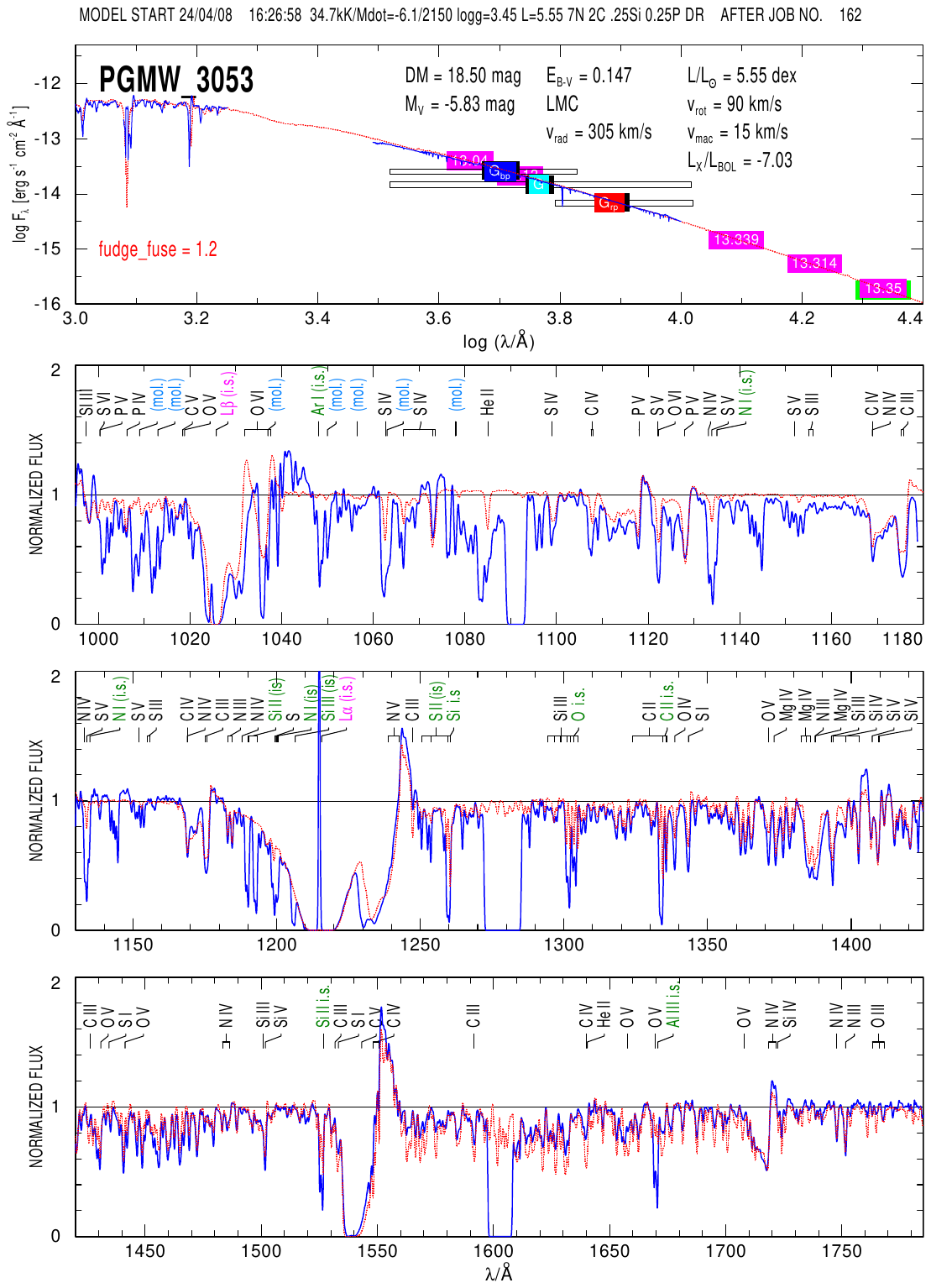}
\par\end{centering}
\caption{
PoWR model for the star PGMW\,3053. The observed spectrum is shown by a blue line and the model by a red dashed line. (first panel) SED with photometric magnitudes (colour boxes). The UV spectra better constrain the \ebv\ and \lstar\ of the star (also indicated at the upper right, among other parameters). The width of the Gaia DR3 filters (blue, central and red arms) are also indicated; (second panel) \fuse/MRDS UV spectrum normalized to the continuum model; (third and fourth panel) \hst/\cosi\ UV spectrum normalized to the continuum model.
The terminal velocity (\vinf) is determined from the blue edge of the \civ\ line in absorption.
Nitrogen enhancement is determined with \niii\ at 1183 and 1185~\AA, also considering these ions in the optical range, and the carbon abundances by using \ciii\ at 1175 and 1176~\AA\ in the UV.
Particularly sensitive to X-rays are the \nvuv\ and \ovi\ features.
Also, ISM absorption features by the Hydrogen Lyman lines were considered in our modeling.
Interstellar (i.s.) atomic, molecular and metal lines in absorption are indicated.
The legend "fudge" means that the \fuse\ spectrum needed to be scaled by 1.2, just for comparison purposes in the 2nd panel.
There is a gap of around 10~\AA\ in the observations around 1090, 1280, and 1605~\AA, where no key lines are present.
See text for details.
}
\label{fig:n11-18_xshootu_sed}
\end{figure*}

\begin{figure*}
\begin{centering}
\includegraphics[trim={0 2.1cm 0 2.0cm},clip,width=0.97\linewidth]{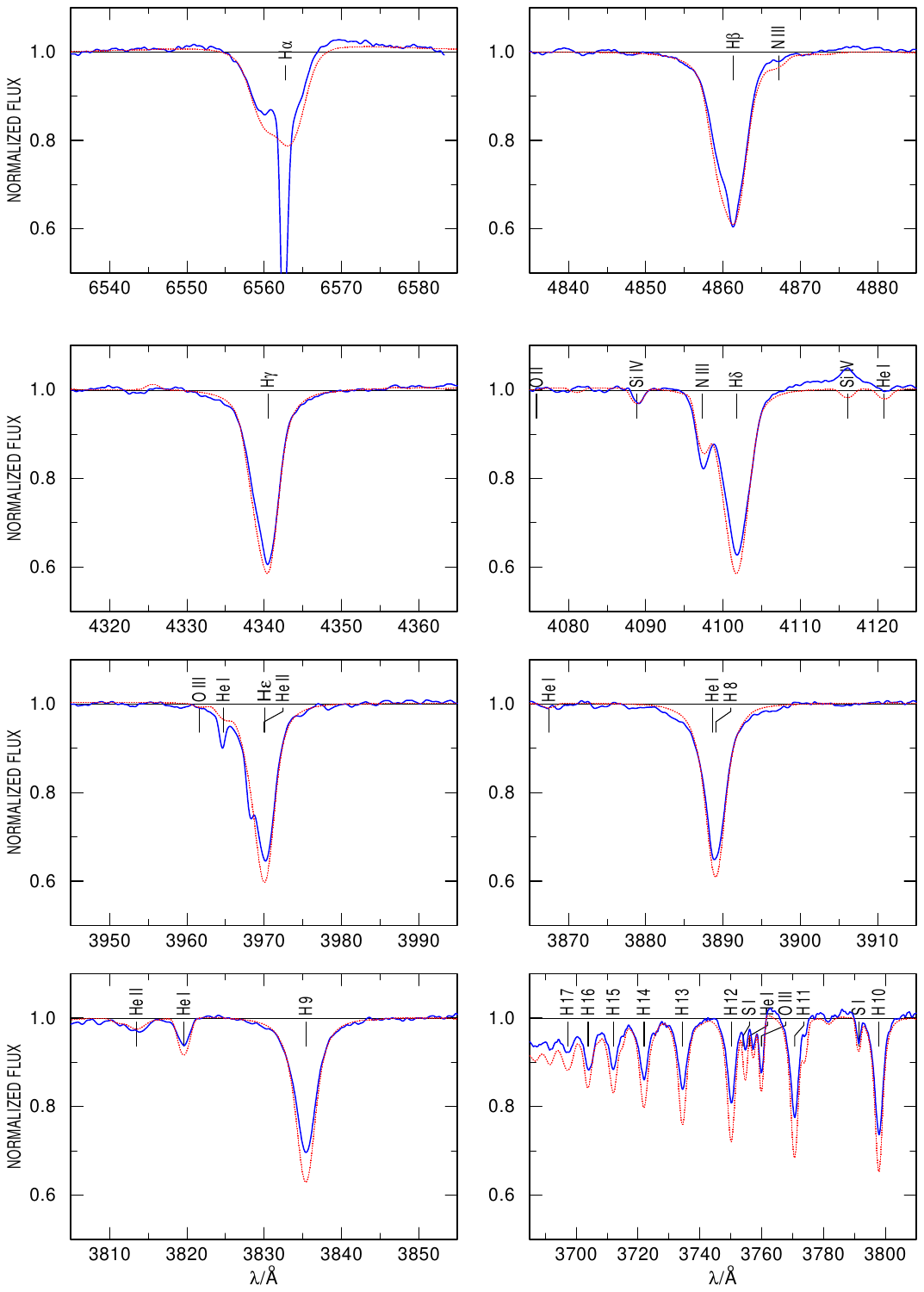}
\par\end{centering}
\caption{
PoWR model for PGMW\,3053. Several windows are displayed to show the Balmer lines present in the observed spectrum in detail: \ha, \hb, \hg, \hd, \hep\ and H10--H17, in comparison with the final model of the star. The surface gravity (\logg) parameter of the star is determined from the wings of the H lines. In this particular case, \hg\ is the optimal line to use, given it is not blended by other lines (e.g., \hb, \hd, \hep), and it is also more intense than H8--H17.
\ha\ is the most intense H line, however is sensitive to the stellar winds and therefore is considered to determine mass-loss rate ($\dot{M}$). The \ha\ line is affected by a strong feature in absorption, most likely by the nebular emission subtraction. The observed spectra are shown by a blue line and the PoWR model by a red dashed line. The most important lines are identified.
}
\label{fig:n11-18_xshootu_balmer}
\end{figure*}

\begin{figure*}
\begin{centering}
\includegraphics[trim={0 2.1cm 0 2.0cm},clip,width=0.97\linewidth]{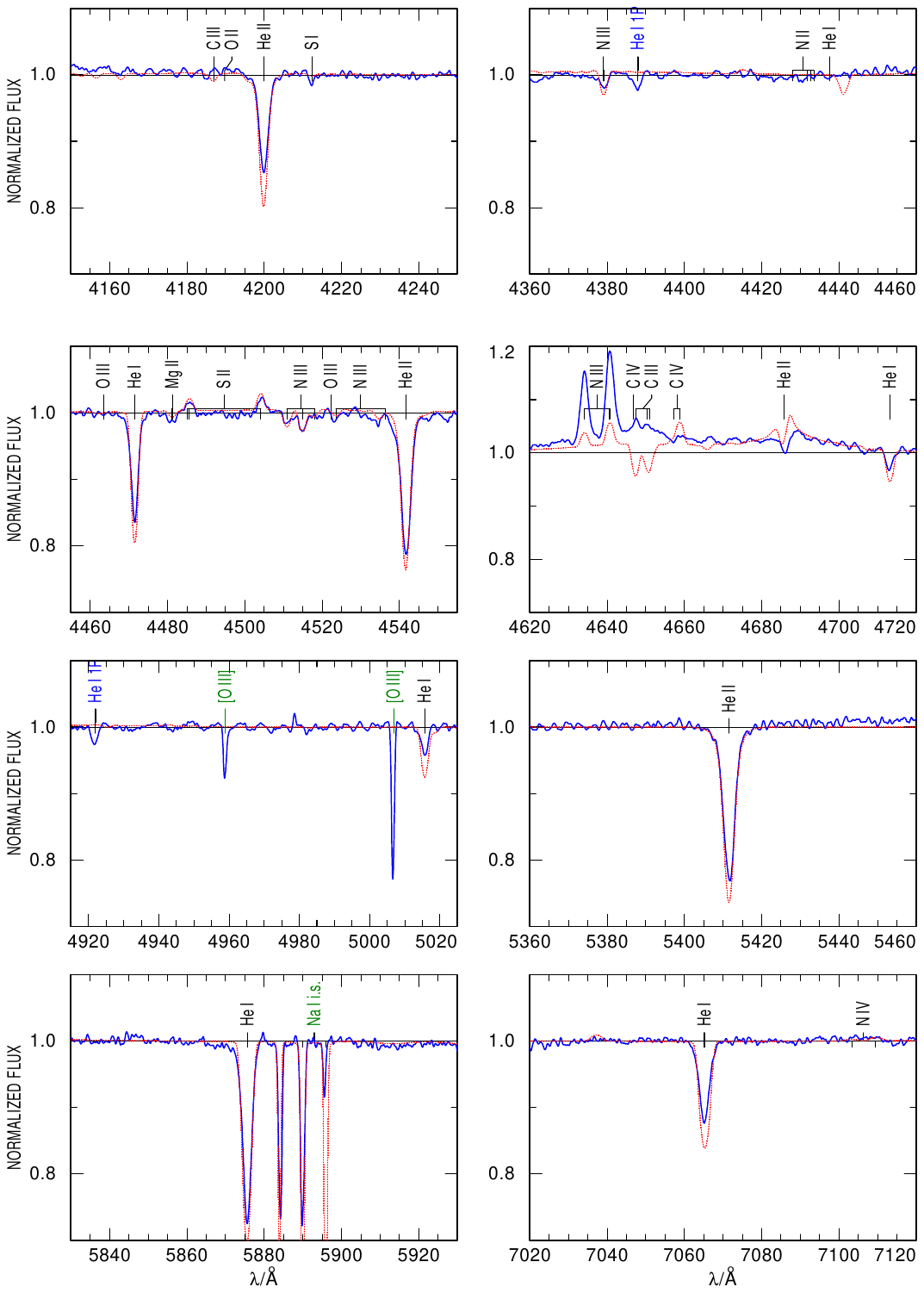}
\par\end{centering}
\caption{
PoWR model for PGMW\,3053. Several windows are displayed to show the most important \hei\ and \heii\ lines, as well as some metals, in comparison with the final model of the star. The temperature (\tstar) of the star is determined by modeling the \hei-\heii\ ratios
(e.g., \hei\,$\lambda$4471.5, and \heii\,$\lambda$4541.6). Singlet lines of \hei\ (1s2p P), at 4387.9 and 4921.9~\AA, identified with blue labels were not used for diagnostic.
Non-photospheric lines (e.g., Na) and features in absorption (\oiiin), most likely due to nebular emission subtraction, are indicated with green labels.
\heiiwr\ is crucial to determine the mass-loss rate ($\dot{M}$) of the star.
Rotational broadening and chemical abundances, like nitrogen enhancement, is determined from metal lines in absorption (e.g., \niiia\ and \niiib).
The observed spectra are shown by a blue line and the PoWR model by a red dashed line. The most important lines are identified.
}
\label{fig:n11-18_xshootu_he}
\end{figure*}

\begin{table*}
\begin{center}
\caption[]{
Physical Parameters obtained with PoWR Models for the eight exemplary ULLYSES target O-type stars in N11\,B.
}
\setlength{\tabcolsep}{0.55\tabcolsep}
\begin{tabular}{ccccccccccccc}
\hline\hline
\rule{0cm}{2.2ex} ID   & \tstar\ &\logg\        & $\log L_{\star}$  & $\log\dot{M}$         & \vinf\        & \rstar\ & \mstar\ & \logQH\  &\logQHeI\ &\logQHeII\ & \loglmec\ & $\log$ \Dmom\ \s \\
PGMW & [kK]    &[cm\,s$^{-2}$]& [L$_\odot$] &[M$_{\odot}$~yr$^{-1}$]& [km\,s$^{-1}$]& [\Rsun] & [\msol] &[ph\,s$^{-1}$]&[ph\,s$^{-1}$]&[ph\,s$^{-1}$] & [\lsun] &   \\
(1)  & (2)     & (3)          & (4)         & (5)                   & (6)           & (7)     & (8)     & (9)      & (10)     & (11)      & (12) & (13)      \\
\hline

3053 & 34.7$\pm$0.2 & 3.5$\pm$0.1  & 5.6$\pm$0.1 & $-6.1\pm0.1$ & 2150$\pm50$ & 16.5 & 28 & 49.2 & 48.1 & 40.9 & 2.7 & $-2.2$ \\
3058 & 40.0$\pm$0.5 & 3.9$\pm$0.1  & 5.4$\pm$0.1 & $-6.5\pm0.1$ & 2750$\pm50$ &  9.9 & 25 & 49.1 & 48.2 & 43.5 & 2.4 & $-2.6$ \\
3061 & 42.0$\pm$0.5 & 3.7$\pm$0.1  & 5.6$\pm$0.1 & $-6.0\pm0.1$ & 3200$\pm50$ & 12.0 & 26 & 49.4 & 48.6 & 40.8 & 3.0 & $-2.0$ \\ 
3100 & 38.0$\pm$0.5 & 3.7$\pm$0.1  & 5.5$\pm$0.1 & $-6.2\pm0.1$ & 2350$\pm50$ & 13.0 & 31 & 49.2 & 48.3 & 40.0 & 2.6 & $-2.3$ \\
3120a& 41.0$\pm$0.5 & 4.1$\pm$0.1  & 5.5$\pm$0.1 & $-6.3\pm0.1$ & 2500$\pm50$ & 11.2 & 57 & 49.2 & 48.4 & 40.5 & 2.6 & $-2.4$ \\
3168 & 33.3$\pm$0.2 & 3.5$\pm$0.1  & 5.3$\pm$0.1 & $-6.7\pm0.1$ & 1950$\pm50$ & 13.5 & 21 & 48.8 & 47.5 & 40.8 & 2.0 & $-2.9$ \\
3204 & 42.0$\pm$0.5 & 4.3$\pm$0.1  & 5.4$\pm$0.1 & $-6.6\pm0.1$ & 2400$\pm50$ &  9.5 & 66 & 49.1 & 48.4 & 43.7 & 2.3 & $-2.7$ \\ 
3223 & 34.0$\pm$0.5 & 3.9$\pm$0.1  & 5.5$\pm$0.1 & $-6.7\pm0.2$ & 1900$\pm50$ & 15.3 & 61 & 48.9 & 47.5 & 40.7 & 2.1 & $-2.8$ \\
\hline
\end{tabular}
\label{tab:results1}
\end{center}
(1) PGMW\,$\#$ ID \citep[][]{Parker1992};
(2) effective temperature (\tstar);
(3) surface gravity (\logg);
(4) bolometric luminosity (\lstar);
(5) mass-loss rate ($\log\dot{M}$);
(6) terminal wind velocity (\vinf);
(7) stellar radius (\rstar) obtained from \lstar\ and \tstar, via the Stefan-Boltzmann law;
(8) stellar mass (\mstar);
log of the number of ionizing photons per second for (9) Hydrogen (\QH);
(10) \hei\ (\QHeI); and (11) \heii\ models (\QHeII), considering X-rays;
(12) mechanical luminosity (\lmec$=0.5\dot{M}\varv_\infty^{2}$); proper units in \lsun($=3.826\times10^{33}$~\ergs) are obtained by dividing the result from the \lmec\ equation by \vinf/2c, following the analytical derivation;
(13) log of the modified wind momentum (\Dmom\ [M$_{\odot}$~yr$^{-1}$\,km\,s$^{-1}$\,\Rsun$^{-0.5}$]), defined as $D\equiv \dot{M}$ $\varv_{\infty}$ $R_{*}^{1/2}$ \citep[][]{Kudritzki2000}.
\end{table*}

Although the stellar and wind parameters of the O-type stars can be determined
by specific diagnostic lines separately, it is important to note that these are not independent of each other, and thus all the lines of the spectrum, from the UV to the optical, should be addressed comprehensively.
In this work, we used multi-wavelength observations to model the sample of ULLYSES targets in N11\,B. We analyse spectra from \fuse/MRDS, \hst/\stis, \hst/\cosi\ in the UV, and from VLT/\xshooter\ in the optical.
Given the multi-wavelength spectral coverage, along with information on distance, precise photometric magnitudes, radial velocities, projected rotation, and extinction, we can construct the SED of each star and determine their bolometric luminosity.
Next, we need to model the spectra of the stars.
We start with prior groundwork.
First, we chose a synthetic spectra from the grid of PoWR models in the \tstar-\logg\ plane by \citet[][]{Hainich2019}.
Then, we chose the one with the closest \tstar\ and \logg\ to our spectra.
%This process is iterative by nature.
Afterward, specific lines and their profiles are carefully examined to recalculate new models and better model the key lines for each physical attribute.
This process is iterative. Next, we explain the determination of each parameter.

The effective temperature (\tstar) of the stars is determined by modeling the He lines (e.g., \hei\,$\lambda\lambda$3819.6, 4120.8, 4471.5, 4713.1, 5015.7, 5875.6, 6678.1, 7065.2, and \heii\,$\lambda\lambda$3813.5, 4200.0, 4541.6, 5411.5, 6074.2, 6118.3, 6170.7, 6233.8, 6310.8, 6406.4, 6528.0, 6685.0, 6890.9, 7177.5, 7592.7),
specifically the \hei\ and \heii\ ratio. The pairs of \hei\,$\lambda$3819.6 with \heii\,$\lambda$3813.5, and \hei\,$\lambda$4471.5 with \heii\,$\lambda$4541.6 are particularly useful for this diagnostic.
Except for \heiiwr, which is known to be mostly sensitive to the strength of the stellar winds of the stars.
\heii\ lines in the UV are also considered (e.g., $\lambda$1640.4).
\heii\,$\lambda$1085 is in principle covered by FUSE, however is not useful given other lines in absorption might be present in that range.
Singlet lines of \hei\ (1s2p P), at 4387.9 and 4921.9~\AA, have shown to be problematic in previous studies \citep[see e.g.,][]{Najarro2006}, and the reason is still unclear in the community.
Thus, these lines were not considered for \tstar\ determination.

The gravitational acceleration at the surface of the stars (\logg),
can in principle be determined by modeling the wings of the H-Balmer lines, since they are expected to be pressure-broadened by the Stark effect.
However, care must be taken with \ha, since it is also sensitive to stellar winds.
Therefore, mainly the lines of \hb, \hg, \hd, \hep, and H9--H17, were considered to determine this parameter.
The rotational and non-rotational broadening determined from the metal lines present in the spectrum (see Sec.~\ref{Rotation}) must be considered before determining the surface gravity; otherwise, \logg\ risks overestimation.
Since \tstar\ and \logg\ are intrinsically correlated parameters and cannot be determined independently, they ought to be modeled \emph{simultaneously}.

The main diagnostic lines for the wind mass-loss rate ($\dot{M}$), in the optical range, are typically \ha\ and \heiiwr.
In the UV range \nv\,$\lambda\lambda$1238.8, 1242.8 and \civ\,$\lambda\lambda$1548.2, 1550.8 are particularly sensible to the wind.
The terminal wind velocity (\vinf) is determined from the blue edge of the \civ\,$\lambda$1548.2 in absorption. The obtained parameters of the stars are given in Table~\ref{tab:results1}.

We initially assumed standard LMC abundance values of C, N and O.
However, we need to change these values to better model certain metal lines.
The abundances of the rest of the elements are, in principle, fixed for all the stars.
In the UV, the following lines we used: \ciii\,$\lambda\lambda$1175.3, 1175.7 and \civ\,$\lambda\lambda$1548.2, 1550.8; \niii\,$\lambda\lambda$1183.03, 1184.51 and 1751.7.
When present, the following lines were considered in the optical range: \civ\,$\lambda\lambda$5801, 5812; \niii\,$\lambda\lambda$4097.4 (although it is blended with \hd), 4510.9, 4514.9, 4518.149; and \oiii\,$\lambda$5592.3.
The lines of \niii\,$\lambda\lambda$4634.1, 4640.6 in emission were found to be problematic to model and therefore were not considered in the determination of nitrogen abundance, they are also sensitive to other parameters, like the \tstar-\logg of the stars. The effect of other parameters, like the $\beta$-law, need to be explored to model these features.
The CNO abundances by mass fraction and number are reported in Table~\ref{tab:results2}.
The errors were estimated by varying the abundances in test calculations. Values outside the given error range lead to a clear discrepancy between the model and the observations.
We corroborated that our results do not change using the Low Temperature Dielectronic Recombination (LTDR) approach. We refer the reader to the PoWR code manual (ManPoWR, Dec. 2013), for more information on this issue.

\begin{table}
\begin{center}
\caption[]{CNO abundances for the O-type stars in N11\,B.}
\setlength{\tabcolsep}{0.4\tabcolsep}
\begin{tabular}{ccccccc}
\hline\hline
\rule{0cm}{2.2ex} ID   & \multicolumn{2}{c}{$X_{\rm C}$}       & \multicolumn{2}{c}{$X_{\rm N}$}       & \multicolumn{2}{c}{$X_{\rm O}$} \s \\   
PGMW & mass              & number            & mass              & number            & mass              & number         \\   

 & ($\times10^{-4}$) & ($\times10^{-6}$) & ($\times10^{-5}$) & ($\times10^{-5}$) & ($\times10^{-3}$) & ($\times10^{-4}$) \\
(1)  & (2)               & (3)               & (4)               & (5)               &  (6)              & (7)    \\
\hline

3053 & 9$\pm$5     & 98$\pm$50 & 55$\pm$8 & 5$\pm$1     & 2.3$\pm$0.3 & 1.8$\pm$0.2 \\
3058 & 0.1$\pm$0.1 &  1$\pm$1  & 23$\pm$8 & 2$\pm$1     & 0.5$\pm$0.5 & 0.4$\pm$0.4 \\
3061 & 0.1$\pm$0.1 &  1$\pm$1  & 39$\pm$8 & 4$\pm$1     & 2.3$\pm$0.3 & 1.9$\pm$0.2 \\ 
3100 & 0.9$\pm$0.5 &  10$\pm$5 & 23$\pm$8 & 2$\pm$1     & 2.5$\pm$0.2 & 1.9$\pm$0.2 \\
3120a& 0.9$\pm$0.5 &  10$\pm$5 & 55$\pm$8 & 5$\pm$1     & 5.3$\pm$2.5 & 4$\pm$2 \\
3168 & 9$\pm$5     & 98$\pm$50 & 23$\pm$8 & 2$\pm$1     & 2.3$\pm$0.3 & 1.9$\pm$0.2 \\
3204 & 0.5$\pm$0.5 &  5$\pm$5  & 23$\pm$8 & 2$\pm$1     & 0.6$\pm$0.5 & 0.5$\pm$0.5 \\ 
3223 & 0.1$\pm$0.1 &  1$\pm$1  &  8$\pm$4 & 0.7$\pm$0.4 & 1.3$\pm$0.5 & 1.0$\pm$0.5 \\
\hline
\end{tabular}
\label{tab:results2}
\end{center}
(1) PGMW\,$\#$ ID \citep[][]{Parker1992};
mass fractions by mass and by number for (2, 3) carbon ($X_{\rm C}$);
(4, 5) nitrogen ($X_{\rm N}$), and (6, 7) oxygen ($X_{\rm O}$) are given.
For reference,
standard mass fractions for the LMC are:
$X_{\rm H}=7.375\times10^{-1}$,
$X_{\rm He}=2.579\times10^{-1}$,
%and for CNO:
$X_{\rm C}=4.75\times10^{-4}$,
$X_{\rm N}=7.83\times10^{-5}$,
and $X_{\rm O}=2.64\times10^{-3}$;
the relative abundances by number are:
$X_{\rm H}=9.193\times10^{-1}$,
$X_{\rm He}=8.038\times10^{-2}$,
%and for CNO:
$X_{\rm C}=4.934\times10^{-5}$,
$X_{\rm N}=6.972\times10^{-6}$,
and $X_{\rm O}=2.057\times10^{-4}$ \citep[][]{Trundle2007}.
\end{table}

\begin{table}
\begin{center}
\caption[]{X-ray parameters for the O-type stars in N11\,B.}
\setlength{\tabcolsep}{0.6\tabcolsep}
\begin{tabular}{cccccc}
\hline\hline
\rule{0cm}{2.2ex} ID & \lxlbol\ & \ff\ & \tx\ & \rmin\     & \lxlbol\ \s \\
PGMW & \citep[][]{Naze2014} &     &[MK]      & [\rstar]    &  (here)     \\
(1)  & (2)            & (3)       & (4)      & (5)         & (6)    \\
\hline
3053 & $<-7.1$        &  1.0      & 0.5      & 1.1         & $-7.0$ \\
3058 & $<-7.05$       &  0.3      & 0.5      & 1.1         & $-7.5$ \\
3061 & $-6.57\pm0.11$ &  1.0      & 0.5      & 1.1         & $-6.6$ \\
3100 & $-6.86\pm0.12$ &  1.0      & 0.5      & 1.1         & $-6.8$ \\
3120a& $-6.53\pm0.07$ &  1.0      & 0.5      & 1.1         & $-7.0$ \\
3168 & $<-6.79$       &  0.5      & 0.6      & 1.1         & $-7.0$ \\
3204 & $-6.56\pm0.05$ &  0.01     & 1.0      & 1.1         & $-6.4$ \\
3223 & $<-6.9$        &  1.0      & 0.6      & 1.1         & $-7.1$ \\
\hline
\end{tabular}
\label{tab:xray}
\end{center}
(1) PGMW\,$\#$ ID \citep[][]{Parker1992};
(2)  X-ray to bolometric luminosity ratios reported by \citet[][]{Naze2014};
(3) fraction of electrons in the plasma phase (\ff);
(4) X-ray temperature (\tx);
(5) minimum radius at which shock-wave occurs (\rmin);
(6) \lxlbol\ ratios needed to model UV spectral features.
\end{table}

Furthermore, non-photospheric continuum emission originating from shock-heated plasma in the stellar wind of stars also has an important imprint in the UV range of their spectra.
Some of these features, like \nvuv, and \ovi, are produced by X-rays and therefore need to be included for a complete modeling of the UV spectra. With PoWR, we are capable of doing so.
For this, three parameters were defined:
1) the fraction of electrons in the plasma phase (\ff); 2) the X-ray temperature (\tx); and 3) the minimum radius at which the shockwaves occur (\rmin).
We note that by including X-rays, energy conservation might be compromised.
Therefore, X-rays are included in the modeling process once we have obtained a converged model considered 'final', with its temperature stratification already defined, and reproducing the main stellar and wind features of the star.
We report the ratio between X-ray and bolometric luminosity ($L_\mathrm{X}/L_\mathrm{bol}$) for the O-type stars in Table~\ref{tab:xray}.
Uncertainties are estimated to be around 0.3~dex of the reported values.
As usual, $L_\mathrm{bol}$ is defined as referring to the wavelength range below $\lambda=40$~\AA\ ($\approx0.3$~keV). 

Additionally, a correction for the expected ISM absorption by the hydrogen Lyman lines (\lya\ and \lyb) present in the observed UV range was included in our models.
For this, the color excess for both the Galaxy and the LMC, and radial velocity were considered.
The column density (\nh) of the local ISM was determined by using: \nh$=3.8\times10^{21}$~\cm\ \ebv, following \citet[]{Groenewegen1989}.

An example of the PoWR model for one of the stars in our study, PGMW\,3053, is shown in Fig.~\ref{fig:n11-18_xshootu_sed}.
The first panel shows the contructed SED.
The second panel shows the FUSE UV spectrum normalized to the continuum model.
The key stellar lines are indicated, and when present, nebular emission lines and interstellar (i.s.) atomic and metal lines in absoption \citep[listed in][Tab.4]{Haser1998} are also labeled.
Strong ISM molecular hydrogen absorption lines are particularly present in this range. Their modeling is outside the scope of this work, yet they are also indicated.
The third and fourth panels show the \hst/\cosi\ UV spectrum normalized to the continuum model.
The normalized line spectra of wavelength ranges with key lines of H, He, and metals, are shown in Fig.~\ref{fig:n11-18_xshootu_balmer} and Fig.~\ref{fig:n11-18_xshootu_he}.
The corresponding modeling for the rest of the stars is shown in the Appendix.

\subsection{Evolutionary masses and ages}
\label{bonnsai}

We used the BONNSAI web-service \footnote{\url{https://www.astro.uni-bonn.de/stars/bonnsai}} \citep[][]{Schneider2014} to obtain the predicted evolutionary mass (\mev), age and initial rotation of the stars.
For this, a list of stellar parameters previously determined with PoWR modeling are introduced: \lstar, \tstar, and \logg.
Also the \vrot, determined with the {\it iacob-broad} tool.

BONNSAI assumes main sequence single stars with initial stellar masses (\mini), between 5 and 5000~\Msun, initial rotational velocities (\vini) between 0 and 600~\kms, and a power-law initial mass function (IMF) with slope $\gamma=-2.35$, from \citet[][]{Salpeter1955}.
Stellar models by \citet[][]{Brott2011,Kohler2015} for the LMC were chosen.
The replicated observables with BONNSAI: \tstar, \logg, \lstar, and \vrot, are given in Sec.~\ref{comments} for each star.
The results are reported in Table~\ref{tab:results3}.
These values must be interpreted with caution, as in some cases, the replicated observables do not precisely match the stellar parameters determined with PoWR models, but rather represent a similar set of parameters (the closest values) from their grid of models. Any such differences are discussed later in the text.

\begin{table}
\begin{center}
\caption[]{Stellar parameters obtained with BONNSAI.
}
\begin{tabular}{lccccc}
\hline\hline
\rule{0cm}{2.2ex} ID   & \mini\       & \mev\       & age          & \vini\ \s \\
PGMW & [\msol]      & [\msol]      & [Myr]       & [\kms]     \\
(1)  & (2)          & (3)          & (4)         & (5)        \\
\hline
3053 & 39$\pm3$ & 37$\pm3$ & 3.6$\pm0.2$ & 100$\pm90$ \\
3058 & 34$\pm2$ & 33$\pm2$ & 2.9$\pm0.3$ &  70$\pm190$ \\
3061 & 29$\pm6$ & 26$\pm5$ & 2.6$\pm0.2$ & 460$\pm40$ \\ 
3100 & 37$\pm4$ & 36$\pm3$ & 3.3$\pm0.2$ & 110$\pm60$ \\
3120a& 36$\pm3$ & 35$\pm3$ & 2.3$\pm0.4$ & 150$\pm60$ \\
3168 & 29$\pm3$ & 28$\pm3$ & 4.5$\pm0.3$ &  60$\pm50$ \\
3204 & 35$\pm3$ & 35$\pm3$ & 1.2$\pm0.7$ &  70$\pm50$ \\ 
3223 & 33$\pm3$ & 32$\pm3$ & 3.9$\pm0.4$ & 120$\pm60$ \\
\hline
\end{tabular}
\label{tab:results3}
\end{center}
(1) PGMW\,$\#$ ID \citep[][]{Parker1992};
(2) initial mass (\mini) and; (3) evolutionary (actual) mass (\mev);
(4) age; (5) initial rotational velocity (\vini).
\end{table}

\subsection{Comments on the O-type stars analyzed in this work}
\label{comments}

Next, we will briefly comment on each of the analyzed objects and compare our results with previously reported parameters for some of our stars.

\textbf{PGMW\,3053}. a.k.a. N11\,18, is a mid O-type bright giant star initially classified as O5.5\,I-III(f) by \citet{Parker1992}, later reclassified as O6\,II(f+) by \citet{Walborn1995, Evans2006, Vink2023}, and revised as O6.5 II(f) by PAC (priv. comm.).
With \giraffe\ multi-epoch spectra, we note that the \heiiwr\ line profile is variable within a time window of 1~day (see Fig.~\ref{fig:n11-18_giraffe}).
\citet[][in their Tab.6]{Evans2006} mentioned a possible "\vrad\ or wind variability".
However, the difference in strength of this line due to this variability is rather small and does not have a significant impact in the determined $\dot{M}$.
There are no other available epochs of observations to check for additional periods of probable variability for this line.
The rest of the He lines observed at different time intervals do not vary over periods ranging from days to months and do not exhibit a double-peak profile.
Thus, according to our analysis, no evidence of binarity was found in this object.
No contamination from additional sources was found by inspecting \hst/WFPC2 images in filters F656N and F502N.

Previously, \citet[][in their Fig. 12]{Evans2006} adopted a temperature for this star based on its spectral type and metallicity.
Later, \citet[][]{Martins2024} reported \loglstar$=5.67$, \tstar$=37$~kK and \logg$=3.6$.
Here we determined lower values: \loglstar$=5.55$, \tstar$=34.7$~kK, and \logg$=3.45$. We also report $\log(\dot{M}/\mdot)=-6.1$.
We find that nitrogen is enriched by a factor of 7, having the highest N enrichment in our sample, and by 2 in carbon.
Silicon on the other hand, has to be reduced by 4 to match the \siiva\ line, although not the \siivb\ in emission. We were not able to reproduce the mentioned \siiv\ lines when they are in emission (for any of the stars) with our PoWR modeling.
We also note that the modeled \niii\,$\lambda\lambda$4634.1, 4640.6 lines in emission are not strong enough to reproduce the observed features (see Sec.~\ref{abundances} for discussion).

The \vrot\ parameter is determined from the metal line in absorption \niiia\ (with EW$=50$~m\AA\ and SNR$=200$, in the neighboring continuum),
resulting in \vrot=88~\kms, and \vmac=15~\kms.
Using \siiv\ (EW$=70$~m\AA; SNR$=180$) one obtains \vrot=68~\kms, and \vmac=66~\kms.
We note a discrepancy between the values obtained from different metal lines.
For this star, we select the \vrot\ obtained from the \niiia, given the consistency of the values obtained by the FT and GOF methods (see Fig.~\ref{fig:vrot_n11-18}).

The replicated observables with BONNSAI were: \tstar$=34.7$~kK, \logg$=3.54$, \loglstar$=5.6$, and \vrot$=90$~\kms.
BONNSAI assumes a \loglstar$\sim$0.05~dex and \logg$=0.1$~dex higher than the introduced values. We remark these differences given the predicted \mev\ is higher than the determined \mspec.

\citet[][]{Naze2014} reported an upper limit for its X-ray luminosity of \lxlbol$<-7.1$.
We include X-rays to reproduce certain features in the UV, the \nvuv\ and the \ovi.
Given that the upper limit previously reported for the \lxlbol\ ratio is practically the same value we obtained,
we thoroughly explored the set of three free parameters that could give us an even lower ratio (\ff, \tx, and \rmin).
However, if  \tx\ is reduced by 0.1~MK, then the \nv\ is weaker than the observation, so this option was discarded.
On the other hand, by increasing \ff\, or \rmin, then the \ovi\ 
become more intense or broader than the observed lines.
A similar exercise was done for the rest of the analyzed stars.
Here we report a ratio of \lxlbol$=-7.0$.

\textbf{PGMW\,3058}. a.k.a. N11\,60, is an early O dwarf star, classified as O3\,V((f*)) by \citet{Walborn2004}, and revised without change by PAC (priv. comm.).
According to our analysis of the \giraffe\ spectra, there is no evidence of binarity.
Previously, \citet[][]{Mokiem2007} determined \loglstar$=5.57$, \tstar$=45.7$~kK, \logg$=3.92$ and $\log(\dot{M}/\mdot)=-6.28$.
Based on recent Gaia DR3 photometry \citep[][]{Gaia2016,Gaia2023},
we obtain \loglstar$=5.35$, which is a lower value by 0.2~dex.
Also our temperature is lower by 6~kK (\tstar$=40$~kK), and we get half the mass-loss rate value ($\log(\dot{M}/\mdot)=-6.5$).
The replicated observables with BONNSAI were: \tstar$=39.2$~kK, \logg$=3.9$, \loglstar$=5.38$, and \vrot$=60$~\kms. BONNSAI assumes a slightly lower \tstar, and higher values for \loglstar\ and \logg, affecting the obtained \mev, which is higher than \mspec.
We find that this object is enriched in N by a factor of 3.
C is has to be reduced by 50, O by 5, and Si by 4.

The \vrot\ is determined from the metal line in absorption \nva\ (EW$=190$~m\AA; SNR$=230$),
resulting in \vrot$=66$~\kms, and \vmac$=80$~\kms.
\nvb\ is weaker (EW$=110$~m\AA; SNR$=230$). From this line we obtain
\vrot$=81$~\kms, and \vmac$=35$~\kms.
We did not consider \civa\ or \civb\ given they are observed in emission.
Neither other weaker lines, like the \oiiic.

For this star, \citet[][]{Naze2014} reported \lxlbol$<-7.05$.
We report \lxlbol$=-7.5$, a value below the upper limit formerly suggested, being consistent with X-ray observations, however obtained with a different approach.

\textbf{PGMW\,3061}. a.k.a. N11\,31, is the earliest object in our sample, an O-type giant reclassified from O3\,III(f*) \citep[][]{Parker1992} to ON2\,III(f*) by \citet[][]{Walborn2004,Evans2006}.
According to our analysis of the \giraffe\ spectra, there is no evidence of binarity.
We find that the \heiiwr\ profile of this star is variable in a period of 1~day (see Fig.~\ref{fig:n11-31_giraffe}), that we attribute to wind variability.
However, the difference in strength of this line is negligible and does not affect the determined $\dot{M}$.

Previously, \citet[][]{Mokiem2007} determined \loglstar$=5.84$, \tstar$=45$~kK, \logg$=3.85$, and $\log(\dot{M}/\mdot)=-5.41$.
Later, \citet[][]{Rivero2012} determined \loglstar$=5.9$, \tstar$=47.8$~kK, \logg$=3.95$, and $\log(\dot{M}/\mdot)=-5.7$.
Compared with the later study, we obtain \loglstar$=5.6$, a lower value by 0.3~dex based on Gaia DR3 photometry \citep[][]{Gaia2016,Gaia2023} and UV spectra constraints for its luminosity and extinction; 
we also get \tstar$=42$~kK, \logg$=3.7$, and $\log(\dot{M}/\mdot)=-6.0$, which are lower values from previously reported.
The replicated observables with BONNSAI were: \tstar$=41.3$~kK, \logg$=3.85$, \loglstar$=5.65$, and \vrot$=120$~\kms.
The evolutionary mass is in agreement with the spectroscopic one.
Additionally, N has to be increased by 5,
C is reduced by 10, and Si by 4.

The \vrot\ is determined from the metal line in absorption \nva\ (EW$=660$~m\AA; SNR$=120$) and \nvb\ (EW$=400$~m\AA; SNR$=80$).
Using \nva, \vrot=133~\kms\ and \vmac=82~\kms\ were obtained, while with \nvb, \vrot=113~\kms\ and \vmac=79.9~\kms\ were determined. The average of the two values, \vrot$=120$~\kms\ and \vmac$=80$~\kms, is reported for this star.

\citet[][]{Naze2014} reported a \lxlbol$=-6.6$.
We report the same value, independently reproducing the observations.

\textbf{PGMW\,3100}. a.k.a. N11\,38, is a mid O giant star, reclassified from O6.5\,V(f) \citep{Parker1992} to O5\,III(f+) by \citet{Evans2006}, and revised as O5 III(f) by PAC (priv. comm.).
With \giraffe\ multi-epoch spectra, we find evidence of \vrad\ variability, consistent with it being a binary SB1.
The lines of H, \hei\ and \heii\ in the range of $3870-4040$~\AA\ and the \ha\ window ($6520-6590$~\AA) have to be shifted by the difference in radial velocity of $9$~\kms. In the $4670-4730$~\AA\ range, this \vrad\ difference is $7$~\kms\ (see Fig.~\ref{fig:n11-38_giraffe}).

Earlier, \citet[][]{Mokiem2007} determined \loglstar$=5.69$, \tstar$=41.0$~kK, \logg$=3.72$ and $\log(\dot{M}/\mdot)=-5.81$.
We obtain \loglstar$=5.5$, a lower value by 0.2~dex, based on Gaia DR3 photometry \citep[][]{Gaia2016,Gaia2023}; also a lower temperature (\tstar$=38$~kK), and a lower mass-loss rate ($\log(\dot{M}/\mdot)=-6.25$).
The replicated observables with BONNSAI were: \tstar$=37$~kK, \logg$=3.7$, \loglstar$=5.5$, and \vrot$=110$~\kms.
BONNSAI assumes a lower \tstar by 1~kK.
The evolutionary mass is close to the spectroscopic one.
It has a N enrichment by 3, and Si is reduced by 4.

The \vrot\ is determined from the observed metal line in absorption \civa\ (EW$=260$~m\AA; SNR$=90$), resulting in \vrot$=99$~\kms, and \vmac$=92$~\kms.
\civb\ is weaker (EW$=200$~m\AA; SNR$=110$). From this line we obtain \vrot$=67$~\kms, and \vmac$=110$~\kms. 

\citet[][]{Naze2014} reported a \lxlbol$=-6.86$.
We report \lxlbol$=-6.8$, consistent with X-ray observations.

\textbf{PGMW\,3120a}. This is a mid O dwarf O5.5V((f*)) star \citep[][]{Parker1992}, revised without change by PAC (priv. comm.).
PGMW\,3120 is actually a cluster with at least three massive stars, as can be noted by eye in the \hst/ACS image in Fig.~\ref{fig:hst_zoom}. The slit apertures of \xshooter\ in the optical, and FUSE in the UV, include the three sources. This is not the case for the STIS/\hst\ spectrum covering only PGMW\,3120a. Also, the photometric magnitudes of these sources are presumably for the cluster. Gaia DR2 and DR3 values are displayed in the SED.
One can note that the central V magnitude is not consistent with the blue and red arms, being an outlier.
The Gaia DR2 values are consistent with each other. In  the first panel of Fig.~\ref{fig:master_pgmw3120_sed}, the SED of the source is shown. A luminosity of \loglstar$=5.95$ to match the \xshooter\ spectra corresponds to the luminosity of the clustered stars. However, such luminosity would imply a spectroscopic mass of 160~\Msun.
Therefore, we performed photometry in the F220W image, using standard {\scshape iraf} tasks, on the three sources that make up the cluster in PGMW\,3120.
We found that the three objects, named here as: PGMW\,3120a, PGMW\,3120b and PGMW\,312005c, have practically the same brightness (14.07, 14.07, and 14.09~mag, respectively).
Given this, we can assume that one-third of the luminosity of the cluster corresponds to the luminosity of our star (PGMW\,3120a), being \loglstar$=5.5$. This luminosity would imply a lower spectroscopic mass (57~\Msun).
In the SED of Fig.~\ref{fig:master_pgmw3120_sed}, it can also be noted that there is no need to multiply the level of the flux of the STIS spectrum by any factor. This is consistent with it being the spectrum from a single source. On the other hand, the FUSE spectrum is divided by 3, in order to be at the same level as the STIS spectra, perfectly consistent with it including the flux of three sources of the same type. The \xshooter\ spectra also includes the cluster, as can be noted on Fig.~\ref{fig:hst_zoom}.
There is not \giraffe\ multi-epoch spectra to check for binarity.

We obtain \loglstar$=5.5$, based on Gaia DR3 photometry \citep[][]{Gaia2016,Gaia2023} and above mentioned UV flux constraints; also \tstar$=41$~kK, \logg$=4.1$, and $\log(\dot{M}/\mdot)=-6.3$ for PGMW\,3120a.
The replicated observables with BONNSAI were: \tstar$=41.2$~kK, \logg$=4.0$, \loglstar$=5.4$, and \vrot$=150$~\kms.
The values obtained with PoWR and later introduced in BONNSAI failed to be reproduced, given they are not covered in its parameter space.
Lower values by 0.1~dex are assumed for \loglstar\ and \logg, affecting the predicted \mev, which is lower than \mspec\ almost by half.
N is increased by a factor of 7,
C is reduced by 5, and Si by 4.

The \vrot\ is determined from the observed metal line in absorption \civa\ (EW$=230$~m\AA; SNR$=100$), resulting in \vrot$=143$~\kms, and \vmac$=148$~\kms.
From \civb\ (EW$=200$~m\AA; SNR$=110$) one obtains similar values, \vrot$=162$~\kms, and \vmac$=163$~\kms.
We chose the values from \civa, given the strength of the line and its profile is better defined than the \civb\ line.

\citet[][]{Naze2014} reported a \lxlbol$=-6.53$.
We report \lxlbol$=-7.0$, which is a lower value by 0.5~dex.

\textbf{PGMW\,3168}. a.k.a. N11\,32, is a mid O giant star, classified as O7\,II(f) by \citet{Evans2006}, and revised as O7.5\,III(f) by PAC (priv. comm.).
According to the \giraffe\ spectra, there is no evidence of binarity.
Also, this is a single object at the spatial resolution of \hst/WFPC2 images.
\citet[][]{Mokiem2007} determined \loglstar$=5.43$, \tstar$=35.2$~kK, \logg$=3.45$ and $\log(\dot{M}/\mdot)=-6.1$.
Recently, \citet[][]{Martins2024} reported \loglstar$=5.3$, \tstar$=34.9$~kK, \logg$=3.5$, but not wind parameters.
We also obtain \loglstar$=5.3$, using Gaia DR3 photometry \citep[][]{Gaia2016,Gaia2023}. Previous works had given lower \lstar\ values. %, consistent with \citet[][]{Martins2024};
Compared with \citet[][]{Martins2024}, we obtain a similar value for
\tstar\ ($=33.3$~kK) and the same \logg.
The $\dot{M}$ that we find ($\log(\dot{M}/\mdot)=-6.75$) is lower than the one reported by \citet[][]{Mokiem2007}.
The replicated observables with BONNSAI were: \tstar$=33.2$~kK, \loglstar$=5.3$, \logg$=3.6$, and \vrot$=50$~\kms.
A higher value by 0.1~dex is assumed for \logg, affecting the obtained \mev.
We find that N is increased by a factor of 3, C by 2, and Si reduced by 4.

The \vrot\ is determined from the observed metal line in absorption \siiva\ (EW$=210$~m\AA; SNR$=170$), resulting in \vrot$=55$~\kms, and \vmac$=89$~\kms.
Using another metal line, \niiia\ (EW$=44$~m\AA; SNR$=314$), we obtain a similar value for \vrot$=67$~\kms, however different for \vmac$=26$~\kms.
We prioritize the values obtained with \siiva, given its EW is stronger.
This star has the slowest \vrot\ and the lowest \tstar\ of the analyzed targets.
It is also among the objects with the lowest \logg, and therefore with the lowest mass, \mspec$=27$~\Msun\ of the analyzed stars.

\citet[][]{Naze2014} reported a \lxlbol$<-6.8$ for this object.
We report \lxlbol$=-7.0$, a value below the upper limit suggested, consistent with previous observations.

\textbf{PGMW\,3204}. a.k.a. N11\,48, is a mid O dwarf star, reclassified from O6-O7\,V \citep{Parker1992} to O6.5\,V by \citet{Evans2006}, and revised as O6\,Vz((f)) by PAC (priv. comm.).
Formerly, \citet[][]{Mokiem2007} determined \loglstar$=5.38$, \tstar$=40.7$~kK, \logg$=4.19$ and $\log(\dot{M}/\mdot)=-6.78$.
We obtain the same \loglstar, a higher temperature (\tstar$=42.0$~kK) and surface gravity (\logg$=4.3$), the highest of the sample, with a mass of 66~\Msun\ and $\log(\dot{M}/\mdot)=-6.6$.
The replicated observables with BONNSAI were: \tstar$=42.7$~kK, \loglstar$=5.3$, \logg$=4.17$, and \vrot$=60$~\kms.
Lower values by 0.1~dex are assumed for \loglstar\ and \logg, affecting the obtained \mev, which is lower than \mspec\ almost by half.
It has a N enrichment by 3,
C is reduced by 10, O by 4 and Si by 4.

A \vrot$=58$~\kms, and \vmac$=97$~\kms\ is determined from the \civa\ (EW$=140$~m\AA; SNR$=130$).

\citet[][]{Naze2014} reported a \lxlbol$=-6.6$.
We report \lxlbol$=-6.4$, the highest among the analyzed targets.
According to its \giraffe\ spectra (see Fig.~\ref{fig:n11-48_giraffe1} and \ref{fig:n11-48_giraffe2}) there is no conclusive evidence of binarity.

\textbf{PGMW\,3223}. a.k.a. N11\,13, is a late O dwarf star, among the brightest objects (V$=12.9$~mag) in N11\,B, reclassified from O8.5\,IV \citep{Parker1992} to O8\,V by \citet{Evans2006}, and revised as O8\,Vz by PAC (priv. comm.).
According to our analysis of the \giraffe\ multi-epoch spectra, we confirm binary SB1 features, also noted by \citet[][]{Evans2006}.
The lines of H, \hei\ and \heii\ in the range of $3870-4040$~\AA\ and the \ha\ window ($6520-6590$~\AA) have to be shifted by the difference in radial velocity of $30$~\kms.
In the $4670-4730$~\AA\ range, this \vrad\ difference is $20$~\kms\ (see Fig.~\ref{fig:n11-13_giraffe}).

Previously, \citet{Serebriakova2023} reports \tstar$=33.8$~kK and \logg$=3.6$.
We obtain \loglstar$=5.5$, \tstar$=34.0$~kK, \logg$=3.9$, and $\log(\dot{M}/\mdot)=-6.7$.
The observables with BONNSAI were: \tstar$=35.1$~kK, \logg$=3.58$, \loglstar$=5.4$, and \vrot$=110$~\kms.
The \logg\ is lower by 0.27~dex, affecting the obtained \mev, which is lower than \mspec\ by half.

The \vrot\ of this object is determined from the metal line in absorption \oiiic\ (EW$=215$~m\AA; SNR$=75$),
resulting in \vrot$=108$~\kms, and \vmac$=93$~\kms.

\citet[][]{Naze2014} reported \lxlbol$<-6.9$.
We report \lxlbol$=-7.1$, consistent with the predicted value.

\textbf{Additional objects}.
In order to quantify the total ionizing feedback of the O-type stars in N11\,B, three additional objects in the LMC were analyzed in this work (see Sec.~\ref{feedback}). We introduce them here so that we can use additional stellar and wind parameters for comparison with our objects in N11\,B. We report their main parameters at the Appendix (see Tables~\ref{tab:results4},\ref{tab:results5},\ref{tab:results6}), as well as their respective modeling.

\textbf{Sk $-66^{\circ}$ 171}. For this late O-type supergiant, classified as O9\,Ia \citep{Fitzpatrick1988}, \citet[][]{Martins2024} reported \loglstar$=5.7$, \tstar$=30$~kK, and \logg$=3.0$.
We obtain consistent values for \loglstar($=5.7$), and \tstar$=30.5$~kK, but a higher \logg($=3.3$). We also determine $\log(\dot{M}/\mdot)=-6.0$.

\textbf{Sk $-69^{\circ}$ 50}.
For this mid-O-type supergiant, classified as O7(n)(f)p \citep{Walborn2010}, and revised as O7\,I(n)(f)p by PAC (priv. comm.), \citet[][]{Martins2024} reported \loglstar$=5.45$, \tstar$=35.0$~kK and \logg$=3.4$.
We obtain a close value for \loglstar($=5.4$), a lower \tstar($=33$~kK), and the same \logg($=3.4$). We also determine $\log(\dot{M}/\mdot)=-6.3$.

\textbf{[ELS2006] N11 046}. For this late-O-type dwarf, classified as O9.5\,V \citep{Evans2006}, and revised as O9.7\,III by PAC (priv. comm.), we report \loglstar$=5$, \tstar$=31.5$~kK, \logg$=4.1$ and $\log(\dot{M}/\mdot)=-7.8$.

\section{Discussion}
\label{discussion}

\subsection{Stellar parameters}
\label{Stellar_parameters}

A sequential star formation scenario has been suggested by \citet[][]{Parker1992} to explain an apparent gradient of age, extinction, and even the IMF slope from the central nebular cavity in N11 (named LH\,9) to the surrounding star-forming complexes, among which N11\,B is the most prominent \hii\ region.
In LH\,9, \citet[][]{Parker1992} derived an age of 5~Myr,
and a reddening of \ebv$=0.05$~mag.
The low value of the visual extinction could suggests a low gas density, probably lost due to stellar winds.
WR stars are known for their strong stellar winds \citep[][]{Crowther2007},
and in fact, at the very center of LH\,9 lies an extended source hosting a WR system \citep[WC4+09.5II;][]{Bartzakos2001}, consistent with the low gas density and the age of the region.
On the other hand, N11\,B is thought to be $\sim2$~Myr younger,
%with a smaller IMF slope of $-1.1$,
and with a factor of three higher extinction.
No WR sources have been found in N11\,B.

\begin{figure}
\begin{centering}
\includegraphics[width=1\linewidth]{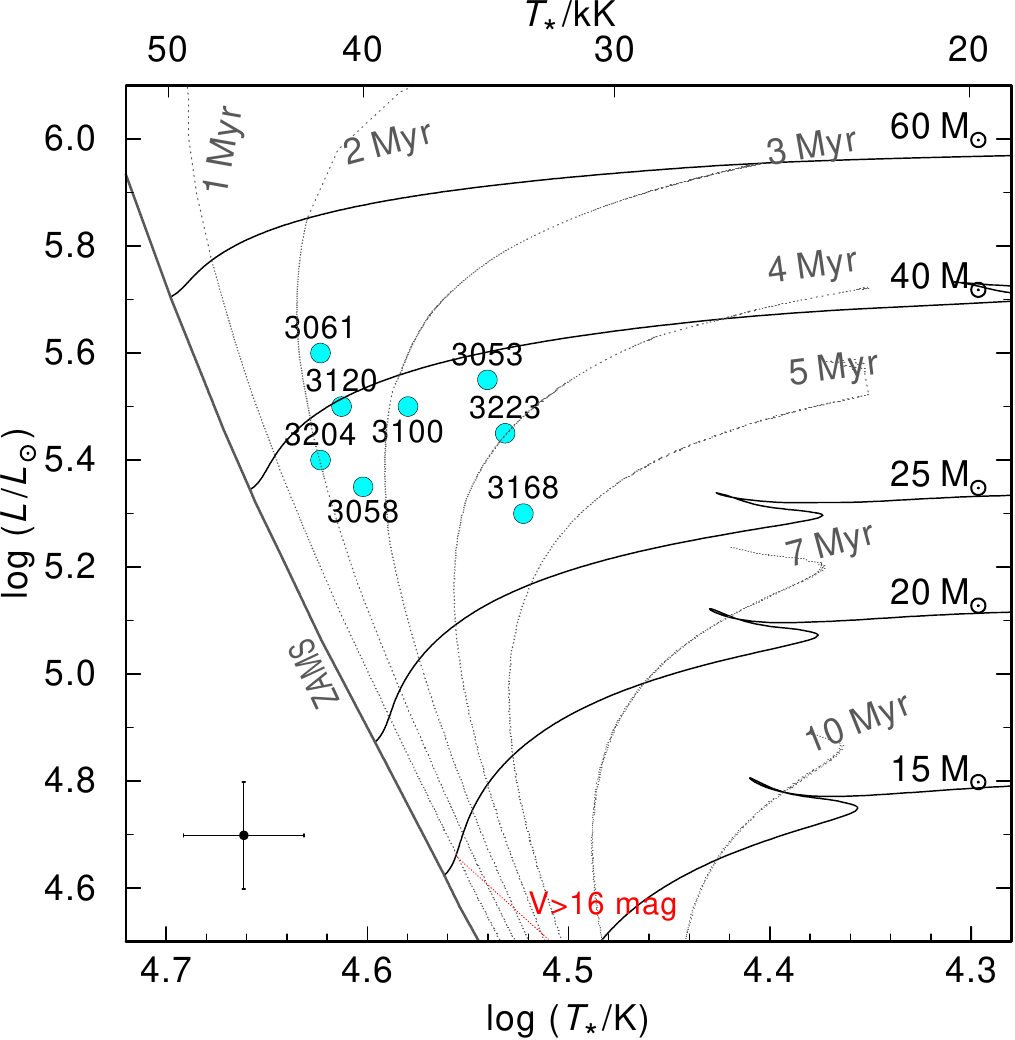}
\par\end{centering}
\caption{
H-R diagram for the sample of ULLYSES target O-type stars in N11\,B.
Evolutionary tracks (continuous lines) and isochrones (dotted lines) from \citet[][]{Brott2011} and \citet[][]{Kohler2015} with LMC composition are displayed.
The ULLYSES targets (cyan circles) are identified with the PGMW\,$\#$ from \citet[][]{Parker1992}. 
The stars have ages between $2-4.5$~Myr, and evolutionary masses of $25-45$~\Msun.
The zero-age main sequence (ZAMS) line, and the completeness threshold of V$>16$~mag are displayed (red line). Typical uncertainties are indicated.
}
\label{fig:hrd}
\end{figure}

Once we obtained \lstar\ and \tstar\ for our targets, we determined the ages and masses by using a Hertzsprung-Russell (H-R) diagram (see Fig.~\ref{fig:hrd}).
Evolutionary tracks and isochrones from \citet[][]{Brott2011, Kohler2015} for the LMC were used.
We find that the O-type stars in N11\,B are located among isochrones for ages of $2-4.5$~Myr.
The young age of this region is consistent with a pre-WR phase ($<4$~Myr).
No SNR have been observed in this region so far.
We report an age difference of $\sim2$~Myr between the stars analyzed in N11\,B and those located in the central region.
Additionally, according to the evolutionary tracks, the stellar masses are between $25-40$~\Msun. We report a median value for extinction of \ebv$=0.19$~mag in N11\,B.

\begin{figure}
\begin{centering}
\includegraphics[width=1\linewidth]{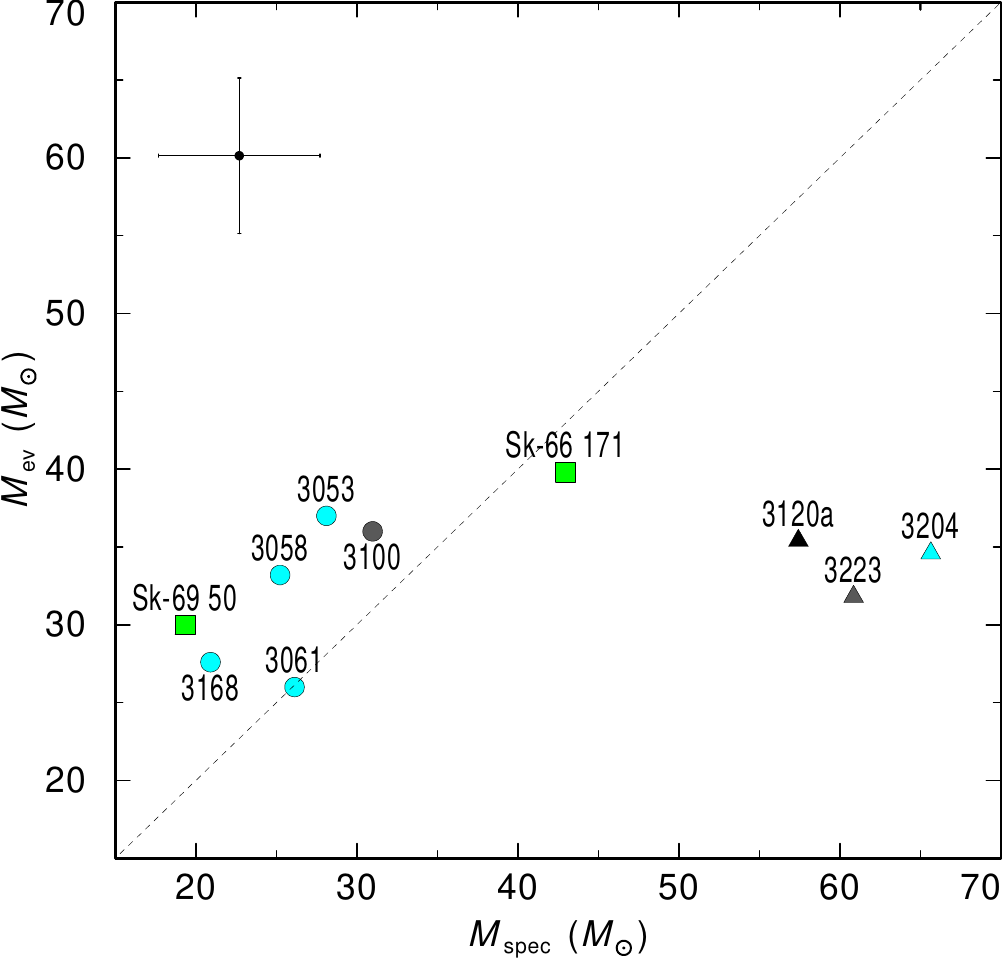}
\par\end{centering}
\caption{ 
Evolutionary (\mev) vs. spectroscopic (\mspec) masses for the analyzed O-type stars in N11\,B.
The \mspec\ values were obtained with PoWR modeling, and the \mev\ values were derived with BONNSAI by introducing PoWR-obtained parameters. The objects are indicated with their PGMW\,$\#$ ID.
In addition, two Sanduleak O-type stars (not in N11\,B), also analyzed here, are shown (green squares).
A one-to-one relation is indicated (dashed line) as a reference.
Triangles represent lower limits (see text for details). Gray color indicates confirmed SB1 binarity, and the member of a cluster is shown in black.
}
\label{fig:M_spec_ev}
\end{figure}

When comparing the spectroscopic masses (\mspec) obtained with PoWR with the evolutionary values (\mev) derived with BONNSAI (see Fig.~\ref{fig:M_spec_ev}), we notice that the stars do not follow a clear one-to-one relation, particularly above \mspec$\sim55$~\Msun.
Stars with \mspec$\sim20-45$~\Msun\ have \mev\ between $25-40$~\Msun.
However, for some stars, \mev\ is higher than \mspec\ by $\sim10$~\Msun. This is the case for PGMW\,3053, PGMW\,3168, PGMW\,3100, and PGMW\,3058, with PGMW\,3061 (and Sk $-66^{\circ}$ 171) being the exception.
On the other hand, stars with \mspec\ of $55-65$~\Msun\ have \mev\ among $30-40$~\Msun. The values for \mev\ are lower than \mev\ by $\sim20$~\Msun. This is the case for PGMW\,3120, PGMW\,3204, and PGMW\,3223.
We note that PGMW\,3120 was found to be a member of a cluster of at least three stars, and PGMW\,3223 is likely a binary SB1.
This is not the case for PGMW\,3204, the third outlier in the plot, where no evidence of binarity has been found in GIRAFFE spectra.
On the other hand, PGMW\,3100 is also a binary SB1, but the mass discrepancy is rather small.
By carefully examining the replicated values of BONNSAI, one notices that \logg\ and/or \lstar\ are underestimated for these three objects, contributing to obtaining lower \mev. The reason for this is the parameter space considered in the models used to predict the masses.

However, we also notice that a discrepancy between \mev\ and \mspec\ has been reported for O-type stars \citep[e.g.,][]{Herrero1992,Martins2012,Mahy2015,Ramachandran2018}, and even for B-types \citep{Bernini2024}. 
The mass discrepancy obtained in this study presents yet another instance of this problem, which remains unsolved so far. In that regard,
\citet[][]{Mahy2020} suggested that stars that suffer from interaction present this discrepancy.
The fact that it is well observed and documented in the Magellanic Clouds reveals that it cannot be attributed solely to the determination of luminosities.

\subsection{Wind parameters}
\label{Wind_parameters}

\begin{figure}
\begin{centering}
\includegraphics[width=1\linewidth]{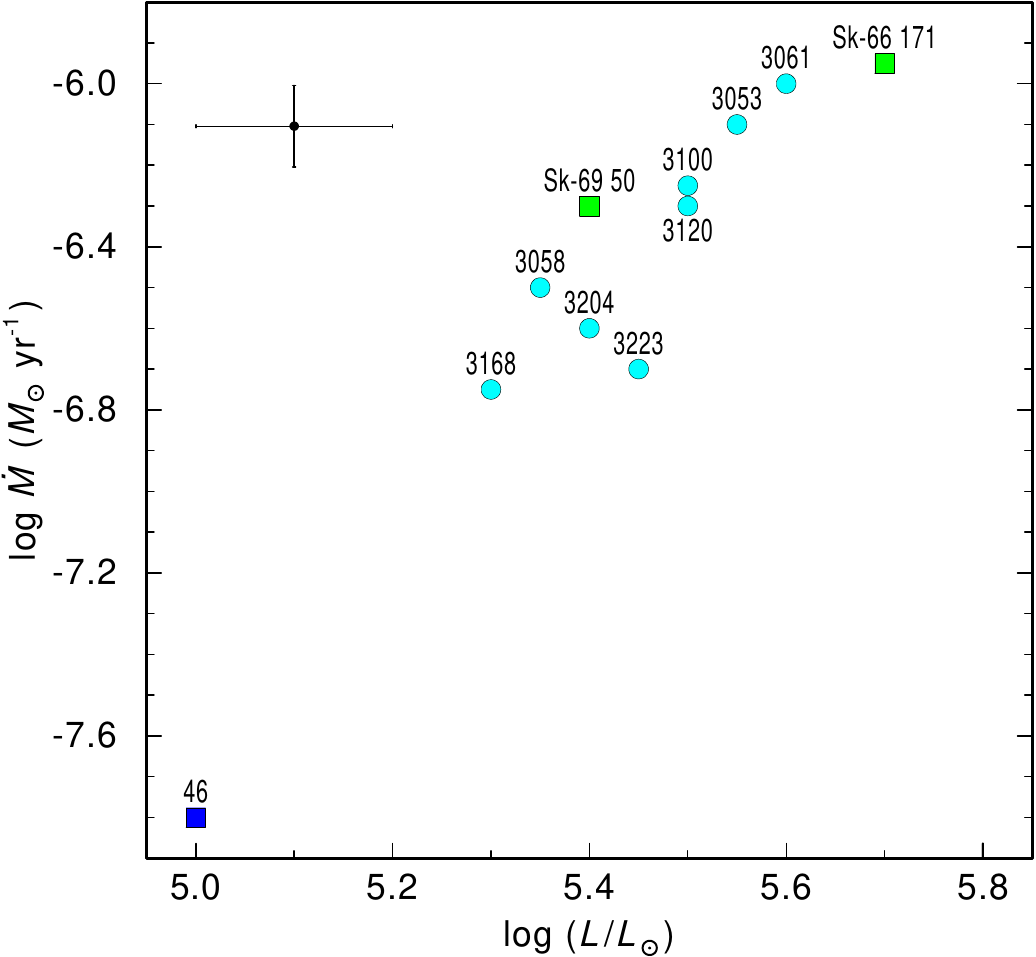}
\par\end{centering}
\caption{
Mass-loss rate ($\log\dot{M}$) vs. bolometric luminosity ($\log L_{\star}$) for the O-type stars analyzed in N11\,B.
In addition, the stars Sk $-66^{\circ}$ 171 and Sk $-69^{\circ}$ 50 (not in N11\,B) are also shown as references (green squares), as well as N11 046 (blue square).
An increasing trend between $\dot{M}$ and $\log L_{\star}$ is observed.
Notations are the same as in Fig.~\ref{fig:M_spec_ev}.
Typical uncertainties are indicated.
}
\label{fig:logL_logmdot}
\end{figure}

Determining the wind parameters of O-type stars is critical for establishing their feedback contribution and investigating the ecology of their local environment.
In this study, we determine the mass-loss rate and other key wind parameters obtained not only in the optical range, but particularly from novel UV spectra from ULLYSES. The P-Cygni lines in the UV allow us to obtain the $\dot{M}$ parameter, together with the \ha\ and \heiiwr\ lines.
In Fig.~\ref{fig:logL_logmdot}, we note an increasing trend between the mass-loss rates and the luminosities of the O-type stars in N11\,B.
We determined $\log\dot{M}$ values in the range of $-6.7$ to $-6.0$~\mdot.
Among our sample of stars, PGMW\,3061 and PGMW\,3053 have the highest mass-loss rates and are also the most luminous objects.
On the other hand, PGMW\,3168 has the weakest winds, being the least luminous star. This star also has the lowest \logg\ and therefore the lowest mass of the sample.
The additional stars analyzed here for the feedback section (Sec.~\ref{feedback}) follow the same trend.

\begin{figure}
\begin{centering}
\includegraphics[width=1\linewidth]{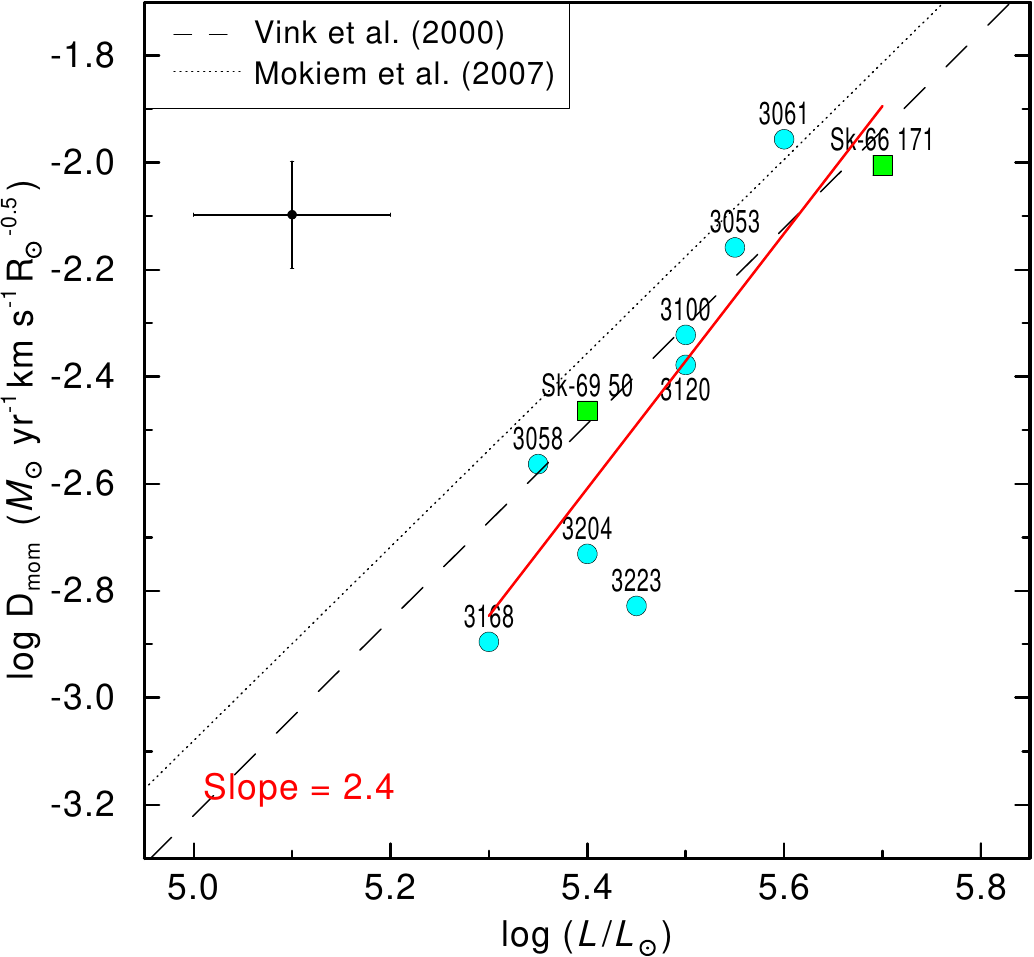}
\par\end{centering}
\caption{
Modified wind momentum (\Dmom) vs. luminosity ($\log L_{\star}$) for the O-type stars analyzed in N11\,B.
An increasing trend is observed.
Two exemplary wind momentum-luminosity relations (WLR) are displayed: the one predicted by \citet[][]{Vink2000} (dashed line) and the empirical relation obtained for LMC OB stars by \citet[][]{Mokiem2007b} (dotted line).
Notations and color-coding are the same as in Fig.~\ref{fig:logL_logmdot}.
Typical uncertainties are indicated.
A linear regression of our values is shown by a red continuous line.
}
\label{fig:logD_logmdot}
\end{figure}

However, in order to discuss the strengths of the stellar winds, it is often more useful to use the modified wind momentum (\Dmom) definition instead of $\dot{M}$ alone.
For this, we used the expression from \citet[][]{Kudritzki2000} for \Dmom, which is defined as $D\equiv \dot{M} \varv_{\infty} R_{*}^{1/2}$.
By doing that, we can compare the parameters from the modeling of our observations with the theoretical wind-momentum-luminosity relation (WLR) predicted by \citet[][]{Vink2000}, and also the empirical WLR for LMC OB stars by \citet[][]{Mokiem2007b}.
In  Fig.~\ref{fig:logD_logmdot}, we plot \Dmom\ vs luminosity for the analyzed O-type stars in N11\,B.
There is a clear increasing trend between these two parameters.
Our observations are consistent with the WLR predicted by \citet[][]{Vink2000}.
The empirical WLR for LMC OB stars by \citet[][]{Mokiem2007b} is 0.2~dex above our results.
Additionally, a linear regression in the form \( y = a + b \cdot x \) of our values is shown, with \( a = -15.5 \) and slope \( b = 2.4 \). Three objects (PGMW\,3168, PGMW\,3204, and PGMW\,3223) lie below the WLR by \citet[][]{Vink2000}. In this work, we confirmed that PGMW\,3223, which shows the highest deviation from the relation, is an SB1 binary (see Sec.~\ref{comments}). However, the other two objects are likely single.
We aim to include additional ULLYSES targets being analyzed within the collaboration to report a robust WLR for the O-stars in the LMC.

Also displayed are the O-type stars Sk $-66^{\circ}$ 171 and Sk $-69^{\circ}$ 50 in the LMC as a reference.
The physical parameters obtained with PoWR for Sk $-66^{\circ}$ 171 and Sk $-69^{\circ}$ 50 are reported in Table~\ref{tab:results4} with the purpose of quantifying the ionizing photons for stars of their sub-types.
These two stars also follow the WLR relation by \citet[][]{Vink2000}, like the O-type stars in N11\,B.

\subsection{X-ray luminosities}
\label{X-ray}

In binary systems, X-rays are expected to be produced either by the collision of stellar winds or by the accretion of the wind onto a compact object \citep[][]{Puls2008}.
The X-ray observations of massive stars in \citet[][]{Oskinova2005} indicate that the correlation between bolometric and X-ray luminosity, known for single O-type stars, also applies to O+O and WR+O binaries.
There is currently no conclusive evidence to determine whether binary stars exhibit X-ray emission in excess compared to single stars.
Low luminosities do not necessarily indicate the absence of a secondary companion. Also, there is still the possibility that the companion is in a quiescent state.
%Nidia says: Many binaries show phase-locked X-ray variations (e.g. works by M. Corcoran).

Using XMM-Newton observations, \citet[][]{Naze2004} reported diffuse emission and X-rays sources associated with the massive stars in N11\,B.
Later, with the more sensitive and higher spatial resolution of the Chandra X-ray Observatory, \citet[][]{Naze2014} investigated the point sources associated with the O-type stars in N11 and determined \lxlbol\ values between $-6.5$ and $-7$. Although several values are upper limits, these results led them to conclude that these stars are highly magnetic or colliding-wind binary systems. Additionally, \citet[][]{Crowther2022} showed that OB and WR stars in the Galaxy follow a relation of \lx$=10^{-7}$~\lbol, which is compatible with \citet[][]{Naze2009}.

We have included X-rays in our PoWR models to reproduce available spectral features in the UV.
X-ray continuum emission from shock-heated plasma is expected to originate in the winds of the stars.
The O-type stars analyzed here are not strong X-ray point sources according to \citet[][]{Naze2014}.
We list their \lxlbol\ values in Table~\ref{tab:xray}.
\citet[][]{Naze2014} reported upper limits for four of our objects: PGMW\,3053, PGMW\,3058, PGMW\,3168, and PGMW\,3223.
Here, we report their \lxlbol\ values obtained by our spectroscopic modeling approach, independently of previous X-ray observations.
Except for one source, we obtained values below the upper limits reported by \citet[][]{Naze2014}.
The values for the rest of the sample are consistent with previous observations.
We report the ratio between X-ray and bolometric luminosity \lxlbol\ for the O-type stars in Table~\ref{tab:xray}, including the assumed \ff, \tx, and \rmin\ parameters.
We should also keep in mind the variable nature of the X-ray sources, but we still lack the necessary multi-epoch UV spectra to study this matter.
However, we conclude that the sample of massive stars in N11\,B analyzed here are not strong X-ray emitters. We find typical \lxlbol\ values for our targets in the range of $-7.0$ to $-6.6$, with the earliest star in N11\,B, PGMW\,3061, having \lxlbol$=-6.6$, making it the brightest X-ray emitter in the region.

\subsection{Chemical abundances}
\label{abundances}

It is commonly assumed that the faster the star rotates, the higher the rotational mixing,
resulting in a higher N abundance in the surface of the star.
However, for a sample of early B-type stars in the LMC, \citet[][]{Hunter2008} already reported evidence contradicting this assumption.
Among their findings, they show highly nitrogen-enriched slow rotators ($<50$~\kms) and nitrogen-unenriched fast rotators ($\sim300$~\kms), thus challenging the often-invoked rotational mixing hypothesis.

Interestingly, \citet[][]{Maeder2009} suggest that not only rotation plays a role in mixing but also age, mass, and metallicity.
\citet[][]{Petrovic2005} have proposed a scenario where nitrogen-unenriched fast rotators may be explained by a close binary companion increasing the rotation rate of the star.
On the other hand, \citet[][]{Wolff2007} suggested that this correlation originates early in the star formation process due to the magnetic locking of the star to the accretion disk, known as the fossil field hypothesis. However, we lack observations of magnetic field strength for our sample of stars. % (check).
Previously, \citet[][]{Brott2011} tested the rotational mixing hypothesis by simulating LMC massive stars. However, due to the lack of reported nitrogen abundances for stars with temperatures greater than 35~kK, meaning that the O-types were practically excluded from their study, their results would apply only to early B-type stars.

\begin{figure}
\begin{centering}
\includegraphics[width=1\linewidth]{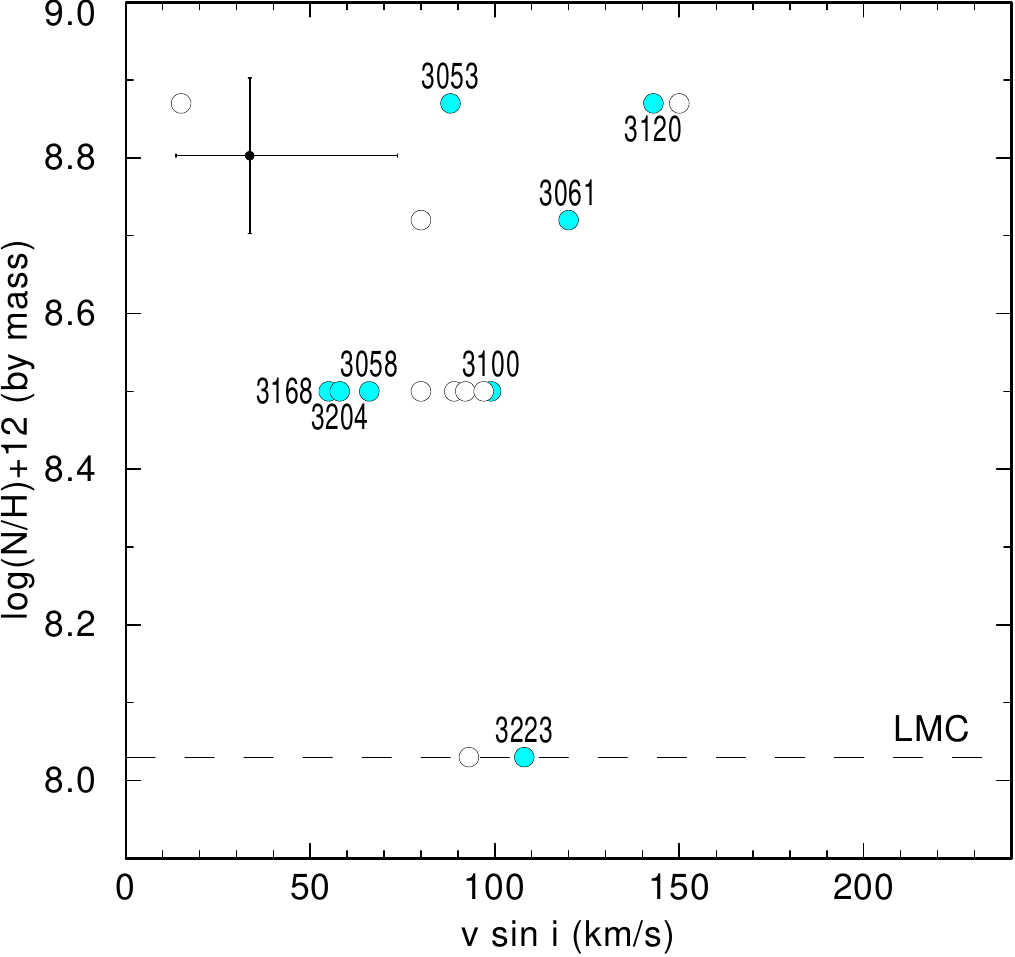}
\par\end{centering}
\caption{
Surface nitrogen abundances ($\log\,(N/H)+12$) by mass vs. projected rotational velocity (\vrot) for the O-type stars analyzed in N11\,B.
Most of the objects are nitrogen-enriched, up to a factor of seven.
A standard nitrogen abundance in the LMC ($\log\,(N/H)+12=6.88$) is indicated as a reference.
With a median value of \vrot$=100$~\kms, none of the stars are considered slow or fast rotators. % ($55<$\vrot$<143$~\kms).
The non-rotational broadening for each star (\vmac) is also indicated with an empty circle. No trend is apparent. Typical uncertainties are indicated.
}
\label{fig:N_vs_vrot}
\end{figure}

The chemical abundances of C, N, and O for the analyzed O-type stars in N11\,B are reported in Table~\ref{tab:results}.
None of the studied objects can be considered fast rotators ($>300$~\kms), nor are they slow rotators.
We note that the majority of the Galactic O-type stars analyzed by \citet[][]{Holgado2018} have \vrot\ $<120$~\kms, with the distribution of \vrot\ reaching a maximum around $40-80$~\kms.
Here, we report rotational velocities in the range between $55-143$~\kms, with a median value of \vrot\ $=100$~\kms, consistent with findings by \citet[][]{Ramirez2013} in 30 Dor.
None of the determined \vrot\ of our stars is higher than $150$~\kms. Despite this fact, half of the targets do display N enrichment, up to a factor of 7, corresponding to 3.5~\zsun. This is the case for PGMW\,3053 and PGMW\,3120.
The surface nitrogen abundances that we find are comparable to the findings for LMC O-type stars by \citet[][]{Ramachandran2018}, showing a nitrogen enrichment between 1-10 for most of their objects.
In Fig.~\ref{fig:N_vs_vrot}, we compare the surface nitrogen abundances ($\log\,(N/H)+12$) by mass vs \vrot\ for the analyzed objects.
Also in this figure, we indicate the non-rotational contribution ( \vmac) to the broadening for each star.
According to these results, we cannot conclude that the nitrogen surface abundances of the stars are modified by the effects of rotational mixing.

We note that we were not able to model \siiva\ and \siivb\ when they are in emission.
In our models, we obtain these features in absorption.
However, reducing the \siiv\ abundance by a factor of 4 decreases the strength of the lines in absorption.
We take note of this result for future analysis efforts.

\subsection{Macroclumping}
\label{macroclumping}

For the sake of simplicity, stellar winds of massive stars have been assumed to be homogeneous and stationary. Using PoWR models on UV observations, \citet{Oskinova2016} point towards highly inhomogeneous stellar winds.
Phosphorus lines in the UV, \pva\ and \pvb\ in particular, are considered mass-loss rate diagnostic features \citep[][]{Fullerton2006}.
However, there has been an observed discrepancy between the values obtained using these phosphorus lines in the UV and those obtained using only the optical diagnostic features.
\pv\ tends to appear weaker than expected based on its $\dot{M}$ determined using optical lines \citep[e.g.,][]{Massa2003, Oskinova2016}.
In order to make sense of this discrepancy, \citet{Massa2003} points to three possible scenarios:
either 1) the $\dot{M}$ is lower than the values obtained by using optical features (\heiiwr\ and \ha); 2) the \emph{assumed}  phosphorus abundance is overestimated; and/or 3) the winds are strongly clumped.
We explored these scenarios.

The effects of including macroclumping to solve this discrepancy have been discussed in \citet{Oskinova2016}.
Here, we use the macroclumping approach for the stars with \fuse\ spectra: PGMW\,3058, PGMW\,3223 and PGMW\,3120.
We confirm the effect of including macroclumping in the modeled \pv\ lines.
However, we also noted that other spectral features are affected, indicating that this approach requires further investigation.

\citet{Massa2003} also remark that there are \emph{not direct} measurements of the phosphorus abundance in the LMC, and that the Ne burning history may cause the trend in abundance to differ for phosphorus in the LMC.
Thus, in this work, we considered this scenario worth exploring.
We find that by reducing the abundances by a factor of four, the modeled \pv\ lines can match the observed features for two of our objects with FUSE spectra: PGMW\,3058 and PGMW\,3223.
We also noted this effect in Sk $-66^{\circ}$ 171 and Sk $-69^{\circ}$ 50.
This exercise has been  explored in the past. Notably, \citet{Bouret2012} also had to decrease the phosphorus abundance to model the spectral lines with CMFGEN.
Interestingly, the model for PGMW\,3120, as well as for N11 046, does not require either macroclumping or lower phosphorus abundance to match the \pv\ features in the UV.

We should consider that if the iron abundances are also assumed to be around half solar, there would be no reason to speculate further on lower phosphorus abundances, as the two elements would be expected to follow the same trend.
However, we take note of this result, which deserves further exploration with ULLYSES targets in the LMC.

\subsection{Stellar feedback}
\label{feedback}

The main contributors to the feedback in star-forming regions are expected to be the hot and luminous OB stars, but particularly the WR population \citep[see][]{Crowther2019}, and SNe remnants.
In N11\,B, there are no WR stars or even candidates, nor are there any known SNe yet.
This could be attributed to the young age of the region ($2-4.5$~Myr).
Thus, N11\,B can be considered an exemplary environment to study the feedback contribution solely by its massive O-type stellar population.
In this region, we find that two stars display the highest mass-loss rates: PGMW\,3053 and PGMW\,3061, both with a $\log(\dot{M}/\mdot)=-6.0$, and the highest \Dmom\ is found in PGMW\,3061 (see Figs.~\ref{fig:logL_logmdot}, \ref{fig:logD_logmdot}).

Individual ionizing photons and the mechanical energy of the analyzed ULLYSES targets are reported in Table~\ref{tab:results}.
The total amount of ionizing photons produced by the 25 O-type stars known in N11\,B is also estimated. For this, the ionizing feedback for each of the O-type stars, based on their spectral type (see Table~\ref{tab:feedback}), was considered to quantify the total ionizing feedback in the region.
Stars of different spectral types in N11\,B, particularly later subtypes, were not among the ULLYSES targets.
We tackle this problem using Sk $-69^{\circ}$ 50, Sk $-66^{\circ}$ 171, and N11 046.
Although these stars are not located in the region studied here, they share spectral classifications with some of the objects.
Sk $-69^{\circ}$ 50, with a spectral type O7(n)(f)p, was used to estimate the ionizing photons of the only O7V star in N11\,B: PGMW\,3102.
Sk $-66^{\circ}$ 171, with a spectral type O9\,Ia, was used to estimate the ionizing photons of the O9.5III star PGMW\,3045.
N11 046, with a spectral type O9.5 V, was used to estimate the ionizing photons of five O9.5V stars, namely: PGMW\,3063 (O9V), PGMW\,3115 (O9V), PGMW\,3103 (O9.5), PGMW\,3016 (O9.5), and PGMW\,3042 (O9.5Vn).
Two stars, PGMW\,3173 and PGMW\,3264, are reported with uncertain spectral types between O4-O6V and O3-O6V, respectively.
For these, the ionizing photons of an intermediate spectral type, PGMW\,3120a (O5.5V), were used.
Including these stars allows for estimating the total ionizing photon fluxes $\sum$\QH, $\sum$\QHeI, and $\sum$\QHeII\ produced by the 25 O-type stars in N11\,B.

\begin{table}
\begin{center}
\footnotesize
  \caption{Individual ionizing photons of the O-type stars in N11\,B.}
\setlength{\tabcolsep}{0.15\tabcolsep}
  \label{tab:feedback}
\begin{tabular}{lcccccc}
\hline\hline
\rule{0cm}{2.2ex} ID$^{a}$           & SpC     & \logQH\ & \logQHeI\ & \multicolumn{2}{c}{\logQHeII} \\
PGMW             &                   &      &       &  no-XR    & XR \\
                 &                   &[ph\,s$^{-1}$]& [ph\,s$^{-1}$]& [ph\,s$^{-1}$] & [ph\,s$^{-1}$]  \\
 (1)             & (2)               & (3)  & (4)   & (5)    & (6)   \\
\hline
3061$^{\dagger}$ & ON2 III(f*)$^{c}$  & 49.4 & 48.6  & 40.8   & 41.7  \\ %c
3058$^{\dagger}$ & O3 V((f*))$^{b}$   & 49.1 & 48.2  & 43.5   & 43.5  \\%u
3209             & O3 III(f*)+OB$^{a}$& 49.4 & 48.5  & 43.8   & \dots \\
%\hline
3120a$^{\dagger}$& O5.5V((f*))$^{b}$ & 49.2 & 48.4  & 40.5   & 41.3  \\ %u
3120b            & O5.5 V?            & 49.2 & 48.4  & 40.5   & \dots \\
3120c            & O5.5 V?            & 49.2 & 48.4  & 40.5   & \dots \\
3100$^{\dagger}$ & O5 III(f)$^{b}$   & 49.2 & 48.3  & 40.0   & 41.2  \\ %u
%\hline
3053$^{\dagger}$ & O6.5 II(f)$^{b}$  & 49.2 & 48.1  & 39.5   & 40.9  \\ %update
3224             & O6 III$^{a}$	     & 49.2 & 48.1  & 39.5   & \dots \\
3070             & O6 V$^{a}$	     & 49.1 & 48.4  & 43.7   & \dots \\
3204$^{\dagger}$ & O6 Vz((f))$^{b}$  & 49.1 & 48.4  & 43.7   & 44.1  \\ %u
3073             & O6.5 V$^{a}$	     & 49.1 & 48.4  & 43.7   & \dots \\
3126             & O6.5 V$^{a}$	     & 49.1 & 48.4  & 43.7   & \dots \\
%\hline
3168$^{\dagger}$ & O7.5 III(f)$^{b}$	 & 48.8 & 47.5  & 39.0   & 40.8  \\ %u
3102             & O7 V$^{a}$	     & 49   & 47.5  & 38.5   & \dots \\
3223$^{\dagger}$ & O8 Vz$^{b}$         & 48.9 & 47.5  & 38.7   & 40.7 \\ %u
3089             & O8 V$^{a}$	     & 48.9 & 47.5  & 38.7   & \dots \\
3123             & O8.5 V$^{a}$	     & 48.9 & 47.5  & 38.7   & \dots \\
3045             & O9.5III$^{a}$	 & 49   & 47    & 38.5   & \dots \\
3063             & O9V$^{a}$	     & 48.1 & 45.9  & 39.4   & \dots \\
3115             & O9V$^{a}$	     & 48.1 & 45.9  & 39.4   & \dots \\
3103             & O9.5:IV:$^{a}$    & 48.1 & 45.9  & 39.4   & \dots \\
3016             & O9.5:V$^{a}$      & 48.1 & 45.9  & 39.4   & \dots \\
3042             & O9.5Vn$^{c}$      & 48.1 & 45.9  & 39.4   & \dots \\
3173             & O4-O6V$^{a}$	     & 49.2 & 48.3  & 40.0   & \dots \\
3264             & O3-O6V$^{a}$	     & 49.2 & 48.3  & 40.0   & \dots \\
\hline
\multicolumn{2}{l}{Total ionizing budget:} &  $\log$ ($\sum$\QH) & $\log$ ($\sum$\QHeI) & $\log$ ($\sum$\QHeII)  & $+$XR  \\
                 &                   & 50.5 & 49.6  & 44.4   & 44.8   \\
\hline
\end{tabular}
\end{center}
1) PGMW\,$\#$ ID \citep[][]{Parker1992};
2) spectral type classification (SpC);
log of the number of ionizing photons per second for (3) H (\QH);
(4) \hei\ (\QHeI); and (5) \heii\ models (\QHeII) without considering X-rays; and \QHeII\ considering X-rays ($+$XR).
$^{(a)}$\citet[][]{Parker1992};
$^{(b)}$PAC priv. comm.;
$^{(c)}$\citet[][]{Evans2006}.
${\dagger}$ Stars analyzed in this work to quantify the total Q of the O-type stars in N11\,B.
For the remaining stars, the Q value from the object with the closest spectral type is assumed.
The total ionizing photons by the 25 O-type stars in N11\,B are estimated.
\end{table}

We estimate that the combined ionizing photon fluxes from the analyzed stars in N11\,B are: \QH$=3.0\times10^{50}$~ph\,s$^{-1}$, \QHeI$=3.5\times10^{49}$~ph\,s$^{-1}$, and \QHeII$=2.7\times10^{44}$~ph\,s$^{-1}$.
If X-rays are taken into account, then \QHeII$=5.9\times10^{44}$~ph\,s$^{-1}$, a factor of about 2 higher than without considering X-rays.
We note that this estimate relies on the assumption that our stars share the physical properties of a given spectral type.
The total \lmec\ is $\sum$\lmec$=2900$\lsun, with PGMW\,3061, the earliest type of the sample, being the main contributor to the total amount of energy by almost half of the total mechanical luminosity, with \lmec$=960$\lsun.
PGMW\,3058, an O3 V star, turned out to be the main contributor in \QHeII, with \logQHeII$=43.5$~ph\,s$^{-1}$.

%Sugggested by Paul Crowther
The Tarantula Nebula in the LMC is considered the reference for extragalactic \hii\ regions. We compare our results with this region.
\citet{Crowther2019} list the contributions of very massive stars, early O-types, and WR stars to the Lyman continuum feedback in 30 Dor.
Details of the census are reported in \citet{Doran2013}.
This complex is the brightest and most important cluster in the LMC, with 570 O-type stars, 523 B-type stars, and 28 WR stars.
WR stars are the main contributors to ionizing feedback in this region \citep[][see their Tab. 4]{Crowther2019}, each one producing an order of magnitude more ionizing photons than a single O-type star; the latter objects inject \logQH$\sim$49~ph\,s$^{-1}$ into their local environment.
On the other hand, B-type stars are not expected to significantly contribute to ionizing feedback. Their contribution in mechanical energy could have an impact, considering that \lmec\ is driven by $\dot{M}$ and terminal velocity ($=0.5\dot{M}\varv_\infty^{2}$).
B-type stars in the SMC have roughly an order of magnitude lower mass-loss rate and terminal velocities that are 2 to 5 times lower \citep{Bernini2024}.
While these regions host several "examples of stellar exotica" \citep[][]{Crowther2019}, N11\,B is a region with a rather modest OB population, with neither WR stars nor supernovae yet, as mentioned above. However, we see precisely this property as an opportunity to quantify the feedback solely from typical O-type stars of different subtypes.

\citet{Pellegrini2012} reported L(H$\alpha)=1.023\times10^{39}$\ergs\ for the N11 region, named MCELS-L65 or DEM L 34 in their catalogue, making it the second brightest \hii\ region in the LMC, surpassed only by the 30 Dor nebula, MCELS L-328 or DEM L 263, where they reported L(H$\alpha)=4.571\times10^{39}$\ergs.
The luminosity reported by \citet{Pellegrini2012} in N11 corresponds to \QH$=7.27\times10^{50}$~ph\,s$^{-1}$, which is obtained using the expression \QH$=7.1\times10^{11}$L(H$\alpha$)\,s$^{-1}$, given by \citet{Kennicutt1995} to determine the number of Lyman continuum photons under Case B recombination and ionization-bounded nebula assumptions.
Here, we estimated the ionizing budget in N11\,B, which is the brightest star-forming region in N11. We report $\sum$\,\QH$=3.0\times10^{50}$~ph\,s$^{-1}$.
This value is consistent with the total budget of ionizing photons in N11.

\section{Conclusions}
\label{conclusions}

We investigated in detail the stellar parameters, wind features, mechanical, and ionizing feedback of eight ULLYSES targets: benchmark O-type stars with spectral types ranging from O2 to O8 and with different luminosity classes, in the N11\,B star-forming region in the LMC. State-of-the-art PoWR models and other standard analysis tools were used in a homogeneous approach.
Novel UV \hst/\stis\ and \cosi\ high-quality spectra from the ULLYSES project, as well as optical spectra from \xshooter\ at VLT, were analyzed, along with the most recent Gaia DR3 photometry.
Next, we summarize our main findings.

\begin{itemize}
\item We report ages between $2-4.5$~Myr and masses of $30-60$\Msun\ for the O-type stars analyzed in N11\,B.
Such young ages are consistent with the absence of WR stars, the fact that no SNe have been reported in this region yet, and with the hypothesis of a sequential star formation scenario from the center of N11 to the surrounding regions.

\item We show that the \mev\ and \mspec\ do not follow a clear one-to-one relation. This mass discrepancy is consistent with previous findings in the literature \citep[e.g.,][in the Magellanic Clouds]{Herrero1992}.

\item The analyzed O-type stars follow a wind-momentum luminosity relation, which is consistent with previously reported observational relations and with the theoretically predicted relation by \citet{Vink2000}.

\item We investigated whether nitrogen enrichment correlates with \vrot.
The O-type stars analyzed here have rotational velocities ranging from 55 to 143~\kms, with a median value of 100~\kms.
Non-rotational components were considered.
We observe nitrogen enrichment in most of the stars, up to a factor of 7. However, we do not find a trend with \vrot, challenging the hypothesis that nitrogen enrichment is correlated with rotation in massive stars.

\item The effect of 'macroturbulence' on line broadening was found to be non-negligible. On the contrary, \vmac\ is an important contributor and in most cases, it even dominates the line broadening. Our result is consistent with previous findings of \citep[e.g.,][]{SimonDiaz2017, Holgado2018}.
We recommend not ignoring its contribution to the total broadening of a line, and avoiding the use of non-metallic lines, when possible, for determining \vrot.
Additionally, we find that even for the same star, the metal lines give different values compared to those obtained with the Balmer and He lines. Therefore, these lines should be avoided when possible.

\item In the UV range, we observed that reducing the phosphorus abundances by a factor of four results in the modeled \pv\ lines matching the observed lines, similar to the well-studied macroclumping effect.

\item By including X-rays in our PoWR models, we quantified the amount of X-rays required to reproduce important UV spectral features in our stars.
We independently report $L_\mathrm{X}/L_\mathrm{bol}$ ratios, between $-7.5$ to $-6.6$, including for the first time the values for four O-type stars in N11\,B.
These values are consistent with soft X-ray emitters and previous findings \citep[][]{Naze2014, Crowther2022}.

\item The fact that N11\,B is free of exotic objects gives us the opportunity to study the ionizing feedback in a star-forming region exclusively through its O-type stars.
We report total ionizing fluxes of $\log(\sum$\QH$)=50.5$~ph\,s$^{-1}$, $\log(\sum$\QHeI$)=49.6$~ph\,s$^{-1}$, and $\log(\sum$\QHeII$)=44.4$~ph\,s$^{-1}$.
If X-rays are considered, then $\log(\sum$\QHeII$)=44.8$ph\,s$^{-1}$, which represents an additional 0.4~dex of ionization.

\item The total mechanical luminosity of the eight analyzed stars is $\sum$\,\lmec$=2900$~\lsun.
PGMW\,3061, the earliest analyzed star (ON2\,III(f*)), is the main contributor with $\log$ (\lmec/\lsun)$=3.0$.
It is a factor of $3-14$ stronger in \lmec\ compared to the next brightest and faintest sources analyzed, respectively.
PGMW\,3061 is also the main contributor to \QH\ and \QHeI.
PGMW\,3058, an O3V((f*)) star, is the main contributor to \QHeII.
Given the absence of WR stars and SNe remnants, the O-type stars alone provide the feedback in N11\,B.

\end{itemize}

Summarizing, this work is part of a larger project aimed at determining the stellar and wind parameters of the OB stars in the ULLYSES sample.
This study is among the initial steps in using PoWR models to achieve this goal in a homogeneous way.

The next step is to increase the sample size analyzed with PoWR and statistically compare the results with theoretical predictions.
Here, we presented the methodology required for this, including the analysis of novel UV spectra where key wind parameters are found, along with other features to constrain the stellar properties, particularly more reliable luminosities and extinction of the stars with UV fluxes and Gaia photometry.

\begin{acknowledgements}
This study is based on observations collected at the European Southern Observatory under \eso\ program 106.211Z.001.
Observations obtained with the NASA/ESA Hubble Space Telescope were retrieved from the Mikulski Archive for Space Telescopes (MAST) at the Space Telescope Science Institute (STScI). STScI is operated by the Association of Universities for Research in Astronomy, Inc. under NASA contract NAS 5-26555.
VMAGG is funded by the Deutsche Forschungsgemeinschaft (DFG - German Research Foundation), grant number 443790621.
This work has made use of data from the European Space Agency (ESA) mission
{\it Gaia} (\url{https://www.cosmos.esa.int/gaia}), processed by the {\it Gaia}
Data Processing and Analysis Consortium (DPAC,
\url{https://www.cosmos.esa.int/web/gaia/dpac/consortium}). Funding for the DPAC
has been provided by national institutions, in particular the institutions
participating in the {\it Gaia} Multilateral Agreement.
AACS, VR, and MBP are supported by the Deutsche Forschungsgemeinschaft (DFG - German Research Foundation) in the form of an Emmy Noether Research Group -- Project-ID 445674056 (SA4064/1-1, PI Sander) and acknowledge funding from the Federal Ministry of Education and Research (BMBF) and the Baden-Württemberg Ministry of Science as part of the Excellence Strategy of the German Federal and State Governments. AuD acknowledges NASA ATP grant number 80NSSC22K0628 and support by NASA through Chandra Award number TM1-22001B and GO2-23003X issued by the Chandra X-ray Observatory 27 Center, which is operated by the Smithsonian Astrophysical Observatory for and on behalf of NASA under contract NAS8-03060.
ACGM thanks the support from project 10108195 MERIT (MSCA-COFUND Horizon Europe).

\end{acknowledgements}

\bibliographystyle{aa}
\bibliography{gomezgonzalez2024b}

%\section{Appendix}
\begin{appendix}

\section{Log of the observations}
\label{log}

The log of the observations used for the ULLYSES targets in N11\,B analyzed in this work is listed in Table~\ref{tab:spec_obs} and Table~\ref{tab:spec_obs2}.

\begin{table*}
\begin{centering}
\scriptsize
  \caption{Log of the spectroscopic observations used for the analysis of the ULLYSES targets in N11\,B.}
\setlength{\tabcolsep}{1\tabcolsep}
  \label{tab:spec_obs}
\begin{tabular}{ccccccccccc}
\hline\hline
\rule{0cm}{2.2ex} Object & Telescope & Instrument & ID   & $\lambda$   & R    & Date       & Exp. t. & SNR & P.I. & slitsize / PA \\ %& seeing \\
Name        &           &            &            & (\AA) &            & (yy/mm/dd) & (s) & & & (arcsec / deg)  \\
(1)         & (2)       & (3)        & (4)  & (5)        & (6)   & (7)        & (8)        & (9) & (10) & (11)  \\
\hline
PGMW\,3053& \fuse & \fuse MDRS  & FUV &  904--1189 & 15000--20000 & 2001/11/14 & 8377 & & Roman-Duval, J. & $4\times20$  \\
    & \hst & \cosi/G130M & FUV & 1131--1428 & 12000--16000 & 2020/08/08 &  475 &     & Roman-Duval, J. & D$2.5$  \\
    & \hst & \cosi/G160M & FUV & 1418--1790 & 13000--20000 & 2020/08/08 &  699 &     & Roman-Duval, J. & D$2.5$  \\
    & \eso-VLT-U2& \giraffe\/MOS& 1a   & 3849--4048 & 22000 & 2003/12/05 & 6825 & 191 & Smartt, S. & D$1.2$  \\ %s 0.8 
    & \eso-VLT-U2& \giraffe\/MOS& 1b   & 3849--4048 & 22000 & 2003/10/16 & 6825 & 178 & Smartt, S. & D$1.2$  \\
    & \eso-VLT-U2& \giraffe\/MOS& 2a   & 4030--4200 & 30000 & 2003/12/06 & 6825 & 134 & Smartt, S. & D$1.2$  \\ 
    & \eso-VLT-U2& \giraffe\/MOS& 2b   & 4030--4200 & 30000 & 2003/12/05 & 6825 & 136 & Smartt, S. & D$1.2$  \\
    & \eso-VLT-U2& \giraffe\/MOS& 3a   & 4180--4400 & 23000 & 2003/12/06 & 6825 & 133 & Smartt, S. & D$1.2$  \\
    & \eso-VLT-U2& \giraffe\/MOS& 3b   & 4180--4400 & 23000 & 2003/12/06 & 6825 & 136 & Smartt, S. & D$1.2$  \\    
    & \eso-VLT-U2& \giraffe\/MOS& 4a   & 4340--4590 & 20000 & 2003/12/07 & 6825 & 261 & Smartt, S. & D$1.2$  \\   
    & \eso-VLT-U2& \giraffe\/MOS& 4b   & 4340--4590 & 20000 & 2003/12/07 & 6825 & 211 & Smartt, S. & D$1.2$  \\ 
    & \eso-VLT-U2& \giraffe\/MOS& 5a   & 4537--4761 & 23000 & 2003/12/08 & 6825 & 259 & Smartt, S. & D$1.2$  \\ 
    & \eso-VLT-U2& \giraffe\/MOS& 5b   & 4537--4761 & 23000 & 2003/12/07 & 6825 & 236 & Smartt, S. & D$1.2$  \\
    & \eso-VLT-U2& \giraffe\/MOS& 6a   & 6300--6690 & 17000 & 2003/11/12 & 6825 & 265 & Smartt, S. & D$1.2$  \\      
    & \eso-VLT-U2& \giraffe\/MOS& 6b   & 6300--6690 & 17000 & 2003/10/12 & 6825 & 223 & Smartt, S. & D$1.2$  \\
    & \eso-VLT-U3& \xshooter\   & UVB  & 2989--5560 & 6655  & 2020/12/16 &  750 & 185 & Vink, J. S. & $0.8\times11$ /  $-170$ \\ %& 0.435 \\
    & \eso-VLT-U3& \xshooter\   & VIS  & 5337--10200& 11333 & 2020/12/16 &  820 & 121 & Vink, J. S. & $0.7\times11$ /  $-170$ \\ %& 0.45 \\     
\hline
PGMW\,3058& \hst & \cosi/G130M  & FUV & 1130--1428 & 12000--16000 & 2021/04/11 & 2066   &     & Roman-Duval, J. & D$2.5$  \\
    & \hst       & \cosi/G160M  & FUV & 1419--1790 & 13000--20000 & 2021/04/11 & 1996   &     & Roman-Duval, J. & D$2.5$  \\
    & \eso-VLT-U2& \giraffe\/MOS& 1a   & 3849--4048 & 22000        & 2003/12/05 & 6825   & 102 & Smartt, S. & D$1.2$  \\
    & \eso-VLT-U2& \giraffe\/MOS& 1b   & 3849--4048 & 22000        & 2003/10/16 & 6825   & 94  & Smartt, S. & D$1.2$  \\
    & \eso-VLT-U2& \giraffe\/MOS& 2a   & 4032--4203 & 30000        & 2003/12/06 & 6825   & 83  & Smartt, S. & D$1.2$  \\ 
    & \eso-VLT-U2& \giraffe\/MOS& 2b   & 4032--4203 & 30000        & 2003/12/05 & 6825   & 84  & Smartt, S. & D$1.2$  \\
    & \eso-VLT-U2& \giraffe\/MOS& 3a   & 4183--4395 & 23000        & 2003/12/06 & 6825   & 133 & Smartt, S. & D$1.2$  \\
    & \eso-VLT-U2& \giraffe\/MOS& 3b   & 4183--4395 & 23000        & 2003/12/06 & 6825   & 137 & Smartt, S. & D$1.2$  \\     
    & \eso-VLT-U2& \giraffe\/MOS& 4a   & 4340--4589 & 20000        & 2003/12/07 & 6825   & 139 & Smartt, S. & D$1.2$  \\   
    & \eso-VLT-U2& \giraffe\/MOS& 4b   & 4340--4589 & 20000        & 2003/12/07 & 6825   & 127 & Smartt, S. & D$1.2$  \\ 
    & \eso-VLT-U2& \giraffe\/MOS& 5a   & 4537--4761 & 23000        & 2003/12/08 & 6825   & 153 & Smartt, S. & D$1.2$  \\ 
    & \eso-VLT-U2& \giraffe\/MOS& 5b   & 4537--4761 & 23000        & 2003/12/07 & 6825   & 146 & Smartt, S. & D$1.2$  \\
    & \eso-VLT-U2& \giraffe\/MOS& 6a   & 6299--6691 & 17000        & 2003/11/12 & 6825   & 145 & Smartt, S. & D$1.2$  \\      
    & \eso-VLT-U2& \giraffe\/MOS& 6b   & 6299--6691 & 17000        & 2003/10/12 & 6825   & 138 & Smartt, S. & D$1.2$  \\
    & \eso-VLT-U3& \xshooter\   & UVB  & 2989--5560 & 6655  & 2021/01/02 & $2\times1655$ & 150 & Vink, J. S. & $0.8\times11$ / $-175$ \\ %& 0.435 \\
    & \eso-VLT-U3& \xshooter\   & VIS  & 5337--10200& 11333 & 2021/01/02 & $2\times1725$ &  90 & Vink, J. S. & $0.7\times11$ / $-175$ \\ %& 0.435\\            
\hline
PGMW\,3061& \hst & \cosi/G130M  & FUV  & 1131--1429 & 18000 & 2023/04/29 & 2066 & & Roman-Duval, J. & D$2.5$  \\
    & \hst       & \cosi/G160M  & FUV  & 1418--1790 & 19000 & 2023/04/29 & 1996 & & Roman-Duval, J. & D$2.5$  \\
    &\eso-VLT-U2&\giraffe\/MOS&1a&3850--4050 & 22000 &  2003/12/05 & 6825 & 133.8 & Smartt, S. & D$1.2$  \\
    & \eso-VLT-U2& \giraffe\/MOS& 1b  & 3850--4050    & 22000 &  2003/10/16 & 6825 & 122.3 & Smartt, S. & D$1.2$  \\
    & \eso-VLT-U2& \giraffe\/MOS& 2a  & 4030--4200    & 30000 &  2003/12/06 & 6825 & 89.4  & Smartt, S. & D$1.2$  \\ 
    & \eso-VLT-U2& \giraffe\/MOS& 2b  & 4030--4200    & 30000 &  2003/12/05 & 6825 & 108.4 & Smartt, S. & D$1.2$  \\
    & \eso-VLT-U2& \giraffe\/MOS& 3a  & 4180--4400    & 23000 &  2003/12/06 & 6825 & 172.2 & Smartt, S. & D$1.2$  \\
    & \eso-VLT-U2& \giraffe\/MOS& 3b  & 4180--4400    & 23000 &  2003/12/06 & 6825 & 182.7 & Smartt, S. & D$1.2$  \\
    & \eso-VLT-U2& \giraffe\/MOS& 4a  & 4340--4590    & 20000 &  2003/12/07 & 6825 & 189.2 & Smartt, S. & D$1.2$  \\
    & \eso-VLT-U2& \giraffe\/MOS& 4b  & 4340--4590    & 20000 &  2003/12/07 & 6825 & 172.7 & Smartt, S. & D$1.2$  \\    
    & \eso-VLT-U2& \giraffe\/MOS& 5a  & 4537--4761    & 23000 &  2003/12/08 & 6825 & 206.8 & Smartt, S. & D$1.2$  \\
    & \eso-VLT-U2& \giraffe\/MOS& 5b  & 4537--4761    & 23000 &  2003/12/07 & 6825 & 187.7 & Smartt, S. & D$1.2$  \\
    & \eso-VLT-U2& \giraffe\/MOS& 6a  & 6300--6690    & 17000 &  2003/11/12 & 6825 & 202.8 & Smartt, S. & D$1.2$  \\      
    & \eso-VLT-U2& \giraffe\/MOS& 6b  & 6300--6690    & 17000 &  2003/10/12 & 6825 & 176.2 & Smartt, S. & D$1.2$  \\
    & \eso-VLT-U3& \xshooter\   & UVB & 2989--5560    & 9861  &  2009/12/25 & $4\times300$ & 70 & Martayan, C. & $0.8\times11$ / $-67.8$ \\ %& 0.9\\
    & \eso-VLT-U3& \xshooter\   & VIS & 5337--10200   & 18340 &  2009/12/25 & $2\times600$ & 70 & Martayan, C. & $0.8\times11$ / $-67.8$ \\ %& 0.85\\    
\hline
PGMW\,3100& \hst & \cosi/G130M  & FUV & 1131--1428 & 12000--16000 & 2020/09/29 & 1990 & & Roman-Duval, J. & D$2.5$  \\
    & \hst & \cosi/G160M        & FUV & 1418--1790 & 13000--20000 & 2020/09/29 & 1968 & & Roman-Duval, J. & D$2.5$ \\
    & \eso-VLT-U2& \giraffe\/MOS& 1a   & 3850--4050 & 22000 & 2003/12/05 & 6825      & 113  & Smartt, S.  & D$1.2$  \\
    & \eso-VLT-U2& \giraffe\/MOS& 1b   & 3850--4050 & 22000 & 2003/10/16 & 6825      & 104  & Smartt, S. & D$1.2$ \\
    & \eso-VLT-U2& \giraffe\/MOS& 2a   & 4030--4200 & 30000 & 2003/12/06 & 6825      & 77   & Smartt, S. & D$1.2$   \\ 
    & \eso-VLT-U2& \giraffe\/MOS& 2b   & 4030--4200 & 30000 & 2003/12/05 & 6825      & 99   & Smartt, S. & D$1.2$   \\
    & \eso-VLT-U2& \giraffe\/MOS& 3a   & 4180--4400 & 23000 & 2003/12/06 & 6825      & 169  & Smartt, S. & D$1.2$  \\
    & \eso-VLT-U2& \giraffe\/MOS& 3b   & 4180--4400 & 23000 & 2003/12/06 & 6825      & 146  & Smartt, S. & D$1.2$  \\     
    & \eso-VLT-U2& \giraffe\/MOS& 4a   & 4340--4590 & 20000 & 2003/12/07 & 6825      & 158  & Smartt, S. & D$1.2$  \\   
    & \eso-VLT-U2& \giraffe\/MOS& 4b   & 4340--4590 & 20000 & 2003/12/07 & 6825      & 159  & Smartt, S. & D$1.2$  \\ 
    & \eso-VLT-U2& \giraffe\/MOS& 5a   & 4540--4760 & 23000 & 2003/12/08 & 6825      & 193  & Smartt, S. & D$1.2$  \\ 
    & \eso-VLT-U2& \giraffe\/MOS& 5b   & 4540--4760 & 23000 & 2003/12/07 & 6825      & 176  & Smartt, S. & D$1.2$  \\
    & \eso-VLT-U2& \giraffe\/MOS& 6a   & 6300--6690 & 17000 & 2003/11/12 & 6825      & 178  & Smartt, S. & D$1.2$  \\      
    & \eso-VLT-U2& \giraffe\/MOS& 6b   & 6300--6690 & 17000 & 2003/10/12 & 6825      & 172  & Smartt, S. & D$1.2$  \\
    & \eso-VLT-U3& \xshooter\   & UVB  & 2989--5560 & 6655  & 2020/12/15 & 950  & 143 & Vink, J. S. & $0.8\times11$ / 0 \\ %& 0.4\\
    & \eso-VLT-U3& \xshooter\   & VIS  & 5337--10200& 11333 & 2020/12/15 & 1020 & 101 & Vink, J. S. & $0.7\times11$ / 0 \\ %& 0.4\\
    \hline 
PGMW3120     & \fuse& FUV/MDRS    &      &  904--1189 & 20000 & 2003/07/09 & 7354 & 7 & Chu, You-Hua & D$2.5$    \\
%PA 298.55
             & \hst & \stis/E140M & FUV   & 1140--1735 & 45800 & 2000/04/14 & 8595      &    & Roman-Duval, J.  & $0.2\times0.6$ \\
%0.2x0.6
             & \hst & \stis/E230M & NUV   & 1574--2382 & 30000 & 2017/05/28 & 3810      &    & Roman-Duval, J.  & $0.2\times0.2$  \\
%0.2x0.2
             & \eso-VLT-U3 & \xshooter\   & UVB &  2989--5560 & 6655  & 2021/02/25 & 490       & 154  & Vink, J. S. & $0.8\times11$ / $-105$ \\ %& 0.5\\
             & \eso-VLT-U3 & \xshooter\   & VIS & 5337--10200 &11333  & 2021/02/25 & 560       & 101  & Vink, J. S. & $0.7\times11$ / $-105$ \\ %& 0.55\\
\hline
\end{tabular}
\end{centering}
\\
(1) Object; (2) telescope; (3) instrument/technique; (4) identification; (5) spectral range; (6) resolving power; (7) date of observation; (8) exposure time; (9) median signal-to-noise ratio; (10) principal investigator; (11) slit-size and position angle (N to E). %(12) mean seeing.
\end{table*}

%HIERARCH ESO QC SEEING = 0.435 / mean seeing
%HIERARCH ESO TEL IA FWHM = 0.7 / Delivered seeing corrected by airm

\begin{table*}
\begin{centering}
%\small
\scriptsize
  \caption{(continue) Log of the spectroscopic observations used for the analysis of the ULLYSES targets in N11\,B.}
\setlength{\tabcolsep}{1\tabcolsep}
  \label{tab:spec_obs2}
\begin{tabular}{ccccccccccc}
\hline\hline
\rule{0cm}{2.2ex} Object & Telescope & Instrument & ID   & $\lambda$   & R    & Date       & Exp. t. & SNR & P.I. & slitsize / PA \\ %& seeing \\
Name        &           &            &            & (\AA) &            & (yy/mm/dd) & (s) & & & (arcsec / deg)  \\
(1)         & (2)       & (3)        & (4)  & (5)        & (6)   & (7)        & (8)        & (9) & (10) & (11)  \\
\hline
PGMW\,3168 & \hst & \cosi/G130M  & FUV  & 1130--1428 & 12000--16000 &  &  & &  & D$2.5$  \\
    & \hst & \cosi/G160M  & FUV  & 1419--1790 & 19000 & 2023/02/10 & 1964 & & Roman-Duval, J. & D$2.5$  \\
    &\eso-VLT-U2&\giraffe\/MOS& 1  & 4030--4200    & 30000 &  2003/12/06 & 6825 & 113.2 & Smartt, S. & D$1.2$  \\ 
    & \eso-VLT-U2& \giraffe\/MOS&  2  & 4180--4400    & 23000 &  2003/12/06 & 6825 & 191   & Smartt, S. & D$1.2$  \\
    & \eso-VLT-U2& \giraffe\/MOS&  3  & 4340--4590    & 20000 &  2003/12/07 & 6825 & 178.8 & Smartt, S. & D$1.2$  \\   
    & \eso-VLT-U2& \giraffe\/MOS& 4a  & 4540--4760    & 23000 &  2003/12/08 & 6825 & 213.8 & Smartt, S. & D$1.2$ \\ 
    & \eso-VLT-U2& \giraffe\/MOS& 4b  & 4540--4760    & 23000 &  2003/12/07 & 6825 & 195.7 & Smartt, S. & D$1.2$ \\
    & \eso-VLT-U2& \giraffe\/MOS& 5   & 6300--6690    & 17000 &  2003/11/12 & 6825 & 182   & Smartt, S. & D$1.2$ \\
    & \eso-VLT-U3& \xshooter\   & UVB & 2989--5560    & 6655  &  2020/12/20 & $2\times1020$ & 135 & Vink, J. S. & $0.8\times11$ / 0 \\ %& 0.68\\
    & \eso-VLT-U3& \xshooter\   & VIS & 5337--10200   & 11333 &  2020/12/20 & $2\times950$  & 90  & Vink, J. S. & $0.7\times11$ / 0 \\ %& 0.73\\    
\hline  
PGMW\,3204 & \hst & \stis/E140M & FUV & 1141--1729 & 45800 & 2023/05/08 & 7768      & 20  & Roman-Duval, J. & $0.2\times0.2$ / 104.2  \\
%0.2x0.2
%PA 104.24
&\eso-VLT-U2&\giraffe\/MOS&1a&3850--4050& 22000 &  2003/12/05 & 6825 & 116.8 & Smartt, S. & D$1.2$ \\
   & \eso-VLT-U2& \giraffe\/MOS& 1b  & 3850--4050    & 22000 &  2003/10/16 & 6825 & 108.2 & Smartt, S. & D$1.2$  \\
   & \eso-VLT-U2& \giraffe\/MOS& 2a  & 4030--4200    & 30000 &  2003/12/06 & 6825 & 70.4 & Smartt, S. & D$1.2$   \\ 
   & \eso-VLT-U2& \giraffe\/MOS& 2b  & 4030--4200    & 30000 &  2003/12/05 & 6825 & 83.4 & Smartt, S. & D$1.2$   \\
   & \eso-VLT-U2& \giraffe\/MOS& 3a  & 4180--4400    & 23000 &  2003/12/06 & 6825 & 142.7 & Smartt, S. & D$1.2$   \\
   & \eso-VLT-U2& \giraffe\/MOS& 3b  & 4180--4400    & 23000 &  2003/12/06 & 6825 & 146.2 & Smartt, S. & D$1.2$   \\     
   & \eso-VLT-U2& \giraffe\/MOS& 4a  & 4340--4590    & 20000 &  2003/12/07 & 6825 & 152.2 & Smartt, S. & D$1.2$   \\  
   & \eso-VLT-U2& \giraffe\/MOS& 4b  & 4340--4590    & 20000 &  2003/12/07 & 6825 & 139.7 & Smartt, S. & D$1.2$   \\ 
   & \eso-VLT-U2& \giraffe\/MOS& 5a  & 4540--4760    & 23000 &  2003/12/08 & 6825 & 161.1 & Smartt, S. & D$1.2$   \\ 
   & \eso-VLT-U2& \giraffe\/MOS& 5b  & 4540--4760    & 23000 &  2003/12/07 & 6825 & 146.2 & Smartt, S. & D$1.2$   \\
   & \eso-VLT-U2& \giraffe\/MOS& 6a  & 6300--6690    & 17000 &  2003/11/12 & 6825 & 157.2 & Smartt, S. & D$1.2$   \\      
   & \eso-VLT-U2& \giraffe\/MOS& 6b  & 6300--6690    & 17000 &  2003/10/12 & 6825 & 135.2 & Smartt, S. & D$1.2$   \\  
   & \eso-VLT-U3& \xshooter\   & UVB & 2989--5560    & 6655  &  2021/01/28 & 1200 & 134 & Vink, J. S. & $0.8\times11$ / $-50$ \\ %& 0.76\\
   & \eso-VLT-U3& \xshooter\   & VIS & 5337--10200   & 11333 &  2021/01/28 & 1270 &  82 & Vink, J. S. & $0.7\times11$ / $-50$ \\ %& 0.75\\    
\hline 
PGMW\,3223& \fuse & FUV/MDRS    &  &  904--1189 & 15000--20000 & 2002/09/23 & 7106 &  & Roman-Duval, J.  & D$2.5$  \\
    & \hst       & \stis/E140M  & FUV  & 1140--1735 & 45800 & 2017/05/20 & 7768      & & Roman-Duval, J. & $0.2\times0.2$ \\
    & \hst       & \stis/E230M  & NUV  & 1574--2382 & 30000 & 2017/05/13 & 4970      & & Roman-Duval, J. & $0.2\times0.2$  \\
    & \eso-VLT-U2& \giraffe\/MOS& 1a   & 3850--4050 & 22000 & 2003/12/05 & 6825      & 185 & Smartt, S.  & D$1.2$  \\
    & \eso-VLT-U2& \giraffe\/MOS& 1b   & 3850--4050 & 22000 & 2003/10/16 & 6825      & 170 & Smartt, S. & D$1.2$  \\
    & \eso-VLT-U2& \giraffe\/MOS& 2a   & 4030--4200 & 30000 & 2003/12/06 & 6825      & 131 & Smartt, S. & D$1.2$  \\ 
    & \eso-VLT-U2& \giraffe\/MOS& 2b   & 4030--4200 & 30000 & 2003/12/05 & 6825      & 131 & Smartt, S. & D$1.2$  \\
    & \eso-VLT-U2& \giraffe\/MOS& 3a   & 4180--4400 & 23000 & 2003/12/06 & 6825      & 244 & Smartt, S. & D$1.2$  \\
    & \eso-VLT-U2& \giraffe\/MOS& 3b   & 4180--4400 & 23000 & 2003/12/06 & 6825      & 231 & Smartt, S. & D$1.2$  \\     
    & \eso-VLT-U2& \giraffe\/MOS& 4a   & 4340--4590 & 20000 & 2003/12/07 & 6825      & 223 & Smartt, S. & D$1.2$  \\   
    & \eso-VLT-U2& \giraffe\/MOS& 4b   & 4340--4590 & 20000 & 2003/12/07 & 6825      & 251 & Smartt, S. & D$1.2$  \\ 
    & \eso-VLT-U2& \giraffe\/MOS& 5a   & 4540--4760 & 23000 & 2003/12/08 & 6825      & 272 & Smartt, S. & D$1.2$  \\
    & \eso-VLT-U2& \giraffe\/MOS& 5b   & 4540--4760 & 23000 & 2003/12/07 & 6825      & 250 & Smartt, S. & D$1.2$  \\
    & \eso-VLT-U2& \giraffe\/MOS& 6a   & 6300--6690 & 17000 & 2003/11/12 & 6825      & 253 & Smartt, S. & D$1.2$  \\      
    & \eso-VLT-U2& \giraffe\/MOS& 6b   & 6300--6690 & 17000 & 2003/10/12 & 6825 & 227 & Smartt, S. & D$1.2$; 0  \\     
    & \eso-VLT-U3& \xshooter\   & UVB  & 2989--5560 & 6655  & 2020/12/13 & 540 & 153 & Vink, J. S. & $0.8\times11$ / $-45$ \\ %& 0.57 \\
    & \eso-VLT-U3& \xshooter\   & VIS  & 5337--10200& 11333 & 2020/12/13 & 610 & 103 & Vink, J. S. & $0.7\times11$ / $-45$ \\ %& 0.52\\  
\hline
Sk $-69^{\circ}$ 50 & \fuse & FUV/LWRS & & 904--1189& 15000--20000 & 2004/06/13 & 9263 & & Roman-Duval, J.  & D$2.5$  \\
% Aperture: LWRS, low resolution, 30 × 30 arcsec
% XPOSURE 9263
                     & \hst & \stis/E140M & FUV & 1140--1735 & 45800 & 2011/10/11 & 2839 & & Roman-Duval, J.  & $0.2\times0.2$ \\
                     % Aperture: 0.2 x 0.2
    & \eso-VLT-U3& \xshooter\   & UVB  & 2989--5560 & 6655  & 2020/12/10 & 640 & 155 & Vink, J. S. & $0.8\times11$ / $-50$ \\ %& 0.69 \\
    & \eso-VLT-U3& \xshooter\   & VIS  & 5337--10200& 11333 & 2020/12/10 & 710 &  97 & Vink, J. S. & $0.7\times11$ / $-50$ \\ %& 0.68 \\                       
\hline
Sk $-66^{\circ}$ 171 & \fuse & FUV/LWRS    & &  904--1189 & 15000--20000 & 2005/04/08 & 1267 & & Roman-Duval, J.  & D$2.5$  \\
                     & \hst & \stis/E140M & FUV & 1140--1735 & 45800 & 2022/01/28 & 3191 & & Roman-Duval, J.  & $0.2\times0.2$ \\
% Aperture: 0.2 x 0.2
                     & \hst & \stis/E230M & NUV & 1574--2382 & 30000 & 2022/01/28 & 1200 & & Roman-Duval, J.   & $0.2\times0.2$   \\
% Aperture: 0.2 x 0.2
    & \eso-VLT-U3& \xshooter\   & UVB  & 2989--5560 & 6655  & 2021/01/04 & $2\times100$ & 100 & Vink, J. S. & $0.8\times11$ / $-160$ \\ %& 0.58 \\
    & \eso-VLT-U3& \xshooter\   & VIS  & 5337--10200& 11333 & 2021/01/04 & $2\times200$ &  80 & Vink, J. S. & $0.7\times11$ / $-160$ \\ %& 0.56 \\ 
%    SNR rounded
% It has UVES spectra
\hline
N11 046             & \hst & \cosi/G130M & FUV & 1090--1240 & 12000--16000 & 2023/02/13 & 1568 & & Roman-Duval, J. 
 & D$2.5$ \\
% Aperture: PSA
                     & \hst & \stis/E140M & FUV & 1140--1735 & 45800 & 2023/03/14 & 7218 & & Roman-Duval, J.  & $0.2\times0.2$ \\
                     % Aperture: 0.2 x 0.2
    & \eso-VLT-U3& \xshooter\   & UVB  & 2989--5560 & 6655  & 2021/01/02 & 1200 & 151 & Vink, J. S. & $0.8\times11$ / $-40$ \\ %& 0.7 \\
    & \eso-VLT-U3& \xshooter\   & VIS  & 5337--10200& 11333 & 2021/01/02 & 1270 &  87 & Vink, J. S. & $0.7\times11$ / $-40$ \\ % & 0.7\\
\hline
\end{tabular}
\end{centering}
\end{table*}

\section{Photometry}
\label{Photometry}

The photometric values used to construct the SEDs of the analyzed stars in N11\,B are listed in Table~\ref{tab:photometry}.

\begin{table*}
\begin{centering}
\scriptsize
  \caption{Photometric values used to construct the SEDs of the analyzed stars in N11\,B.}
\setlength{\tabcolsep}{0.2\tabcolsep}
  \label{tab:photometry}
\begin{tabular}{ccccccccc}
\hline\hline
\rule{0cm}{2.2ex} Band & \multicolumn{8}{c}{magnitude}                                           \\
   & PGMW\,3223                          & PGMW\,3053              & PGMW\,3061                & PGMW\,3168               & PGMW\,3100               & PGMW\,3204               & PGMW\,3058               & PGMW\,3120 \\
\hline
U1 & 11.998$\pm$0.021$^{a}$           & \dots                & 12.699$\pm$0.221$^{a}$ & 12.58$^{h}$           & 13.007$\pm$0.049$^{a}$& 12.87$^{h}$           &12.901$\pm$0.104$^{a}$ & 11.700$\pm$0.024$^{a}$\\
U2 & 12.87$^{h}$                      & \dots                & 13.16$^{h}$            & \dots                 & 12.89$^{h}$           & \dots                 &13.18$^{h}$            &    \\
B1 & 12.768$\pm$0.052$^{a}$           & 13.04$^{e}$          & 13.595$\pm$0.029$^{a}$ & 13.57$^{h}$           & 13.736$\pm$0.029$^{a}$& 13.85$^{h}$           &14.092$\pm$0.034$^{a}$ & 12.496$\pm$0.074$^{a}$\\
B2 & 12.83$^{h}$                      & \dots                & 13.67$^{h}$            & \dots                 & 13.81$^{h}$           & \dots                 &14.18$^{h}$            &    \\
V1 & 12.896$\pm$0.153$^{a}$           & 13.13$^{e}$          & 13.491$\pm$0.0501$^{a}$& 13.68$^{h}$           & 13.648$\pm$0.029$^{a}$& 14.02$^{h}$           &14.089$\pm$0.031$^{a}$ & 12.466$\pm$0.124$^{a}$\\
V2 & 12.93$^{h}$                      & \dots                & 13.68$^{h}$            & \dots                 & 13.81$^{h}$           & \dots                 &14.24$^{h}$            &  \\
R  & 13.210$\pm$0.10$^{b}$            & \dots                & \dots                  & \dots                 & \dots                 & \dots                 & \dots                 & \dots            \\
G2 &12.9626$\pm$0.0032$^{c}$          & 13.0918$\pm$0.0004   & 13.6624$\pm$0.0003     & 13.6189$\pm$0.0003    & 13.7639$\pm$0.0003    & 13.9583$\pm$0.0008    &14.2088$\pm$0.0003     & 12.8676$\pm$0.0046\\
Gbp&12.8016$\pm$0.0055$^{c}$          & 12.9587$\pm$0.0027   & 13.5410$\pm$0.0024     & 13.4892$\pm$0.0025    & 13.6969$\pm$0.0024    & 13.6853$\pm$0.0035    &13.9913$\pm$0.0056     & 12.6421$\pm$0.0067\\
Grp&12.9646$\pm$0.0021$^{c}$          & 13.1378$\pm$0.0012   & 13.5529$\pm$0.0015     & 13.6293$\pm$0.0010    & 13.6891$\pm$0.0008    & 13.9264$\pm$0.0019    &14.0966$\pm$0.0039     & 12.8296$\pm$0.0025\\
G3$^{k}$$^{l}$ &13.326458$\pm$0.003264&13.078708$\pm$0.002791& 13.639601$\pm$0.002793 & 13.609666$\pm$0.002786& 13.749496$\pm$0.002789& 14.033206$\pm$0.003400&14.198952$\pm$0.002786 & 13.793874$\pm$0.023012\\
Gbp$^{k}$$^{l}$&12.840068$\pm$0.003135&13.017043$\pm$0.003038& 13.586996$\pm$0.003147 & 13.549040$\pm$0.002942& 13.739211$\pm$0.002949& 13.771400$\pm$0.003161&14.087885$\pm$0.005192 & 12.694622$\pm$0.003200\\
Grp$^{k}$$^{l}$&12.978003$\pm$0.003947&13.167009$\pm$0.003868& 13.571906$\pm$0.004226 & 13.663751$\pm$0.003856& 13.704623$\pm$0.003812& 13.975013$\pm$0.004363&14.158316$\pm$0.005903 & 12.860391$\pm$0.004241\\
I  & 12.663$\pm$0.125$^{a}$           & \dots                & 13.405$\pm$0.069$^{a}$ & \dots                 & 13.547$\pm$0.063$^{a}$& \dots                 &13.991$\pm$0.074$^{a}$ & 12.464$\pm$0.163$^{a}$\\
J1 & 13.121$\pm$0.033$^{d}$           &13.339$\pm$0.021$^{d}$& 13.574$\pm$0.029$^{d}$ & 13.786$\pm$0.026$^{d}$& 13.733$\pm$0.035$^{d}$& 14.20$^{i}$           &14.199$\pm$0.060$^{d}$ & 12.946$\pm$0.040$^{d}$\\
J2 & 13.17$^{i}$                      & \dots                & 13.71$^{i}$            & 13.81$^{i}$           & 13.75$^{i}$           & \dots                 &14.70$^{i}$            & 13.08$^{i}$   \\
H1 & 13.194$\pm$0.047$^{d}$           &13.314$\pm$0.025$^{d}$& 13.485$\pm$0.039$^{d}$ & 13.795$\pm$0.044$^{d}$& 13.770$\pm$0.049$^{d}$& \dots                 &14.218$\pm$0.078$^{d}$ & 12.997$\pm$0.048$^{d}$\\
K1 & 13.186$\pm$0.048$^{d}$           &13.428$\pm$0.045$^{d}$& 13.523$\pm$0.048$^{d}$ & 13.734$\pm$0.048$^{d}$& \dots                 & 14.19$^{i}$           &14.152$\pm$0.089$^{d}$ & 13.022$\pm$0.052$^{d}$\\
K2 & \dots                            &13.35$^{i}$           & 13.76$^{i}$            & 13.18$^{i}$           & 13.70$^{i}$           & \dots                 &14.29$^{i}$            & 13.07$^{i}$    \\
\hline
\label{tab:photometry}
\end{tabular}
\\
References: (a) \citet[][]{Bonanos2009};
(b) \citet[][]{Zacharias2012};
(c) \citet[][]{Gaia2018};
(d) \citet[][]{Cutri2003};
(e) \citet[][]{Evans2006};
(f) \citet[][]{Walborn1995};
(g) \citet[][]{Penny2009};
(h) \citet[][]{Parker1992};
(i) \citet[][]{Cioni2011};
(k) \citet[][]{Gaia2016};
(l) \citet[][]{Gaia2023}.
\\
\end{centering}
\end{table*}

\section{PoWR models}
\label{models}

\begin{figure*}
\begin{centering}
\includegraphics[trim={0 5.1cm 0 5.5cm},clip,width=0.53\linewidth]{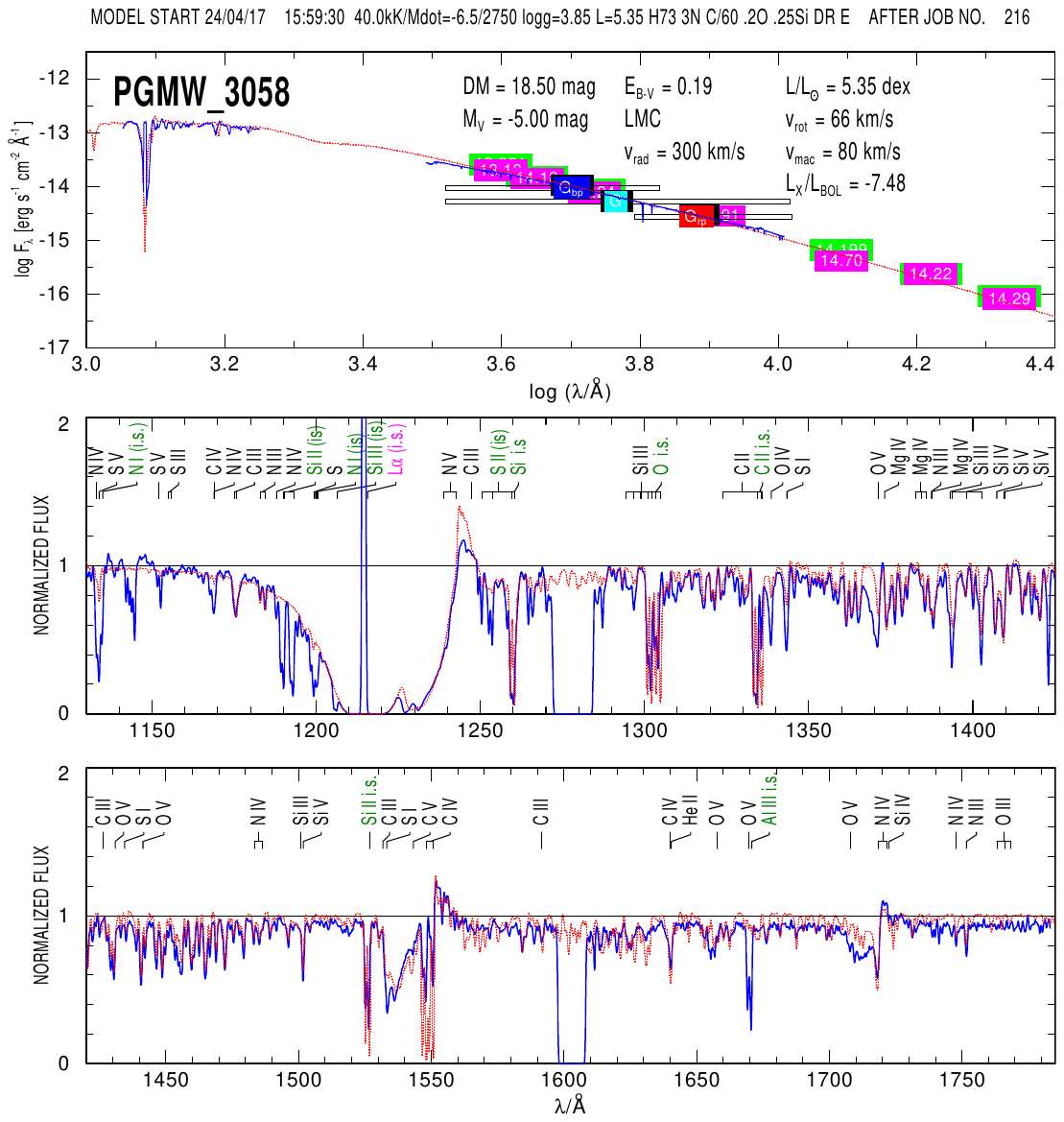}
\par\end{centering}
\caption{
PoWR model for the star PGMW\,3058. The observed spectra is shown by a blue line and the model by a red dashed line.
(1st panel) SED with photometric magnitudes (colour boxes). The UV spectra better constrain the \ebv\ and \lstar\ of the star (also indicated at the upper right, among other parameters);
(2nd and 3rd panel) \hst/\cosi\ UV spectra normalized to the continuum model.
The terminal velocity (\vinf) is determined from the blue edge of the \civ\ line in absorption.
N enhancement is determined with \niii\ at 1183 and 1185~\AA, also considering these ions in the optical range, and C abundances by using \ciii\ at 1175 and 1176~\AA\ in the UV.
ISM absorption features by the H Lyman lines were considered in our modeling.
Interstellar (i.s.) atomic, molecular and metal lines in absorption are indicated.
There is a gap of around 10~\AA\ in the observations around 1280, and 1605~\AA, where no key lines are present.
}
\label{fig:master_n11-60_sed}
\end{figure*}

\begin{figure*}
\begin{centering}
\includegraphics[trim={1.5cm 2.0cm 1.5cm 2.0cm},clip,width=0.43\linewidth]{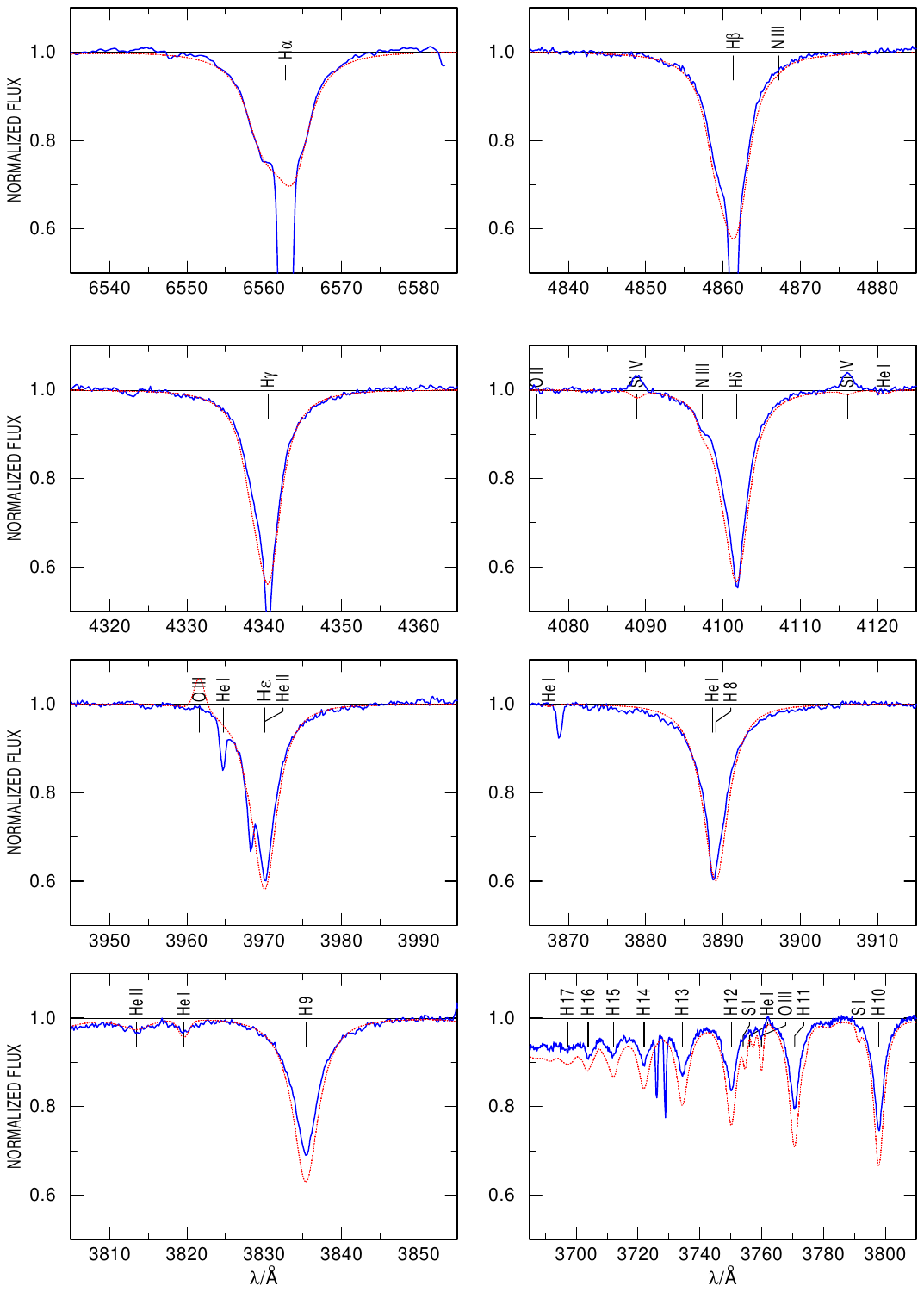}~
\includegraphics[trim={1.5cm 2.0cm 1.5cm 2.0cm},clip,width=0.43\linewidth]{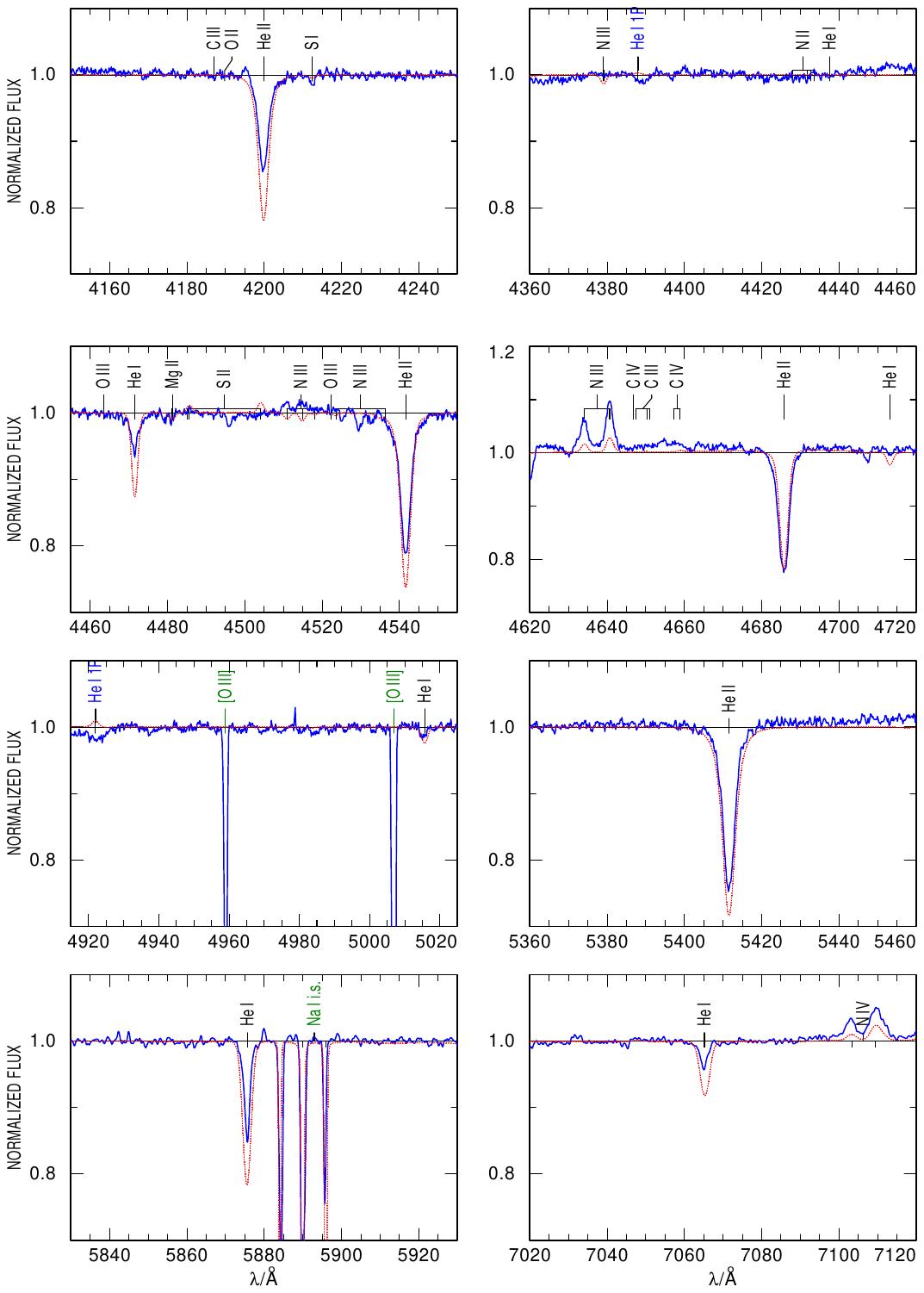}\\
\par\end{centering}
\caption{
PoWR model for PGMW\,3058. ({\it left}) Balmer lines present in the observed spectrum: \ha, \hb, \hg, \hd, \hep\ and H10--H17, in comparison with the final model of the star. The surface gravity (\logg) parameter of the star is determined from the wings of the H lines. In this particular case, \hg\ is the optimal line to use, given it is not blended by other lines (e.g., \hd). \ha\ is considered to determine $\dot{M}$. The \ha\ line is affected by a strong absorption, most likely of nebular origin.
({\it right}) The most important \hei\ and \heii\ lines, as well as some metals, in comparison with the final model of the star. The temperature (\tstar) of the star is determined by modeling the \hei-\heii\ ratios
(e.g., \hei\,$\lambda$4471.5, and \heii\,$\lambda$4541.6). Singlet lines of \hei\ (1s2p P), at 4387.9 and 4921.9~\AA, identified with blue labels were not used for diagnostic. Not photospheric lines are indicated with green labels.
\heiiwr\ is crucial to determine $\dot{M}$.
The observed spectra are shown by a blue line and the PoWR model by a red dashed line. The most important lines are identified.
}
\label{fig:master_n11-60_balmer}
\end{figure*}

\begin{figure*}
\begin{centering}
\includegraphics[trim={0 5.1cm 0 5.5cm},clip,width=0.53\linewidth]{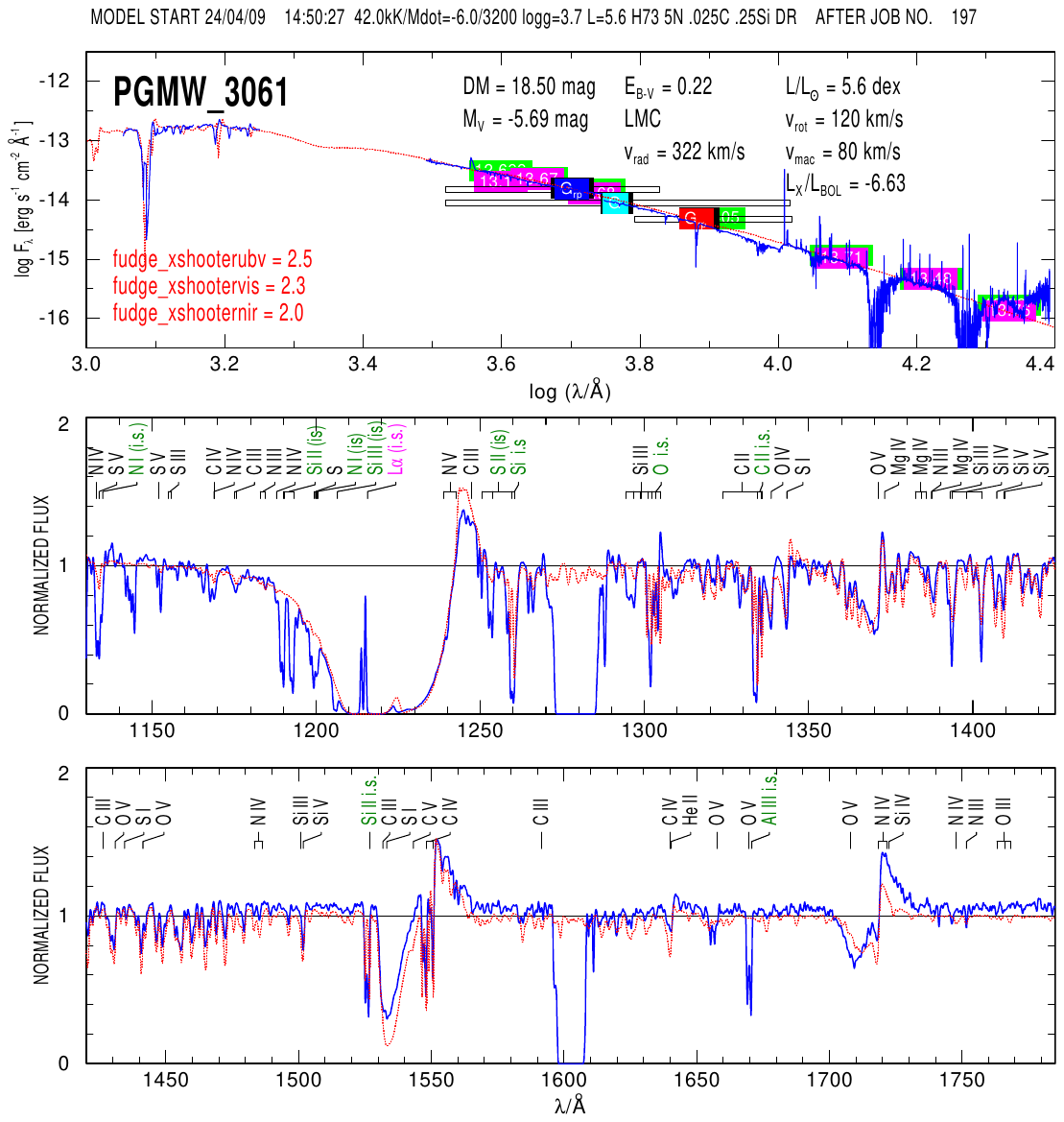}
\par\end{centering}
\caption{
PoWR model for PGMW\,3061. The observed spectra is shown by a blue line and the model by a red dashed line.
(1st panel) SED with photometric magnitudes (colour boxes).
(2nd and 3rd panel) \hst/\cosi\ UV spectra normalized to the continuum model.
Interstellar (i.s.) atomic, molecular and metal lines in absorption are indicated.
The legend "fudge" means that the spectrum needed to be scaled (see text for details).
There is a gap of around 10~\AA\ in the observations around 1280, and 1605~\AA, where no key lines are present.
}
\label{fig:master_n11-31_sed}
\end{figure*}

\begin{figure*}
\begin{centering}
\includegraphics[trim={1.5cm 2.3cm 1.5cm 2.1cm},clip,width=0.43\linewidth]{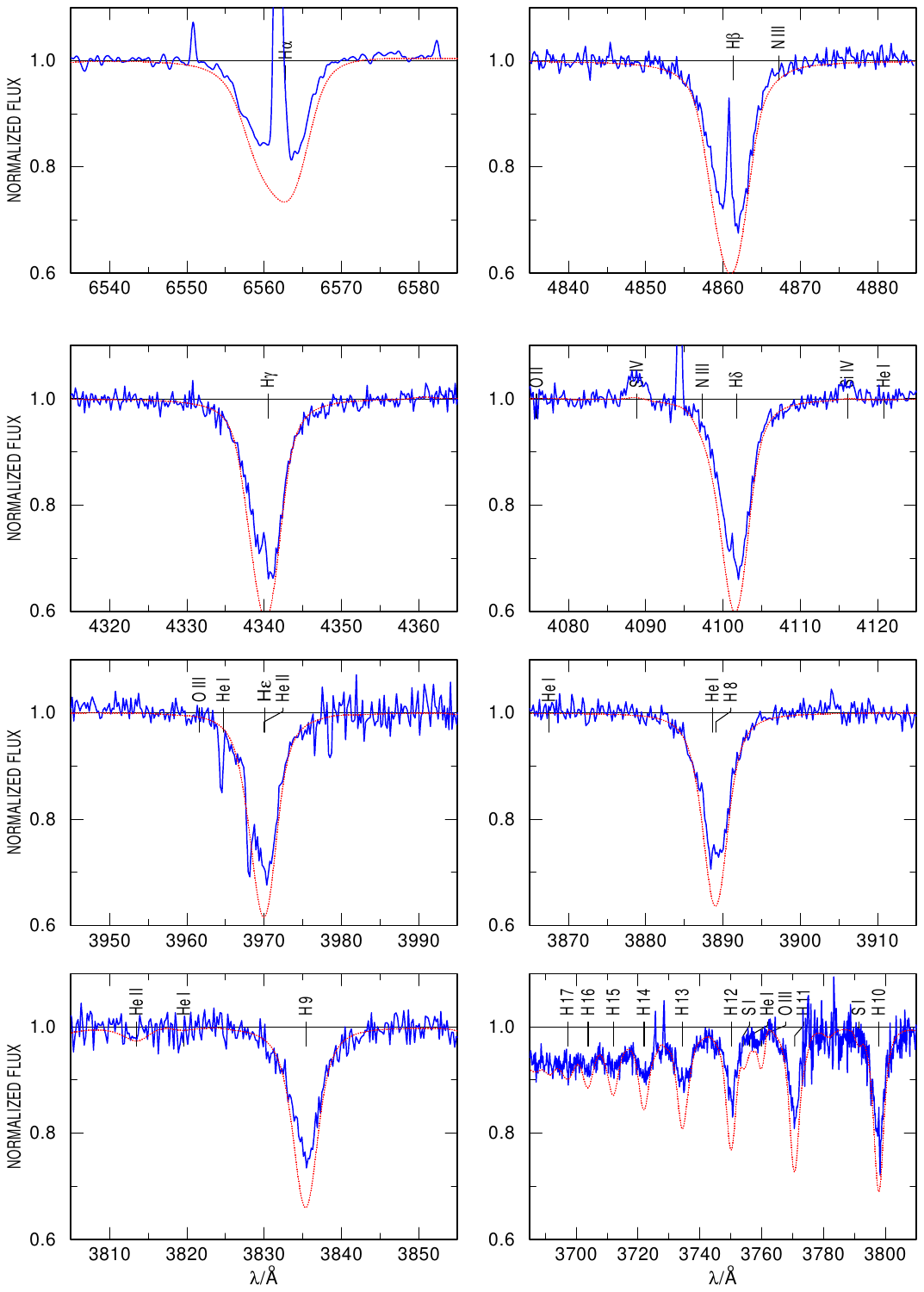}~
\includegraphics[trim={1.5cm 2.3cm 1.5cm 2.1cm},clip,width=0.43\linewidth]{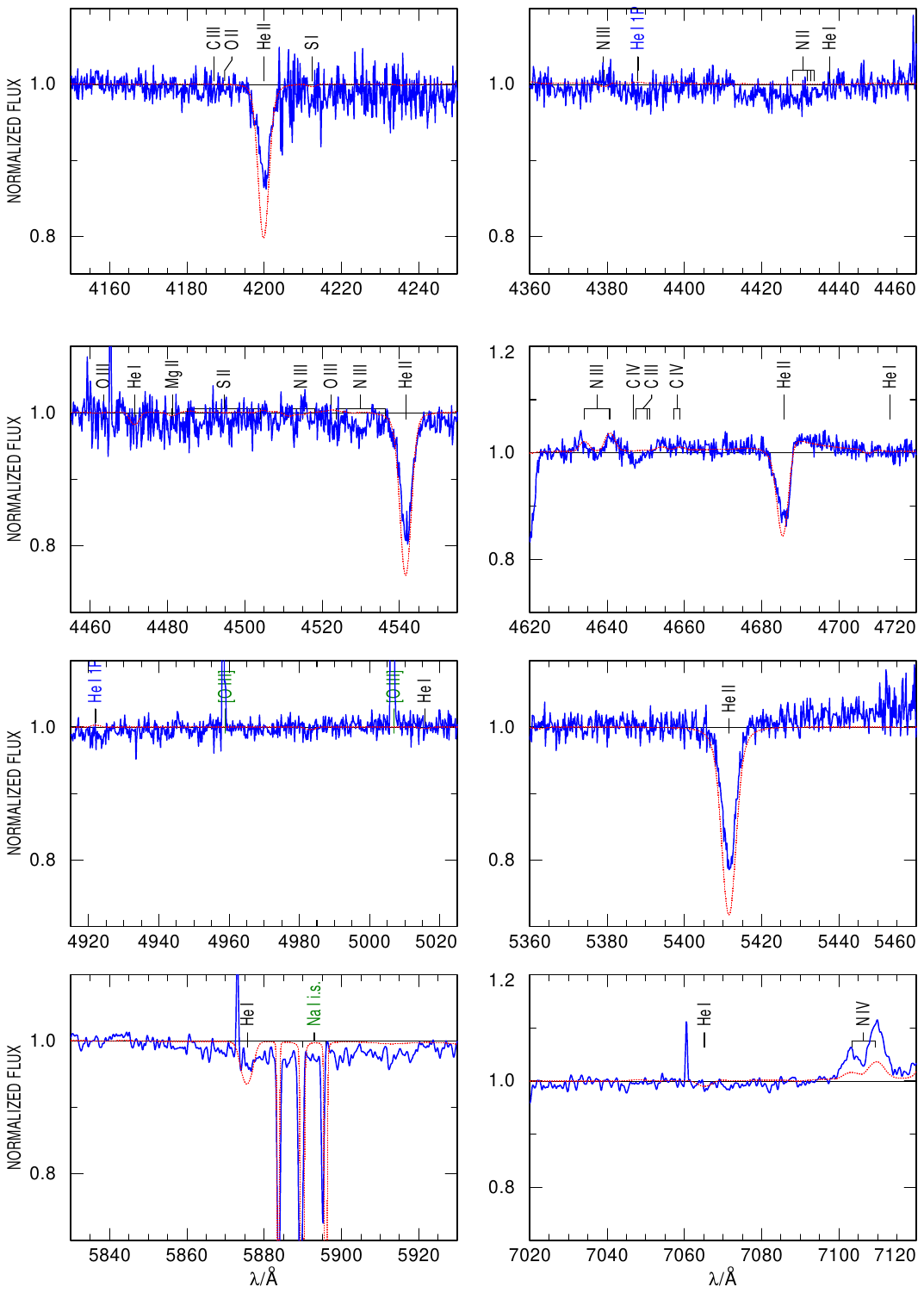}\\
\par\end{centering}
\caption{
PoWR model for PGMW\,3061. ({\it left}) Balmer lines in the observed spectrum: \ha, \hb, \hg, \hd, \hep\ and H10--H17, in comparison with the final model of the star. The \logg\ parameter is determined from the wings of the H lines. \ha\ is considered to determine $\dot{M}$. The \ha\ line is affected by a strong emission, most likely of nebular origin.
({\it right}) The most important \hei\ and \heii\ lines, as well as some metals, in comparison with the final model of the star. The \tstar\ of the star is determined by modeling the \hei-\heii\ ratios. Singlet lines of \hei\ identified with blue labels were not used for diagnostic. Not photospheric lines are indicated with green labels.
\heiiwr\ is crucial to determine $\dot{M}$.
The observed spectra are shown by a blue line and the PoWR model by a red dashed line. The most important lines are identified.
}
\label{fig:master_n11-31_balmer}
\end{figure*}

\begin{figure*}
\begin{centering}
\includegraphics[trim={0 5.1cm 0 5.7cm},clip,width=0.53\linewidth]{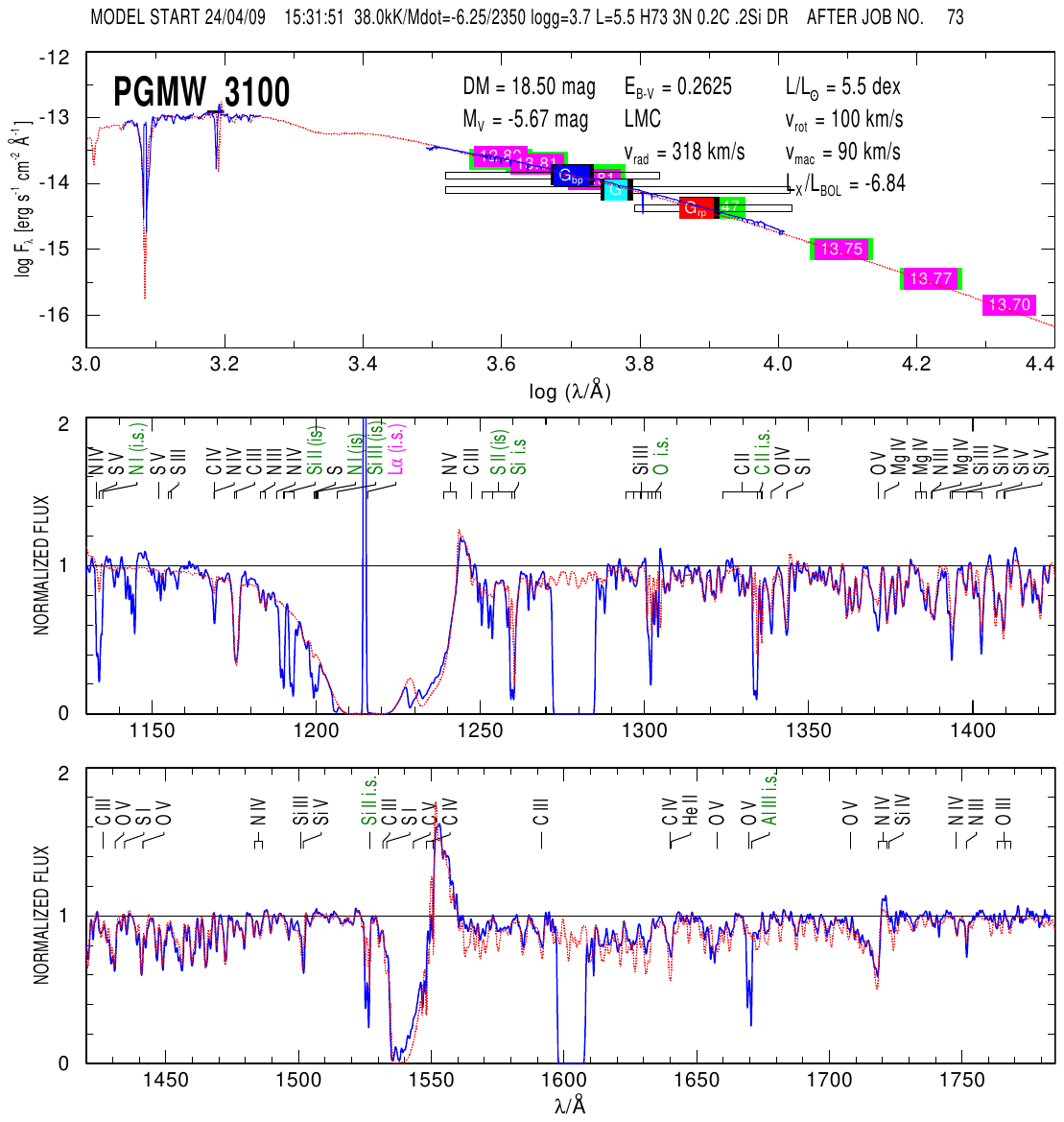}
\par\end{centering}
\caption{
PoWR model for PGMW\,3100. The observed spectra is shown by a blue line and the model by a red dashed line. (1st panel) SED with photometric magnitudes (colour boxes). The UV spectra better constrain the \ebv\ and \lstar; (2nd and 3rd panel) \hst/\cosi\ UV spectra normalized to the continuum model. Interstellar (i.s.) atomic, molecular and metal lines in absorption are indicated.
There is a gap of around 10~\AA\ in the observations around 1280, and 1605~\AA, where no key lines are present.
}
\label{fig:master_n11-38_sed}
\end{figure*}

\begin{figure*}
\begin{centering}
\includegraphics[trim={1.5cm 2.3cm 1.5cm 2.1cm},clip,width=0.43\linewidth]{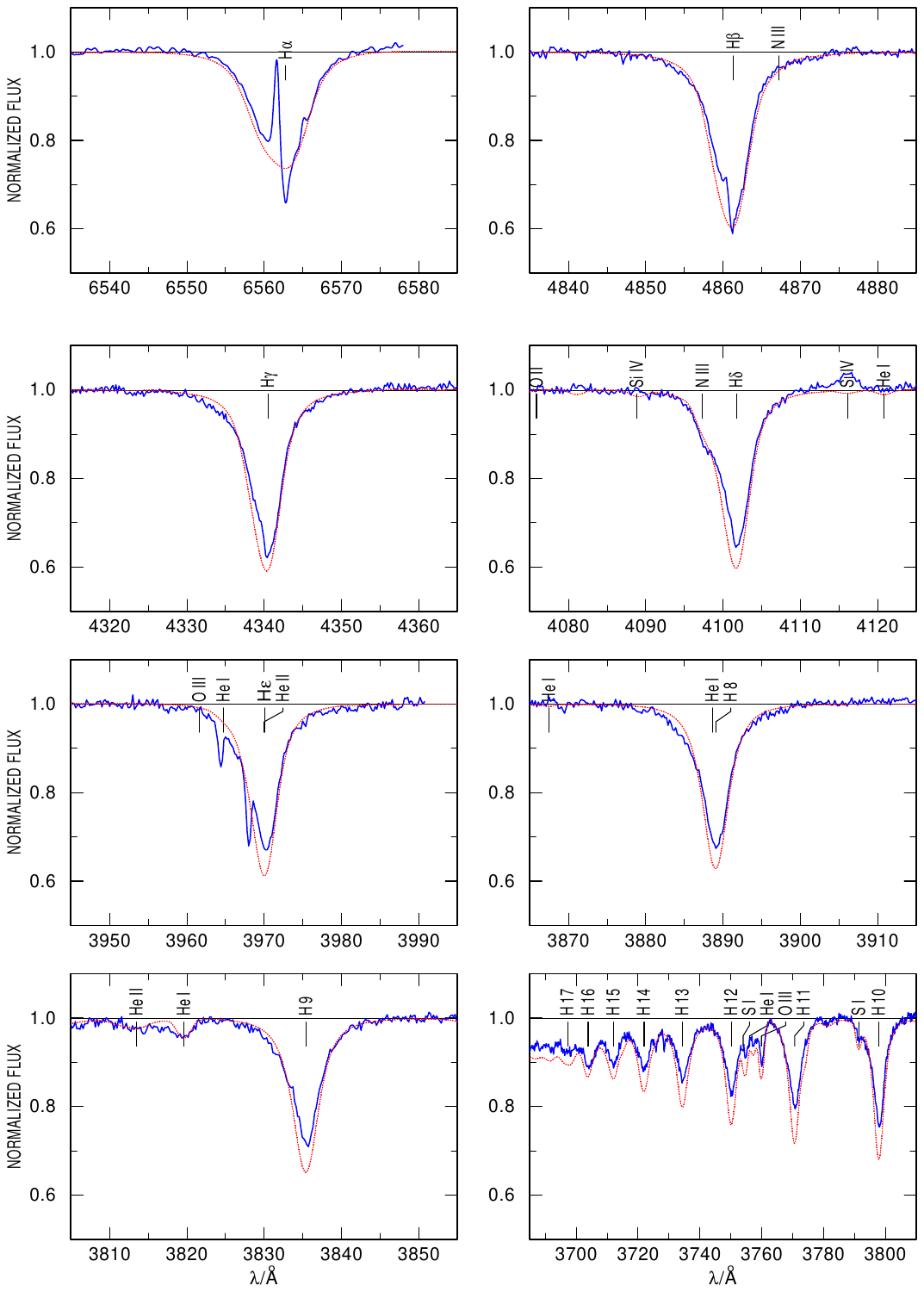}~
\includegraphics[trim={1.5cm 2.3cm 1.5cm 2.1cm},clip,width=0.43\linewidth]{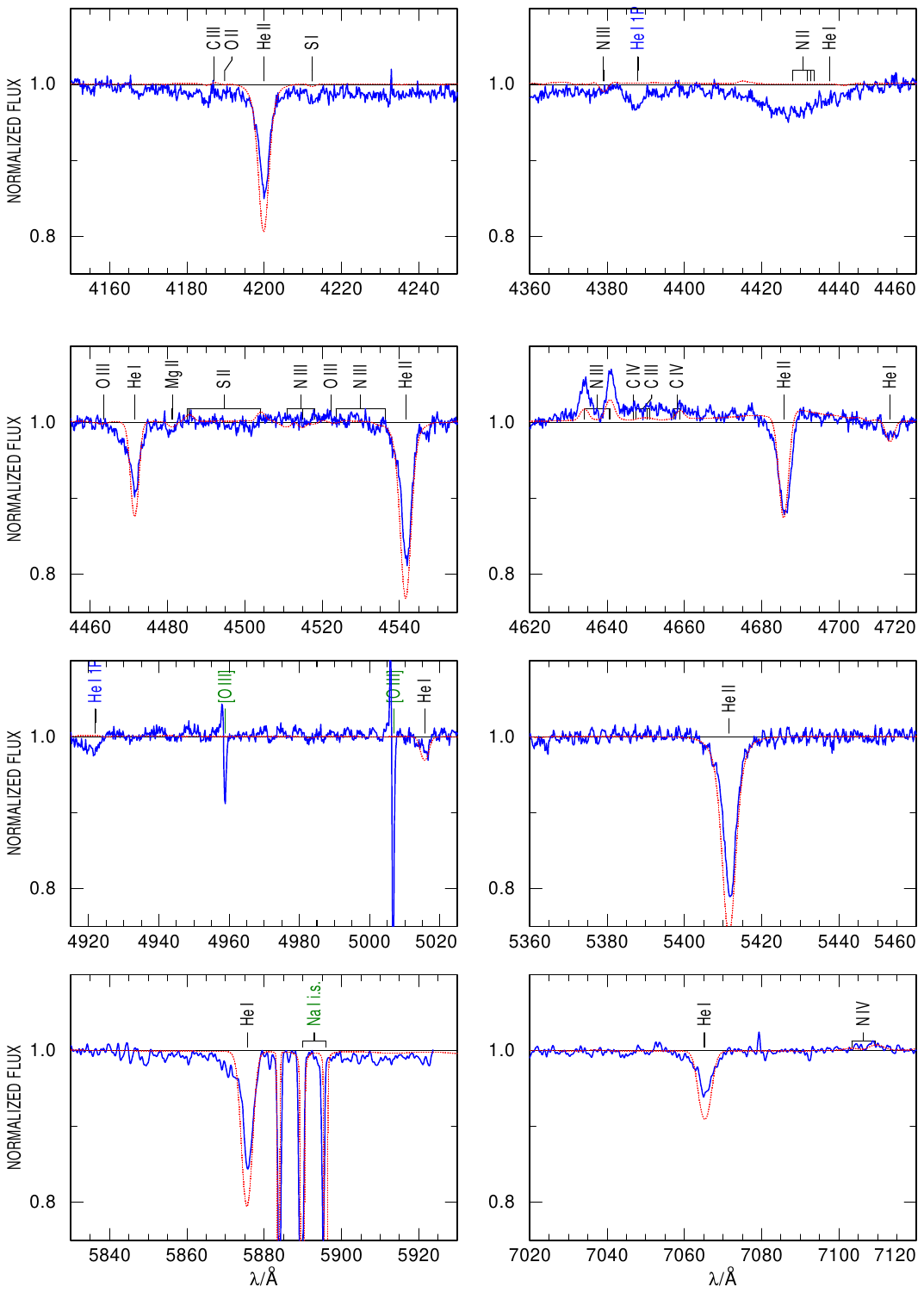}\\
\par\end{centering}
\caption{
PoWR model for PGMW\,3100. ({\it left}) Balmer lines in the observed spectrum: \ha, \hb, \hg, \hd, \hep\ and H10--H17, in comparison with the final model of the star. The \logg\ parameter is determined from the wings of the H lines. \ha\ is considered to determine $\dot{M}$. The \ha\ line is affected by a line, most likely of nebular origin.
({\it right}) The most important \hei\ and \heii\ lines, as well as some metals, in comparison with the final model of the star. The \tstar\ of the star is determined by modeling the \hei-\heii\ ratios. Singlet lines of \hei\ identified with blue labels were not used for diagnostic. Not photospheric lines are indicated with green labels.
\heiiwr\ is crucial to determine $\dot{M}$.
The observed spectra are shown by a blue line and the PoWR model by a red dashed line. The most important lines are identified.
}
\label{fig:master_n11-38_balmer}
\end{figure*}

\begin{figure*}
\begin{centering}
\includegraphics[trim={0 2.1cm 0 2.7cm},clip,width=0.45\linewidth]{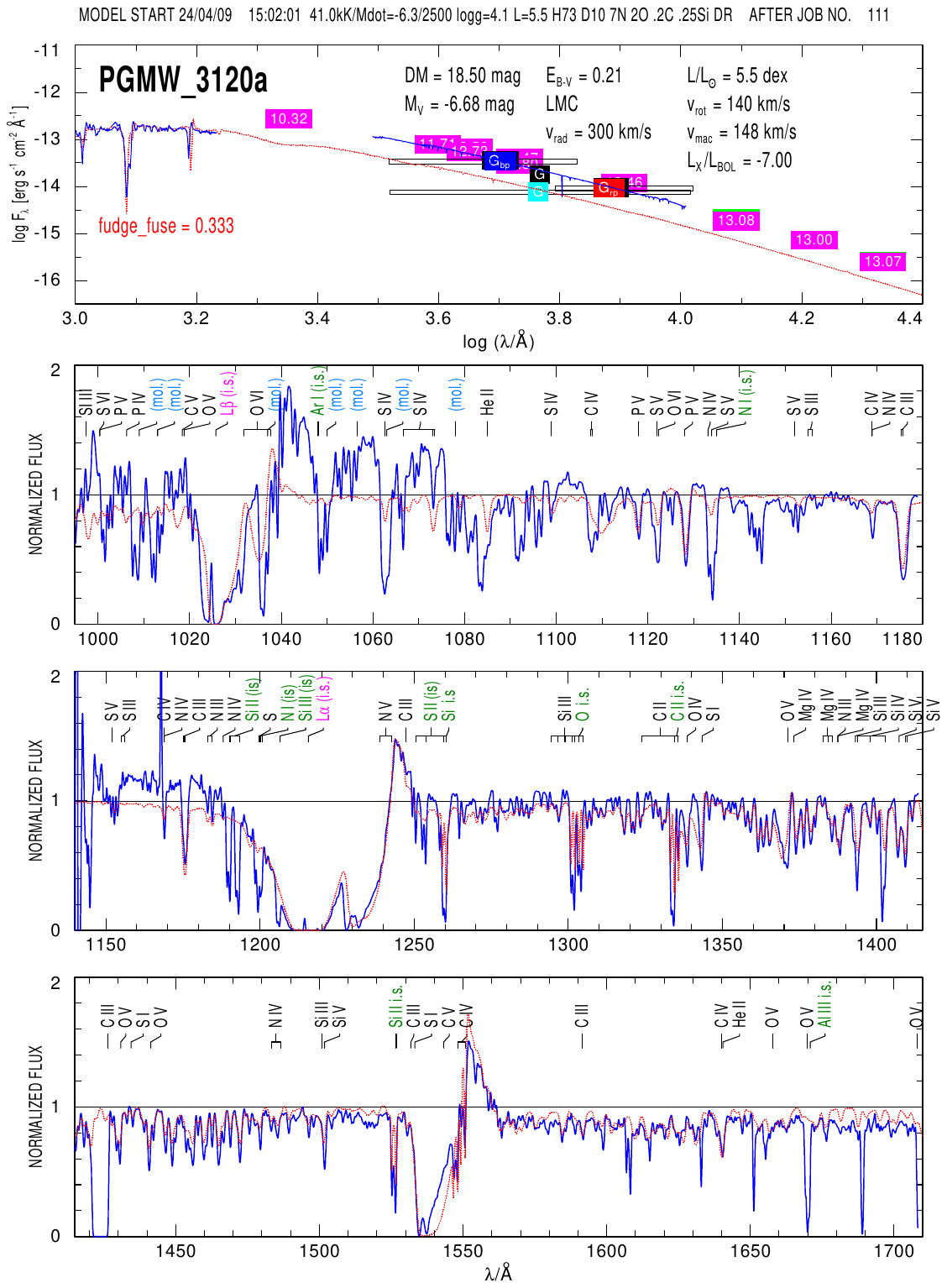}
\par\end{centering}
\caption{
PoWR model for PGMW\,3120. The observed spectra is shown by a blue line and the model by a red dashed line.
(1st panel) SED with photometric magnitudes (colour boxes).
(2nd panel) \fuse/MRDS UV spectra normalized to the continuum model;
(3rd and 4th panel) \hst/\stis\ UV spectra normalized to the continuum model.
Particularly sensitive to X-rays are the \nvuv\ and \ovi\ features.
Interstellar (i.s.) atomic, molecular and metal lines in absorption are indicated.
The legend "fudge" means that the \fuse\ spectrum needed to be scaled (see text for details). 
There is a gap around 1420~\AA, where no key lines are present.
}
\label{fig:master_pgmw3120_sed}
\end{figure*}

\begin{figure*}
\begin{centering}
\includegraphics[trim={1.5cm 2.3cm 1.5cm 2.1cm},clip,width=0.4\linewidth]{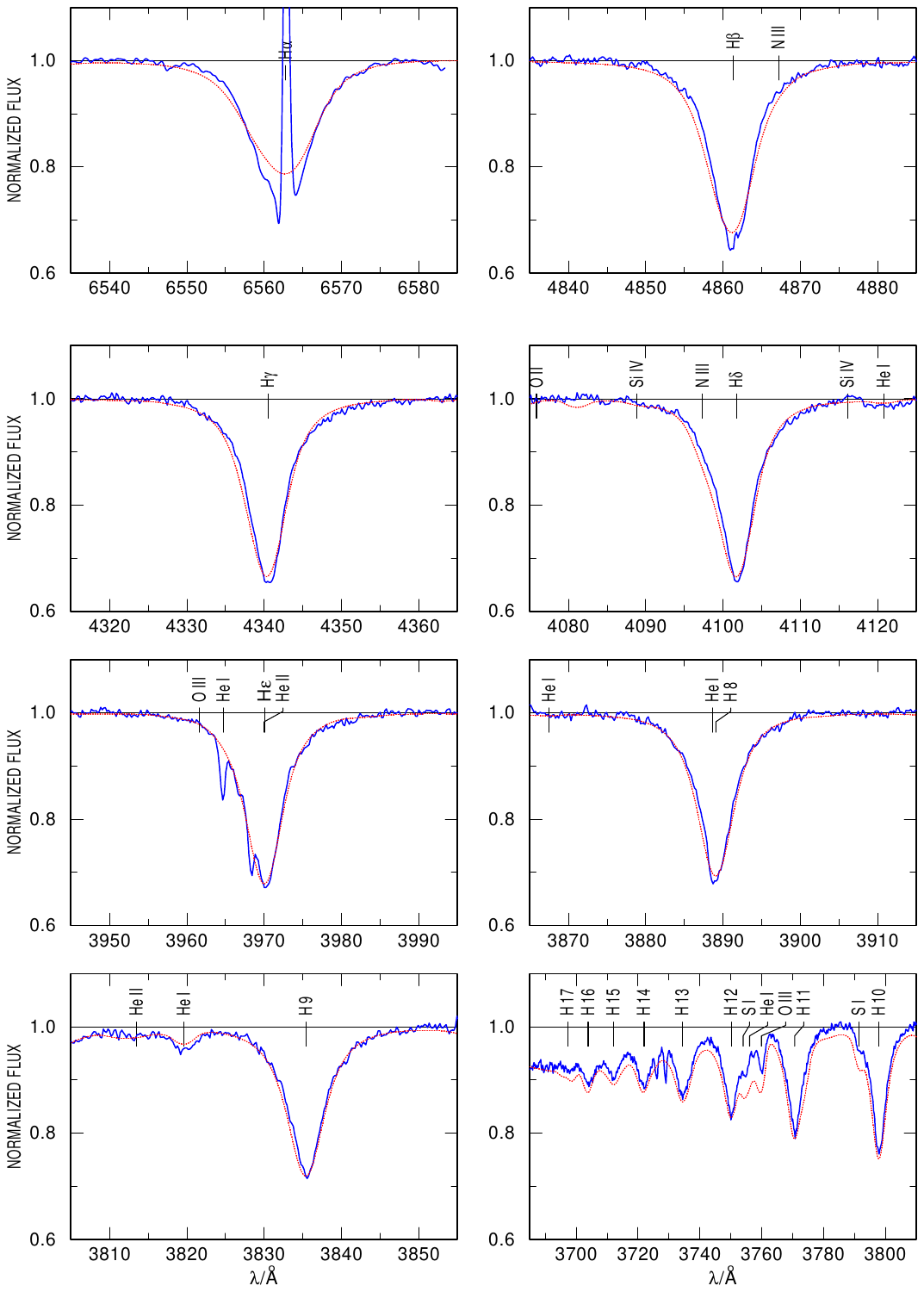}~
\includegraphics[trim={1.5cm 2.3cm 1.5cm 2.1cm},clip,width=0.4\linewidth]{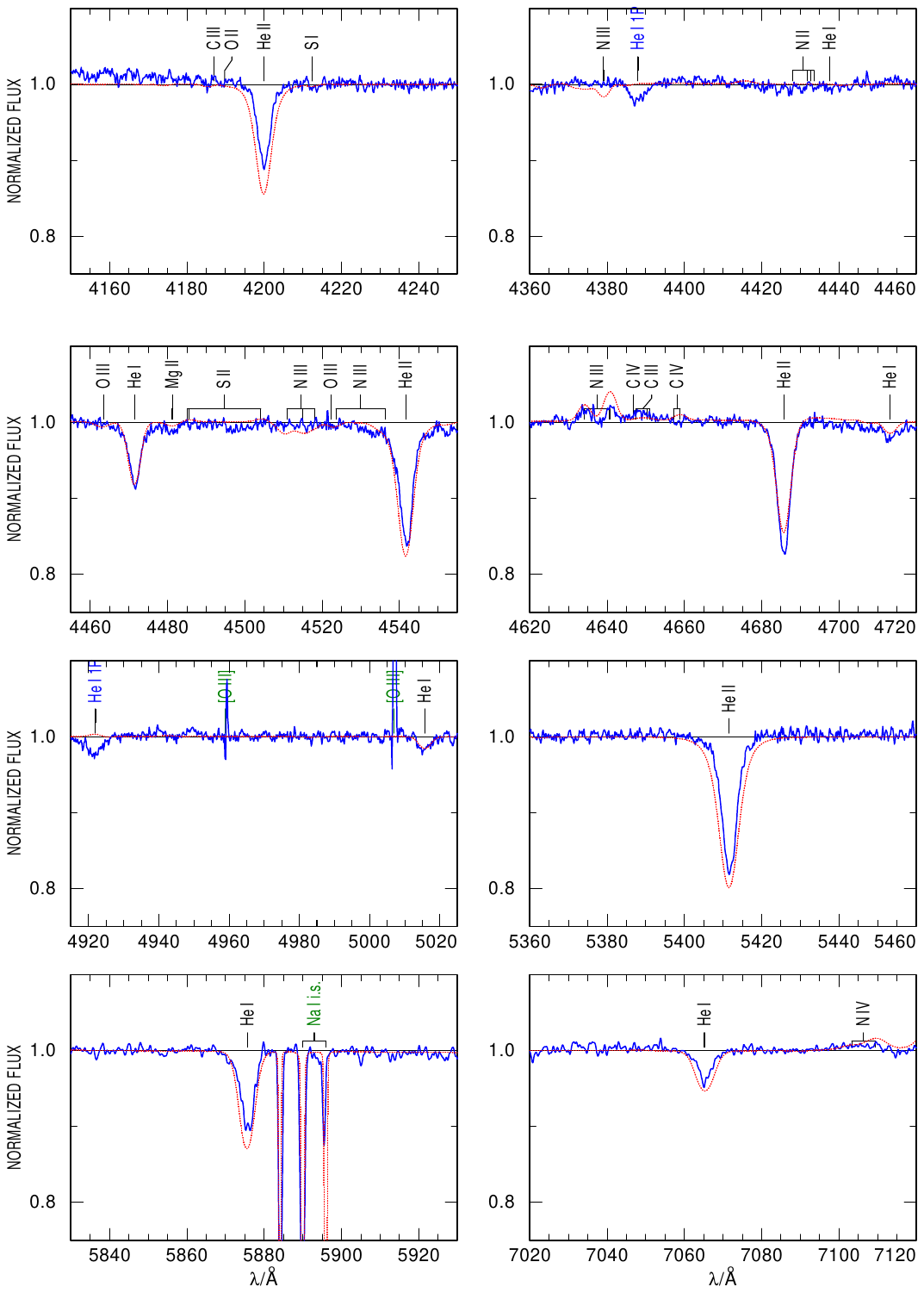}\\
\par\end{centering}
\caption{
PoWR model for PGMW\,3120. ({\it left}) Balmer lines in the observed spectrum in comparison with the final model of the star. The \logg\  is determined from the wings of the H lines. \ha\ is considered to determine $\dot{M}$. The \ha\ line is affected by an emission line, most likely of nebular origin.
({\it right}) The most important \hei\ and \heii\ lines, as well as some metals, in comparison with the final model of the star. The \tstar\ of the star is determined by modeling the \hei-\heii\ ratios. Not photospheric lines are indicated with green labels.
\heiiwr\ is crucial to determine $\dot{M}$.
The observed spectra are shown by a blue line and the PoWR model by a red dashed line. The most important lines are identified.
}
\label{fig:master_pgmw3120_balmer}
\end{figure*}

\begin{figure*}
\begin{centering}
\includegraphics[trim={0 5.0cm 0 5.5cm},clip,width=0.5\linewidth]{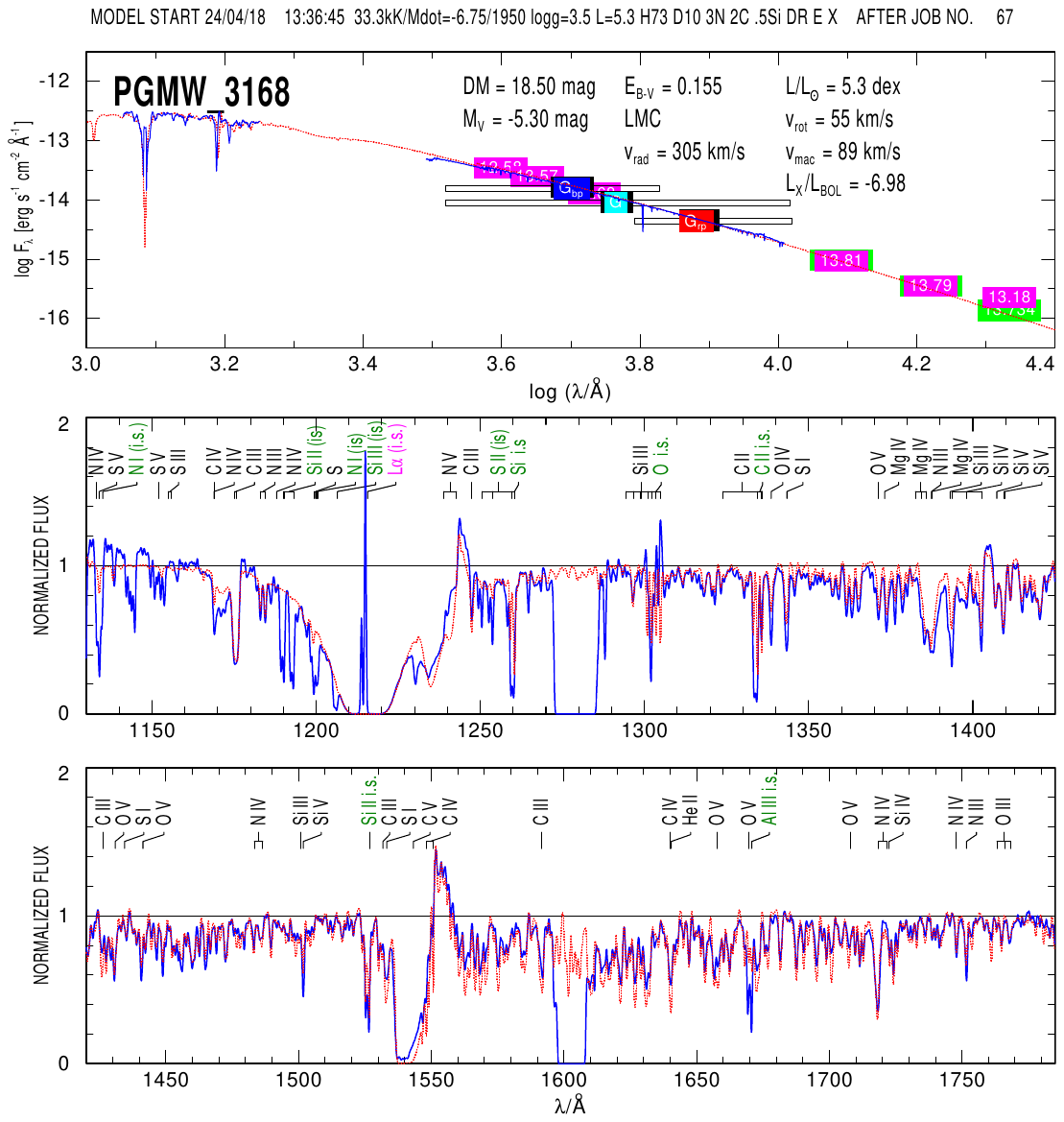}
\par\end{centering}
\caption{
PoWR model for PGMW\,3168. The observed spectra is shown by a blue line and the model by a red dashed line. (1st panel) SED with photometric magnitudes (colour boxes). The UV spectra better constrain the \ebv\ and \lstar; (2nd and 3rd panel) \hst/\cosi\ UV spectra normalized to the continuum model. Interstellar (i.s.) atomic, molecular and metal lines in absorption are indicated.
There is a gap of around 10~\AA\ in the observations around 1280, and 1605~\AA, where no key lines are present.
}
\label{fig:master_n11-32_sed}
\end{figure*}

\begin{figure*}
\begin{centering}
\includegraphics[trim={1.5cm 2.0cm 1.5cm 2.0cm},clip,width=0.43\linewidth]{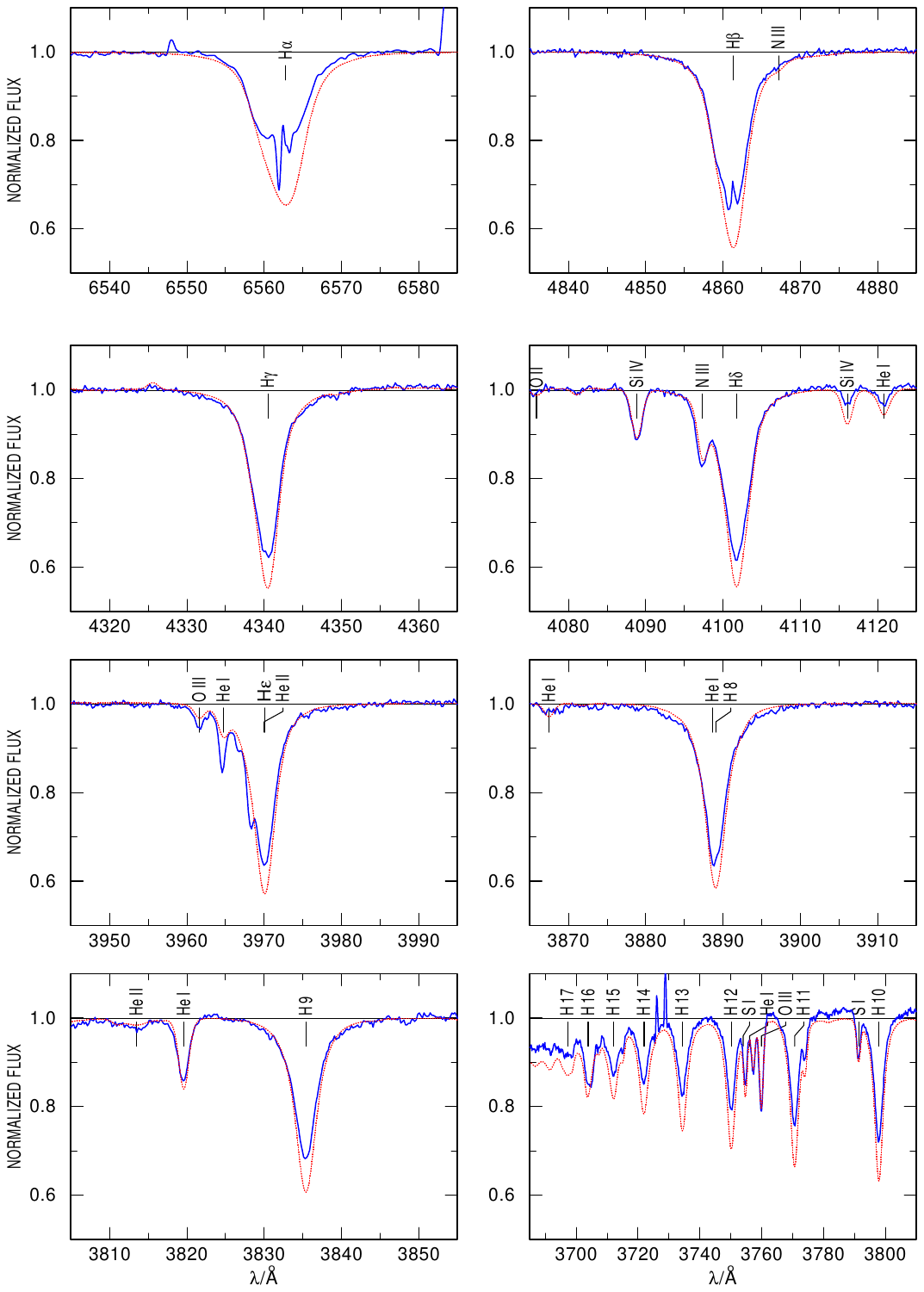}~
\includegraphics[trim={1.5cm 2.0cm 1.5cm 2.0cm},clip,width=0.43\linewidth]{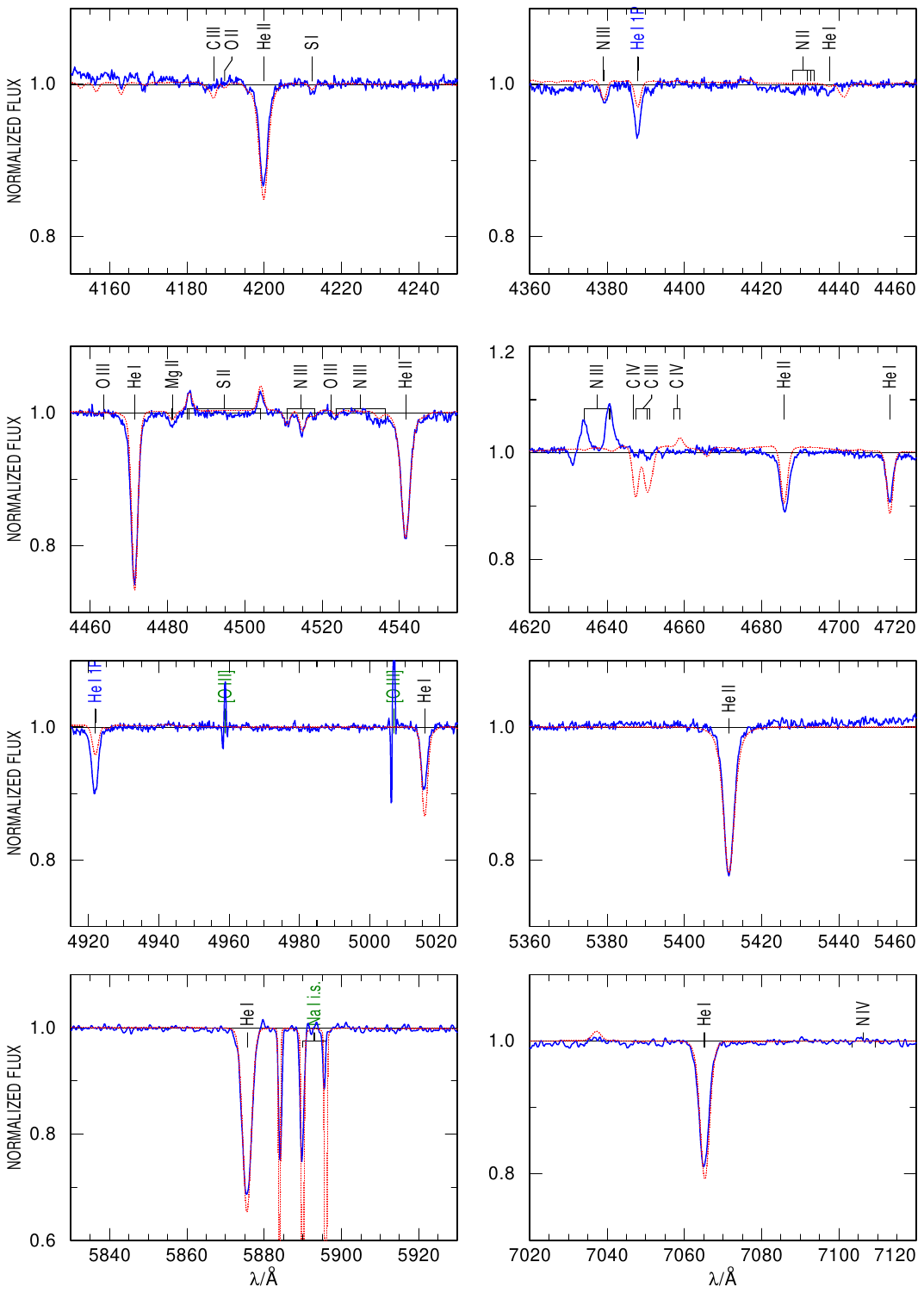}\\
\par\end{centering}
\caption{
PoWR model for PGMW\,3168. ({\it left}) Balmer lines in the observed spectrum in comparison with the final model of the star. The \logg\ is determined from the wings of the H lines. \ha\ is considered to determine $\dot{M}$. The \ha\ line is affected by an emission line, most likely of nebular origin.
({\it right}) The most important \hei\ and \heii\ lines, as well as some metals, in comparison with the final model of the star. The \tstar\ of the star is determined by modeling the \hei-\heii\ ratios. Not photospheric lines are indicated with green labels.
\heiiwr\ is crucial to determine $\dot{M}$.
The observed spectra are shown by a blue line and the PoWR model by a red dashed line. The most important lines are identified.
}
\label{fig:master_n11-32_balmer}
\end{figure*}

\begin{figure*}
\begin{centering}
\includegraphics[trim={0 5.0cm 0 5.5cm},clip,width=0.5\linewidth]{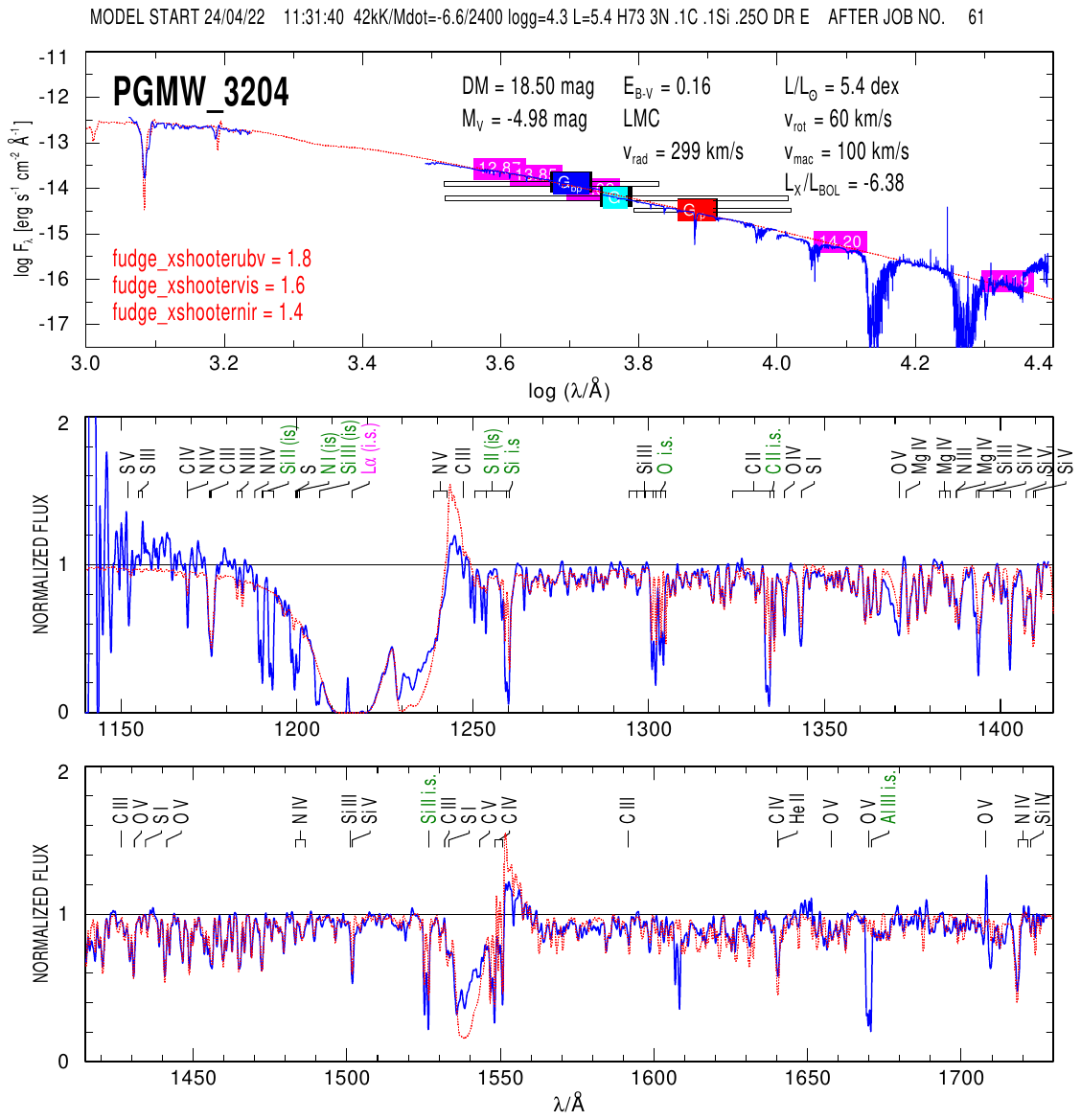}
\par\end{centering}
\caption{
PoWR model for PGMW\,3204.
The observed spectra is shown by a blue line and the model by a red dashed line. (1st panel) SED with photometric magnitudes (colour boxes). The UV spectra better constrain the \ebv\ and \lstar; (2nd and 3rd panel) \hst/\cosi\ UV spectra normalized to the continuum model. Interstellar (i.s.) atomic, molecular and metal lines in absorption are indicated.
There is a gap of around 10~\AA\ in the observations around 1280, and 1605~\AA, where no key lines are present.
}
\label{fig:master_n11-48_sed}
\end{figure*}

\begin{figure*}
\begin{centering}
\includegraphics[trim={1.5cm 2.0cm 1.5cm 2.0cm},clip,width=0.43\linewidth]{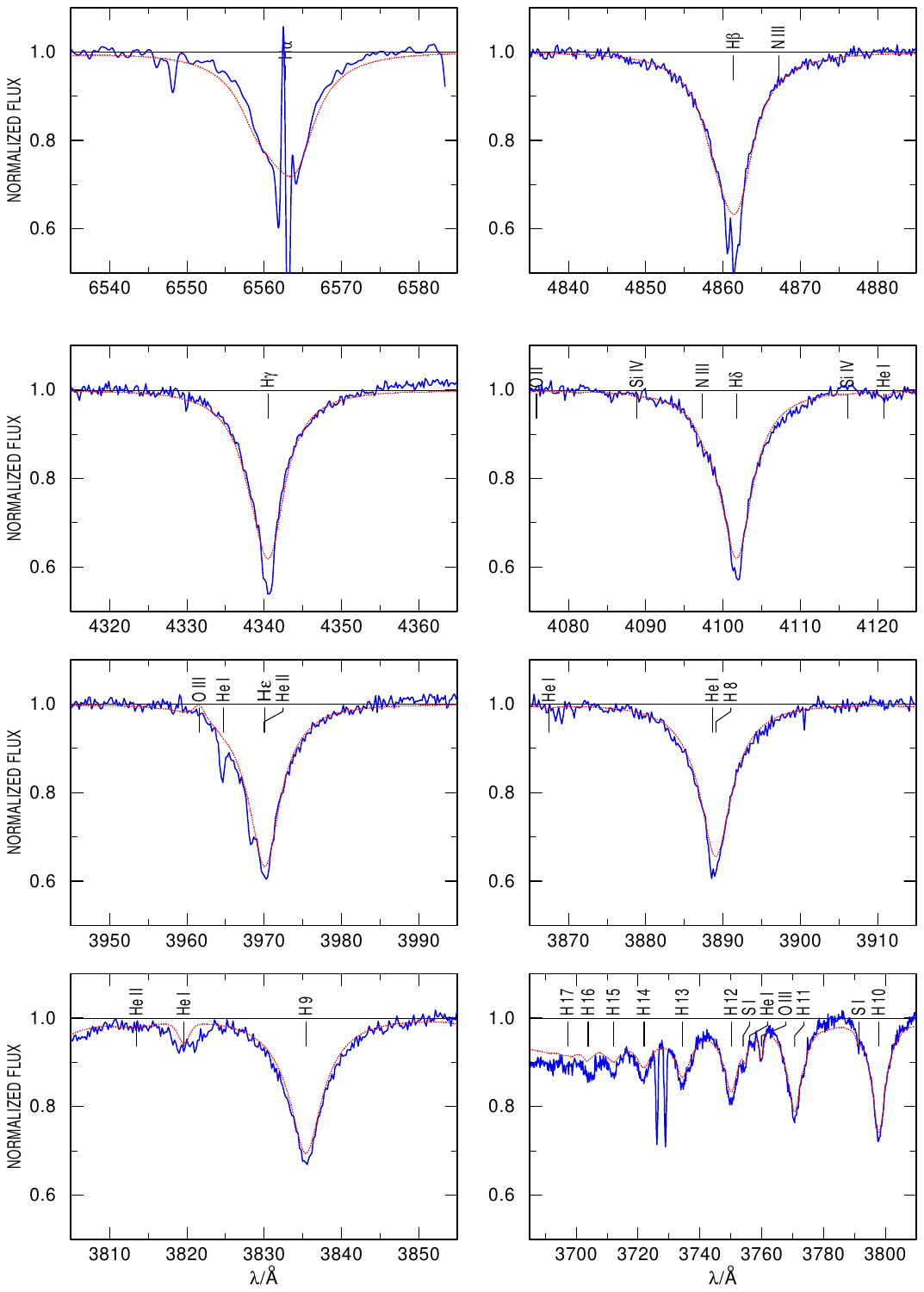}~
\includegraphics[trim={1.5cm 2.0cm 1.5cm 2.0cm},clip,width=0.43\linewidth]{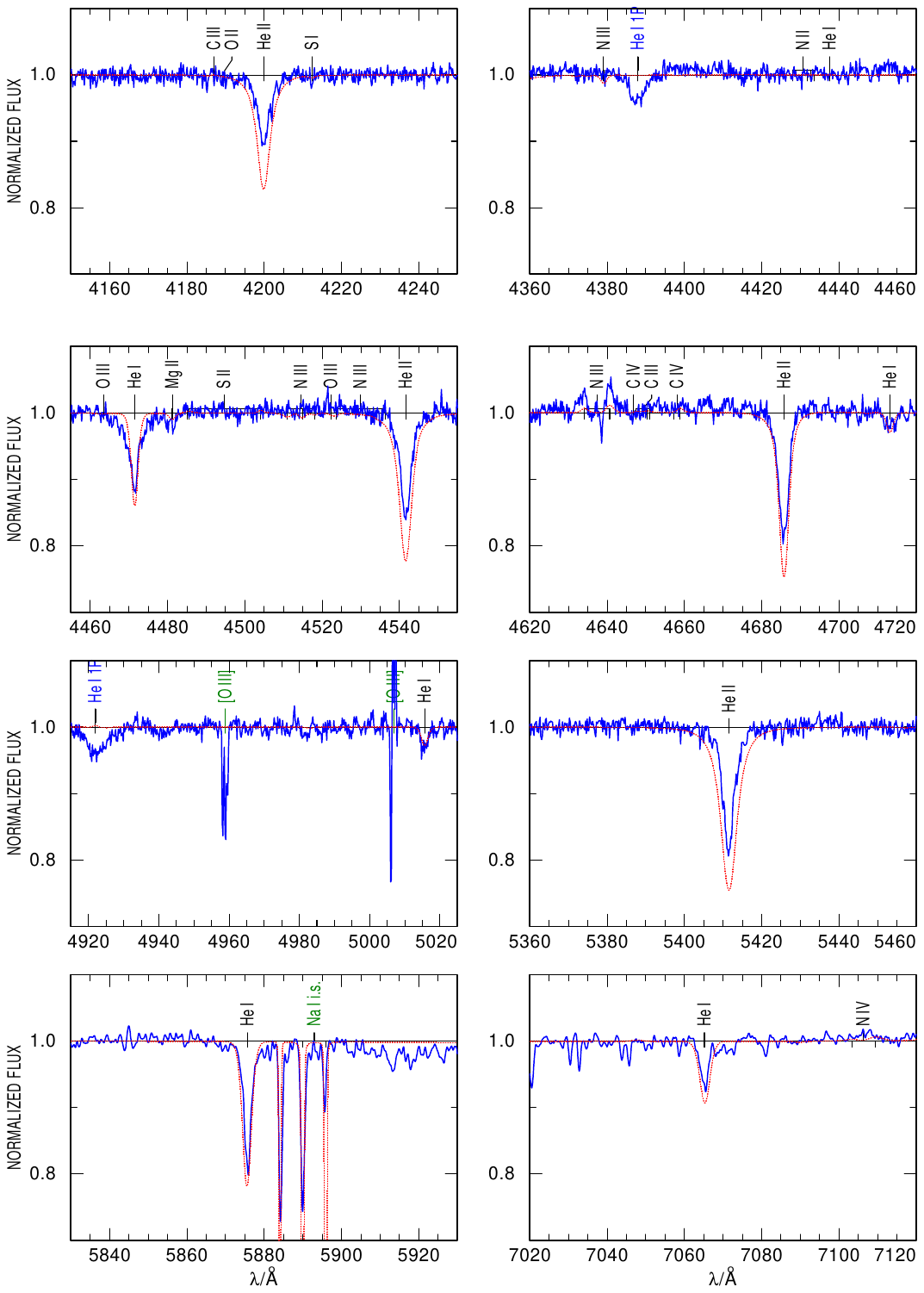}\\
\par\end{centering}
\caption{
PoWR model for PGMW\,3204.
({\it left}) Balmer lines in the observed spectrum in comparison with the final model of the star. The \logg\ is determined from the wings of the H lines. \ha\ is considered to determine $\dot{M}$. The \ha\ line is affected by an emission line, most likely of nebular origin.
({\it right}) The most important \hei\ and \heii\ lines, as well as some metals, in comparison with the final model of the star. The \tstar\ of the star is determined by modeling the \hei-\heii\ ratios. Not photospheric lines are indicated with green labels.
\heiiwr\ is crucial to determine $\dot{M}$.
The observed spectra are shown by a blue line and the PoWR model by a red dashed line. The most important lines are identified.
}
\label{fig:master_n11-48_balmer}
\end{figure*}

\begin{figure*}
\begin{centering}
\includegraphics[trim={0 2.0cm 0 2.7cm},clip,width=0.43\linewidth]{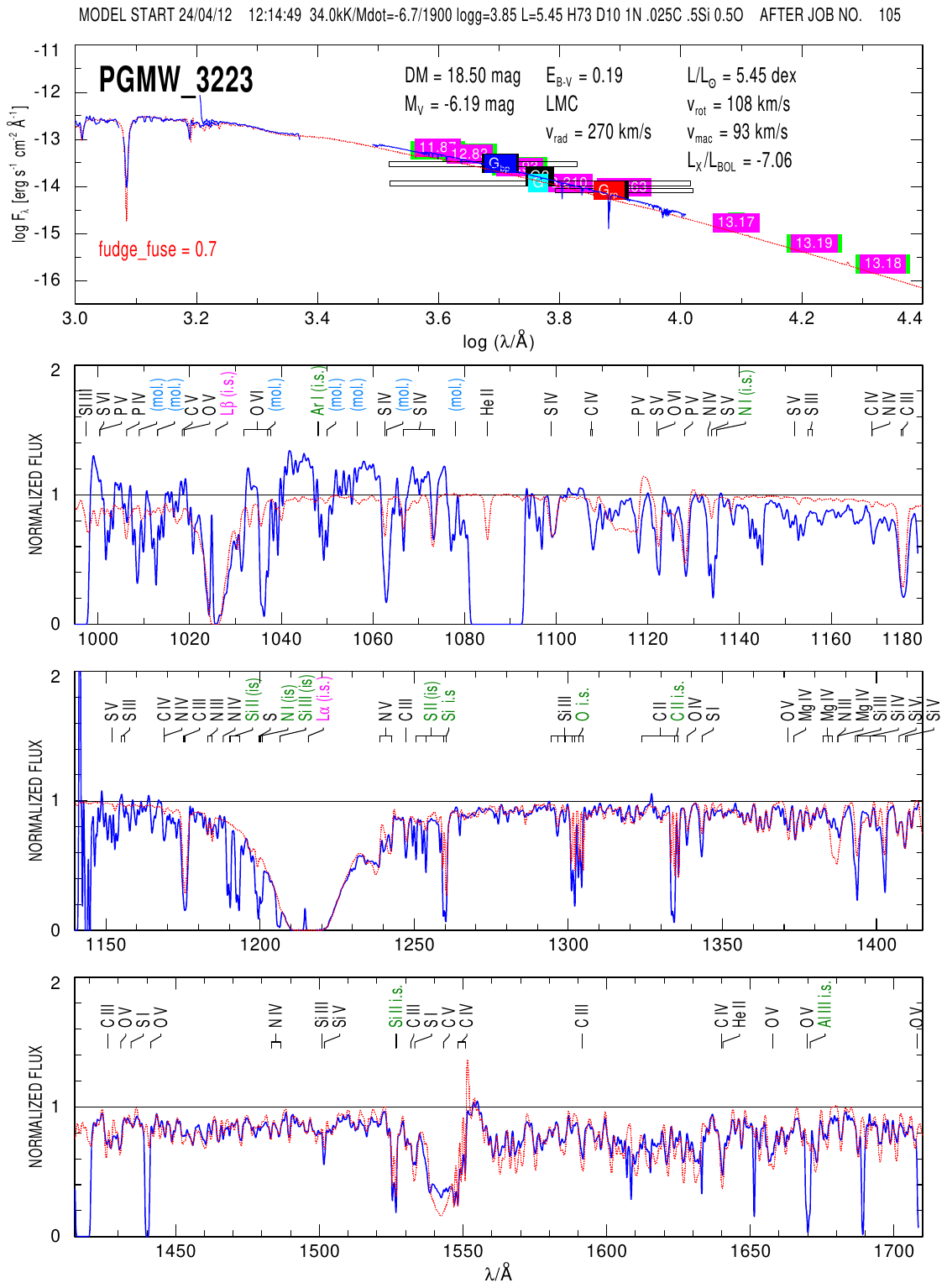}
\par\end{centering}
\caption{
PoWR model for PGMW\,3223. The observed spectra is shown by a blue line and the model by a red dashed line.
(1st panel) SED with photometric magnitudes (colour boxes).
(2nd panel) \fuse/MRDS UV spectra normalized to the continuum model;
(3rd and 4th panel) \hst/\stis\ UV spectra normalized to the continuum model.
Particularly sensitive to X-rays are the \nvuv\ and \ovi\ features.
Interstellar (i.s.) atomic, molecular and metal lines in absorption are indicated.
The legend "fudge" means that the \fuse\ spectrum needed to be scaled (see text for details).
There is a gap around 1090, and 1420~\AA, where no key lines are present.
}
\label{fig:n11-13_xshootu_sed}
\end{figure*}

\begin{figure*}
\begin{centering}
\includegraphics[trim={1.5cm 2.1cm 1.5cm 2.0cm},clip,width=0.4\linewidth]{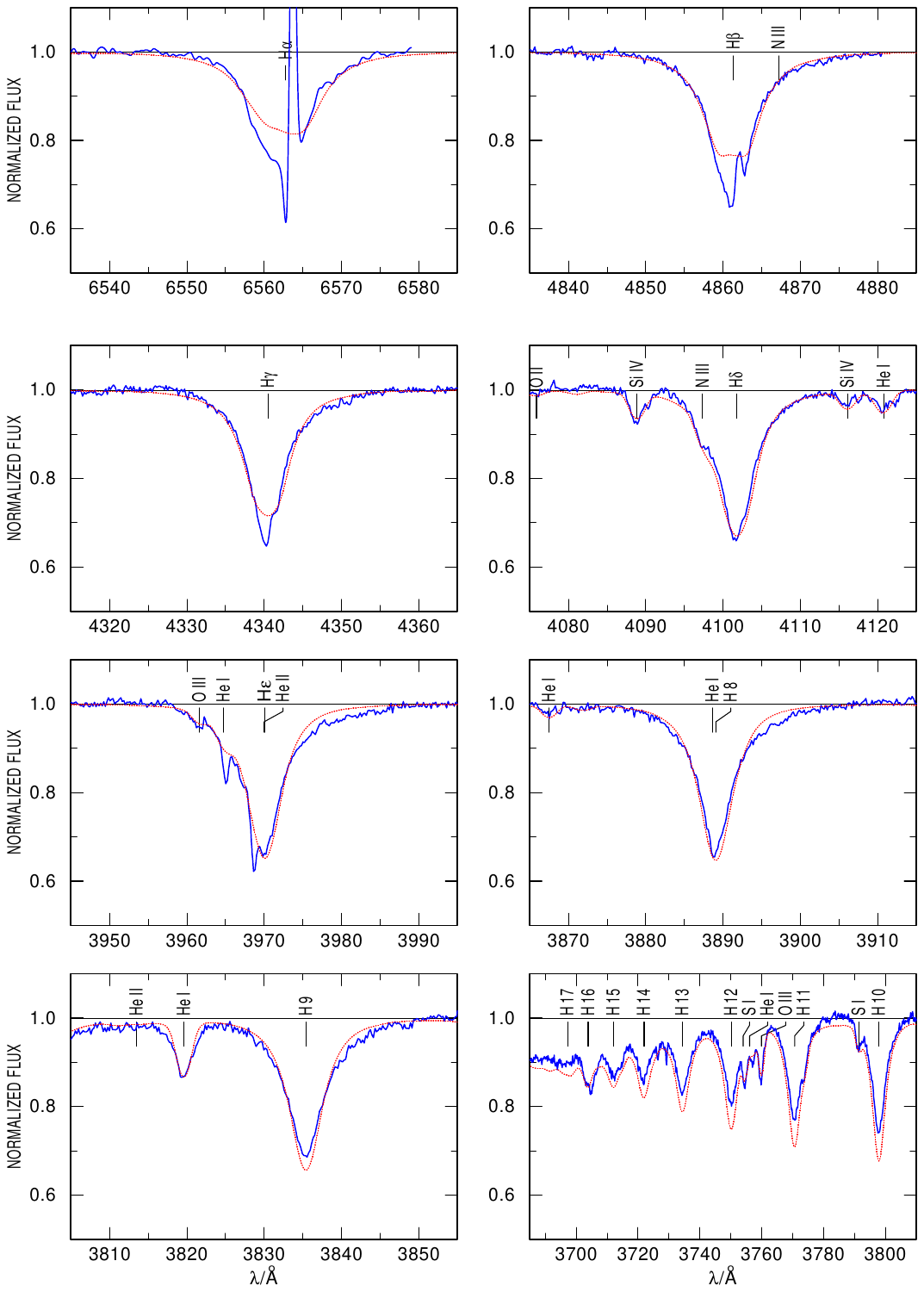}~
\includegraphics[trim={1.5cm 2.1cm 1.5cm 2.0cm},clip,width=0.4\linewidth]{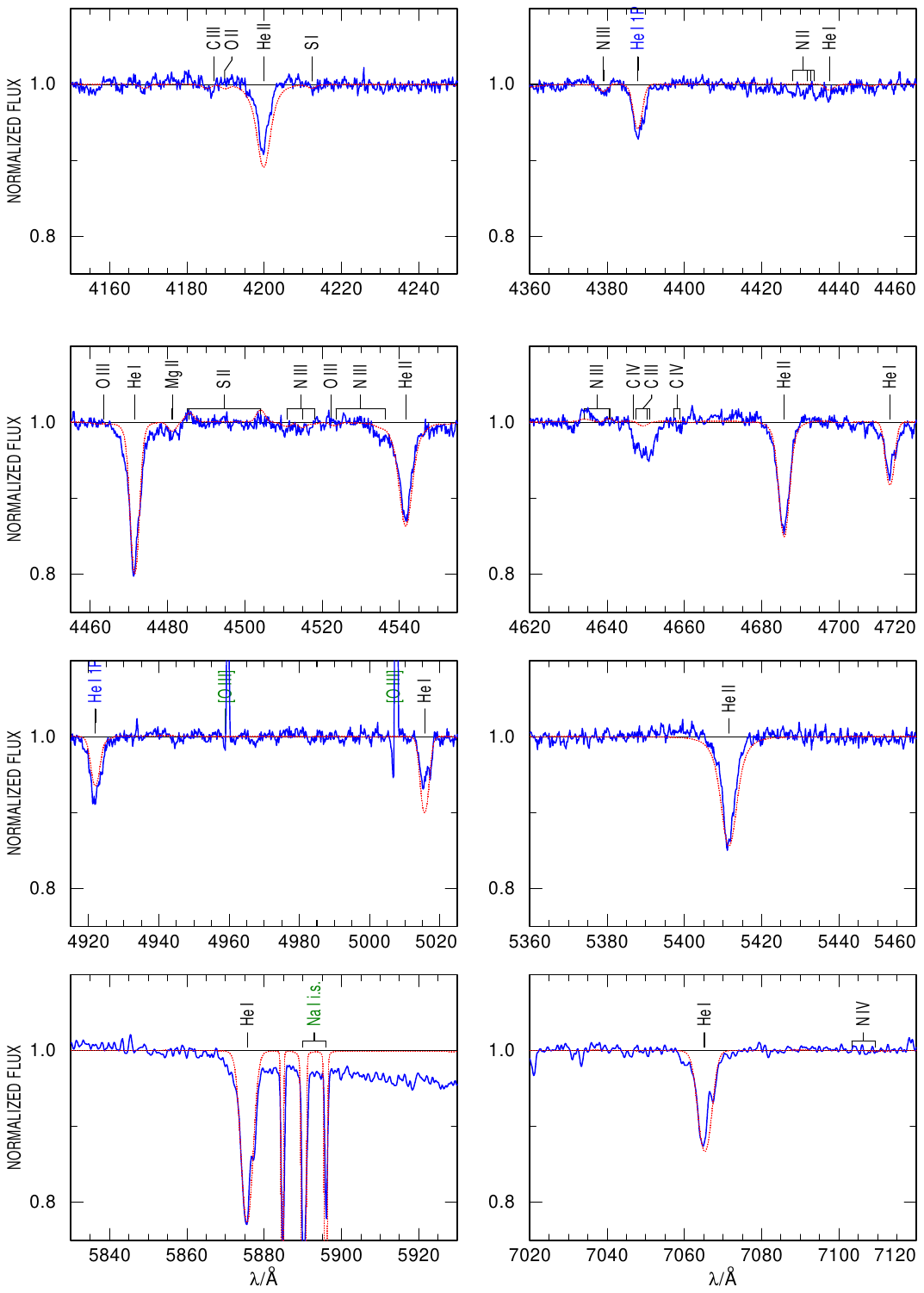}\\
\par\end{centering}
\caption{
PoWR model for PGMW\,3223.
({\it left}) Balmer lines in the observed spectrum in comparison with the final model of the star. The \logg\ is determined from the wings of the H lines. \ha\ is considered to determine $\dot{M}$. The \ha\ line is affected by an emission line, most likely of nebular origin.
({\it right}) The most important \hei\ and \heii\ lines, as well as some metals, in comparison with the final model of the star. The \tstar\ of the star is determined by modeling the \hei-\heii\ ratios. Not photospheric lines are indicated with green labels.
\heiiwr\ is crucial to determine $\dot{M}$.
The observed spectra are shown by a blue line and the PoWR model by a red dashed line. The most important lines are identified.
}
\label{fig:n11-13_xshootu_balmer}
\end{figure*}

\begin{figure*}
\begin{centering}
\includegraphics[trim={0 2.0cm 0 2.5cm},clip,width=0.43\linewidth]{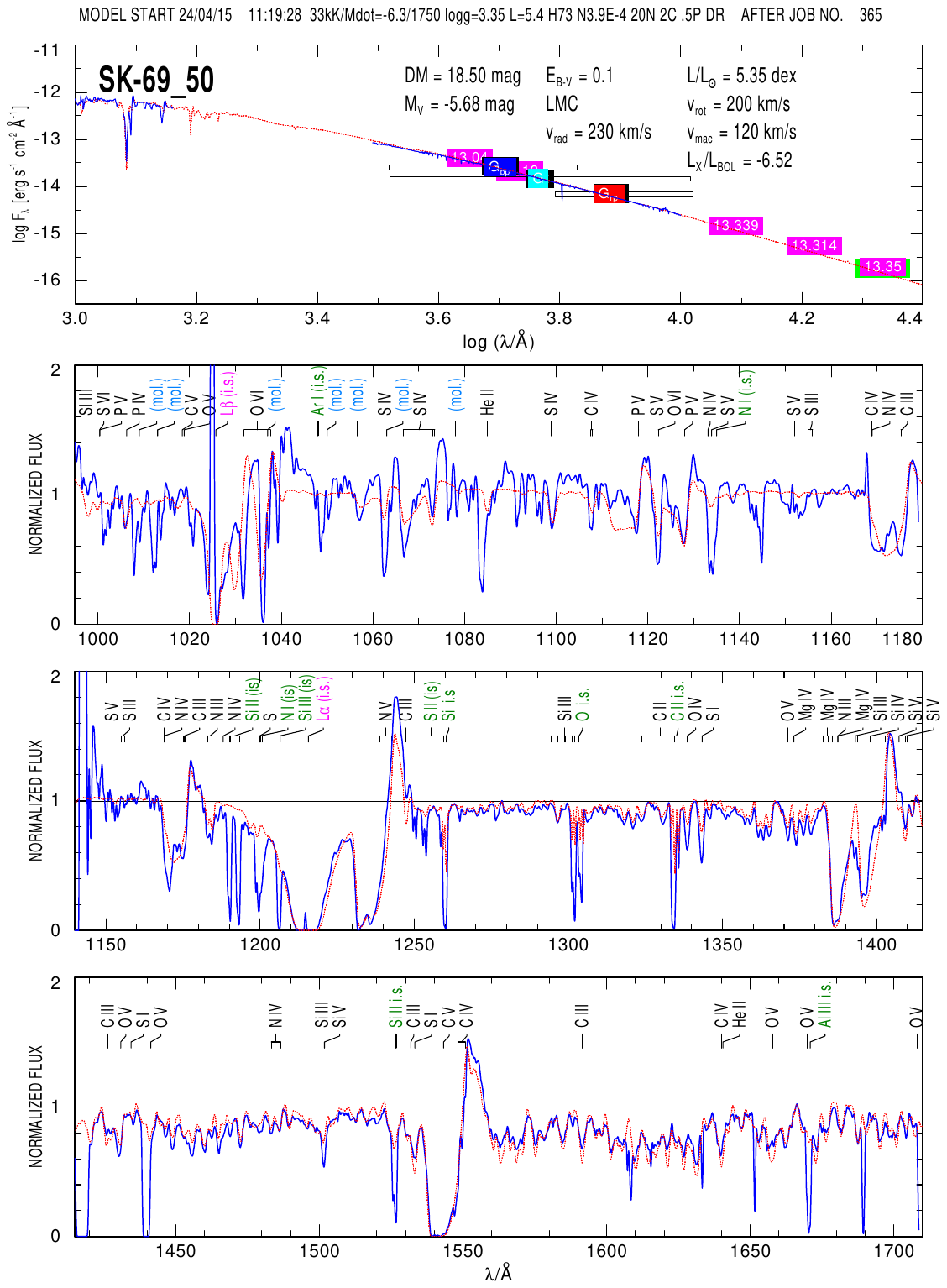}
\par\end{centering}
\caption{
PoWR model for Sk $-69^{\circ}$ 50.
The observed spectra is shown by a blue line and the model by a red dashed line.
(1st panel) SED with photometric magnitudes (colour boxes).
(2nd panel) \fuse/MRDS UV spectra;
and (3rd and 4th panel) \hst/\stis\ UV spectra normalized to the continuum model.
Particularly sensitive to X-rays are the \nvuv\ and \ovi\ features.
Interstellar (i.s.) atomic, molecular and metal lines in absorption are indicated.
}
\label{fig:sk-69d50_xshootu_sed}
\end{figure*}

\begin{figure*}
\begin{centering}
\includegraphics[trim={1.5cm 2.3cm 1.5cm 2.0cm},clip,width=0.41\linewidth]{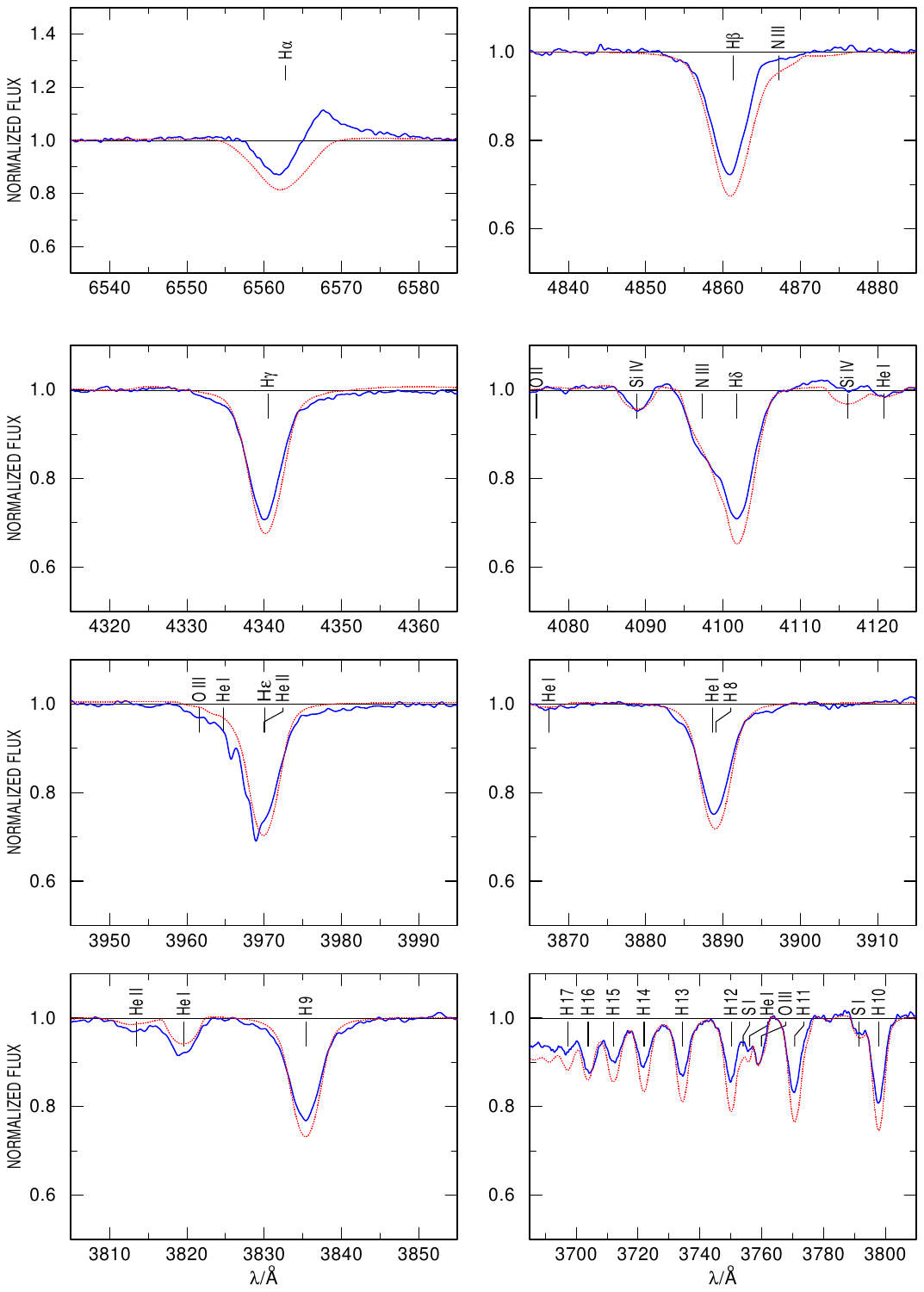}~
\includegraphics[trim={1.5cm 2.3cm 1.5cm 2.0cm},clip,width=0.41\linewidth]{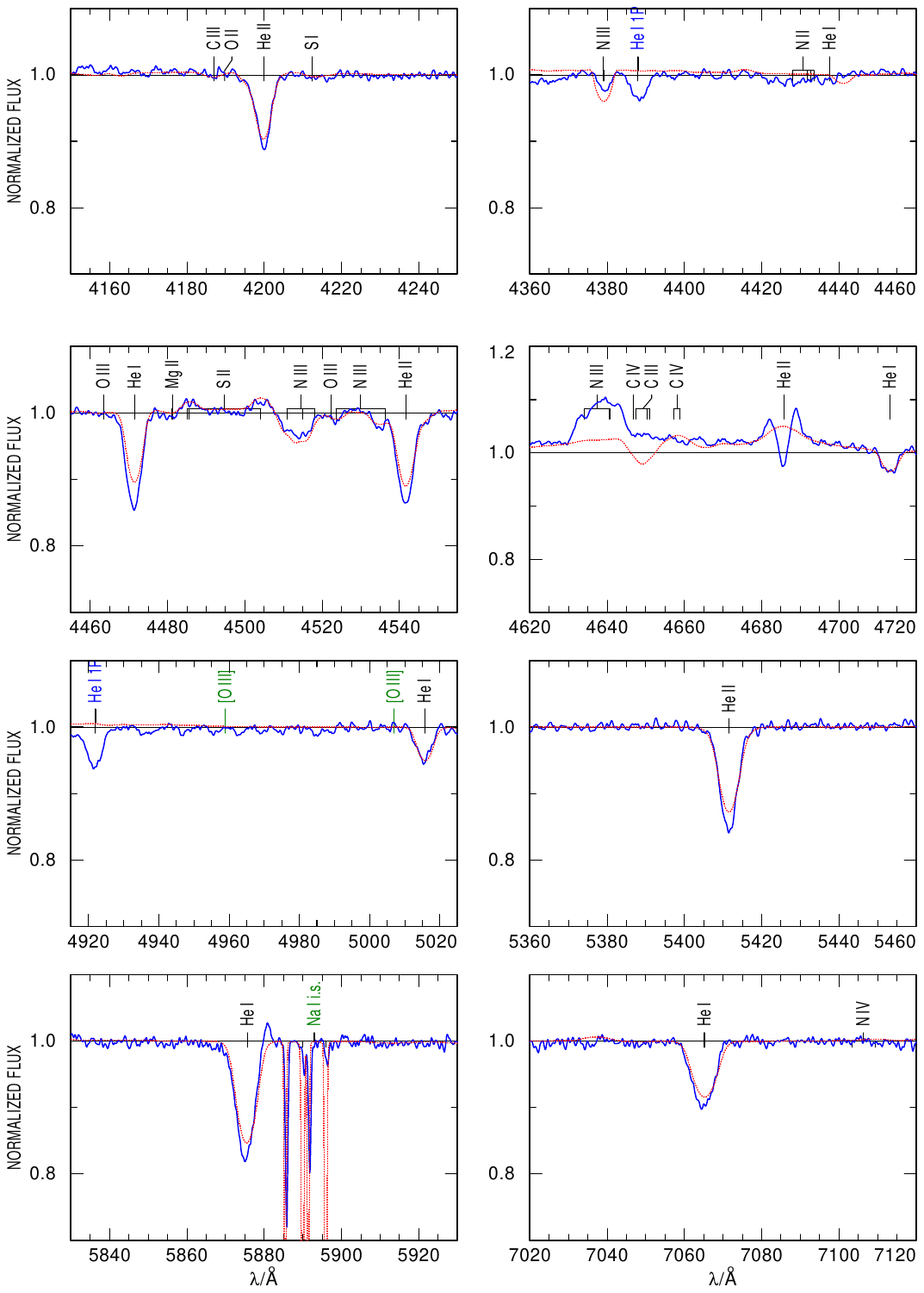}\\
\par\end{centering}
\caption{
PoWR model for Sk $-69^{\circ}$ 50.
({\it left}) Balmer lines in the observed spectrum in comparison with the final model of the star. The \logg\ is determined from the wings of the H lines. \ha\ is considered to determine $\dot{M}$. The \ha\ line is affected by an emission line, most likely of nebular origin.
({\it right}) The most important \hei\ and \heii\ lines, as well as some metals, in comparison with the final model of the star. The \tstar\ of the star is determined by modeling the \hei-\heii\ ratios. Not photospheric lines are indicated with green labels.
\heiiwr\ is crucial to determine $\dot{M}$.
The observed spectra are shown by a blue line and the PoWR model by a red dashed line. The most important lines are identified.
}
\label{fig:sk-69d50_xshootu_balmer}
\end{figure*}

\begin{figure*}
\begin{centering}
\includegraphics[trim={0 2.0cm 0 2.7cm},clip,width=0.43\linewidth]{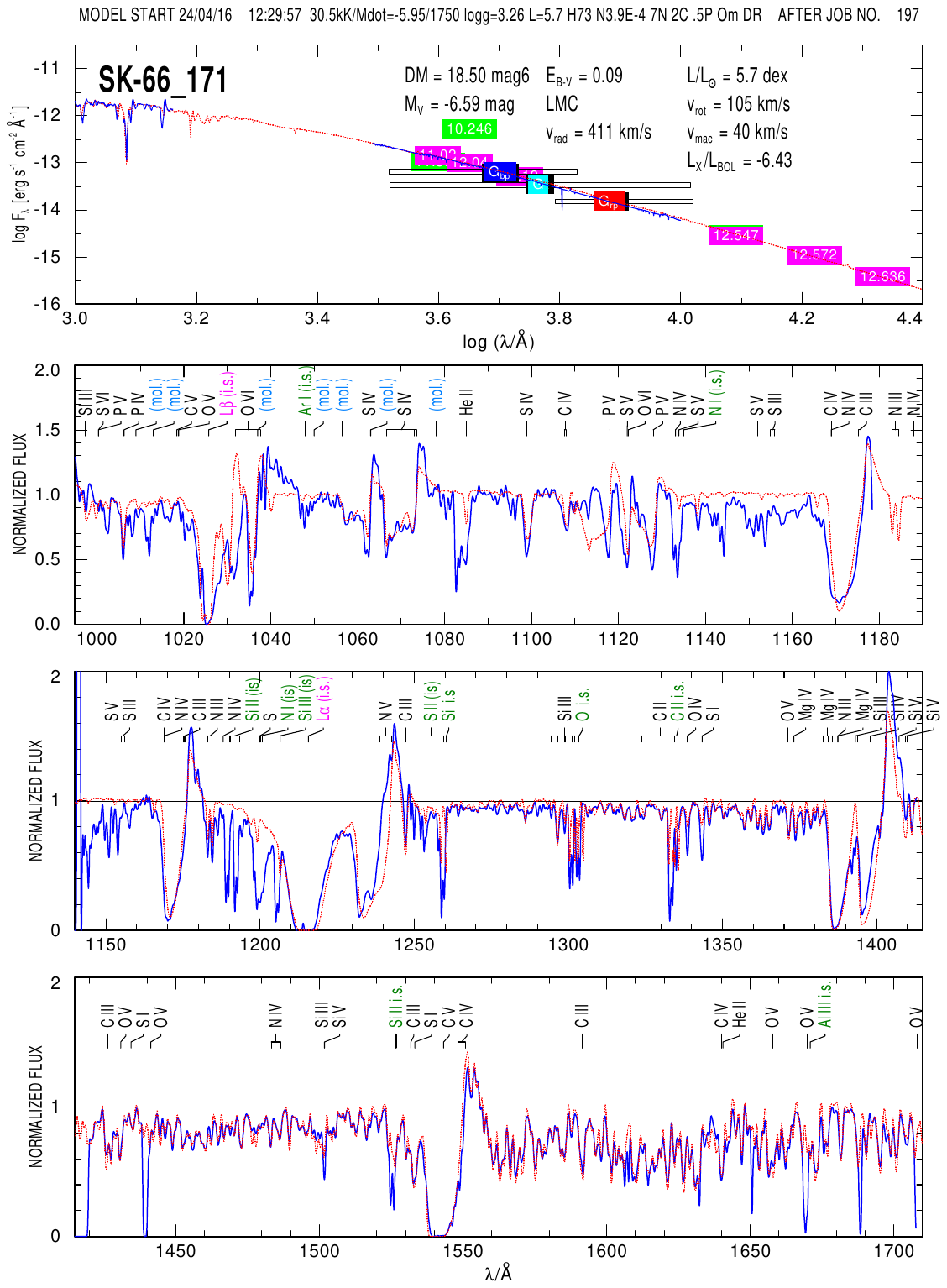}
\par\end{centering}
\caption{
PoWR model for Sk $-66^{\circ}$ 171.
The observed spectra is shown by a blue line and the model by a red dashed line.
(1st panel) SED with photometric magnitudes (colour boxes).
(2nd panel) \fuse/MRDS UV spectra;
and (3rd and 4th panel) \hst/\stis\ UV spectra normalized to the continuum model.
Particularly sensitive to X-rays are the \nvuv\ and \ovi\ features.
Interstellar (i.s.) atomic, molecular and metal lines in absorption are indicated.
}
\label{fig:sk-66d171_xshootu_sed}
\end{figure*}

\begin{figure*}
\begin{centering}
\includegraphics[trim={1.5cm 2.3cm 1.5cm 2.1cm},clip,width=0.41\linewidth]{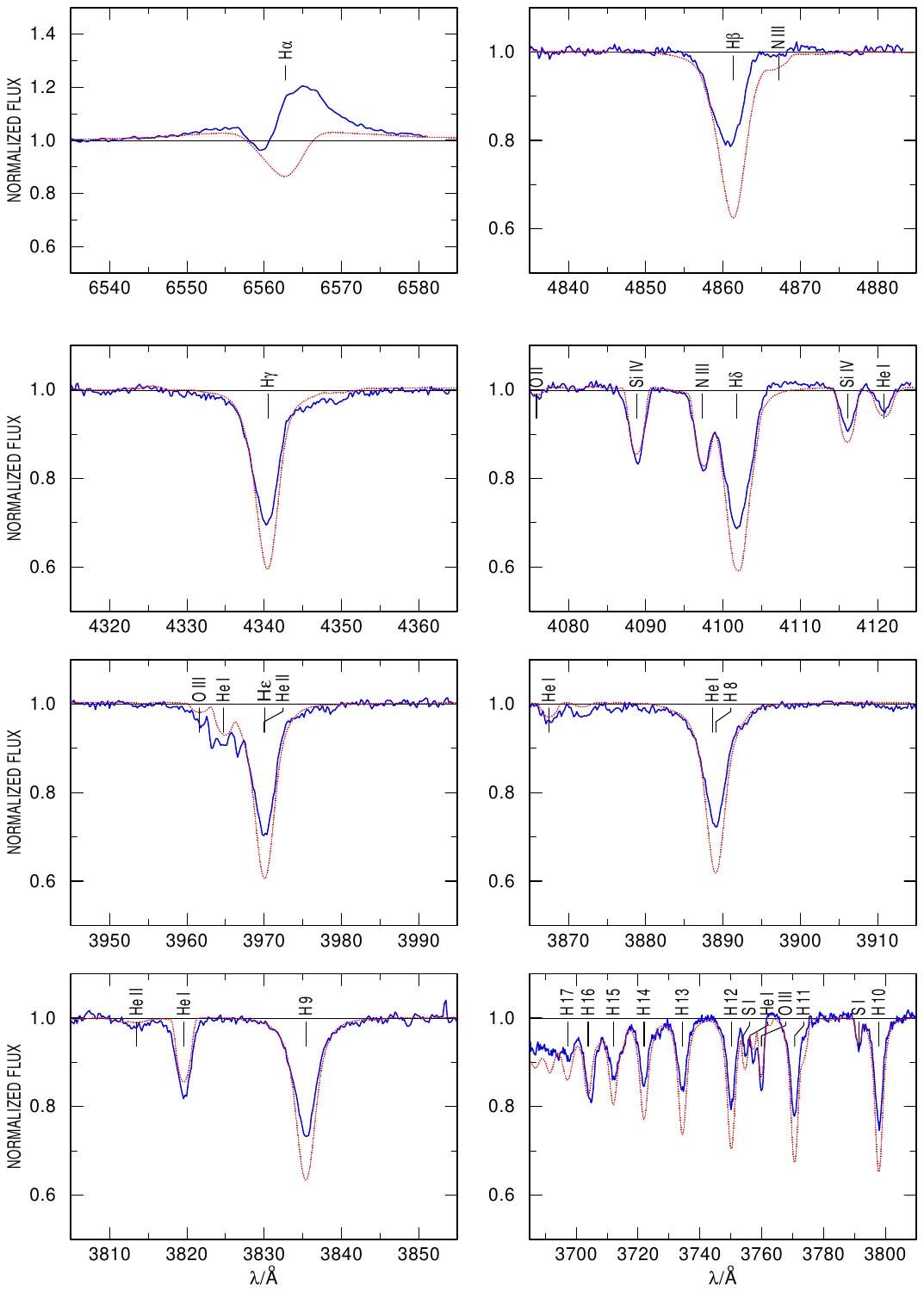}~
\includegraphics[trim={1.5cm 2.3cm 1.5cm 2.1cm},clip,width=0.41\linewidth]{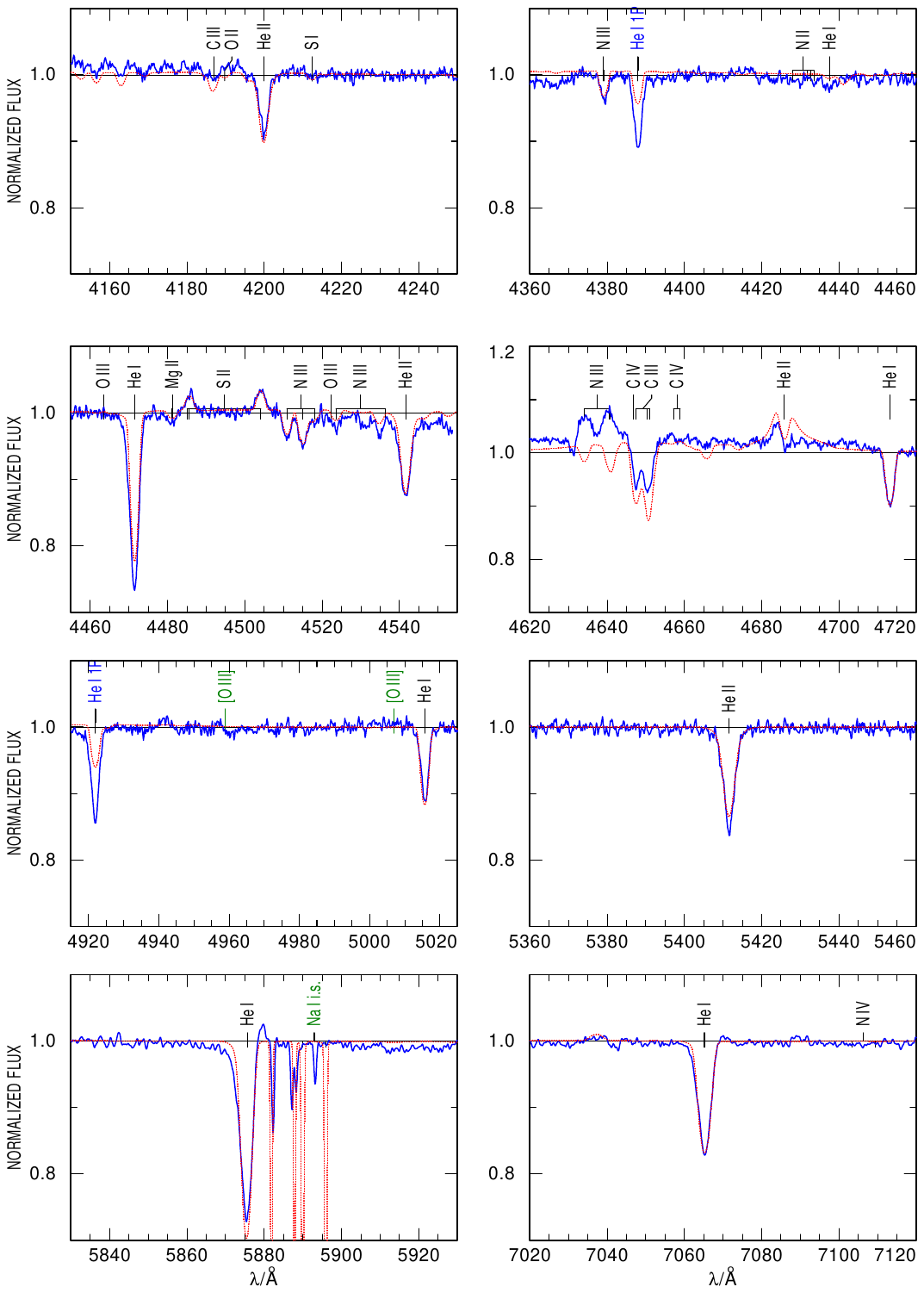}\\
\par\end{centering}
\caption{
PoWR model for Sk $-66^{\circ}$ 171.
({\it left}) Balmer lines in the observed spectrum in comparison with the final model of the star. The \logg\ is determined from the wings of the H lines. \ha\ is considered to determine $\dot{M}$. The \ha\ line is affected by an emission line, most likely of nebular origin.
({\it right}) The most important \hei\ and \heii\ lines, as well as some metals, in comparison with the final model of the star. The \tstar\ of the star is determined by modeling the \hei-\heii\ ratios. Not photospheric lines are indicated with green labels.
\heiiwr\ is crucial to determine $\dot{M}$.
The observed spectra are shown by a blue line and the PoWR model by a red dashed line. The most important lines are identified.
}
\label{fig:sk-66d171_xshootu_balmer}
\end{figure*}

\begin{figure*}
\begin{centering}
\includegraphics[trim={0 2.0cm 0 2.5cm},clip,width=0.43\linewidth]{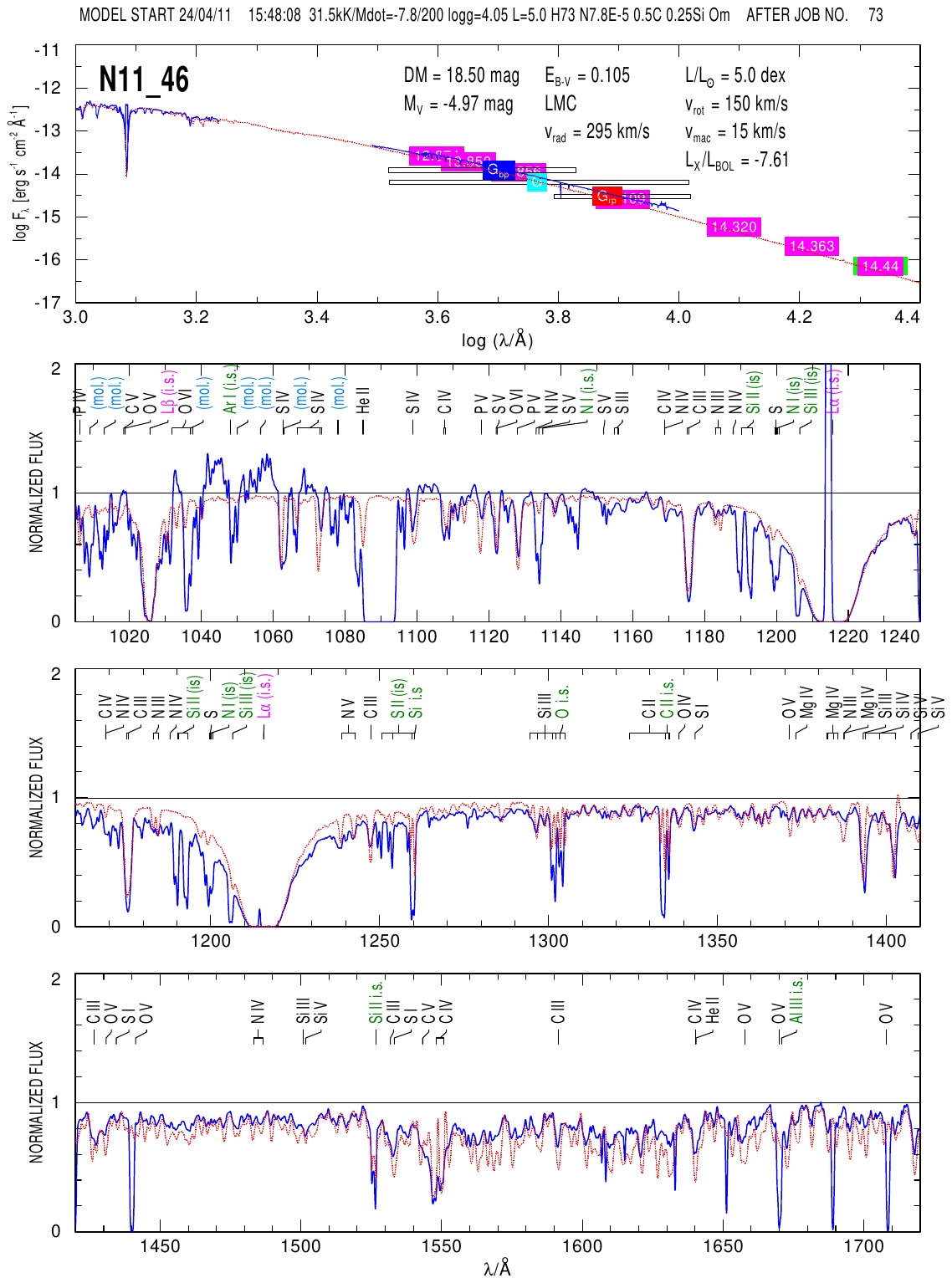}
\par\end{centering}
\caption{
PoWR model for N11\,046.
The observed spectra is shown by a blue line and the model by a red dashed line.
(1st panel) SED with photometric magnitudes (colour boxes).
(2nd panel) \hst/\cosi\ UV spectra; and
(3rd and 4th panel) \hst/\stis\ UV spectra normalized to the continuum model.
Particularly sensitive to X-rays are the \nvuv\ and \ovi\ features.
Interstellar (i.s.) atomic, molecular and metal lines in absorption are indicated.
There is a gap around 1090~\AA, where no key lines are present.
}
\label{fig:n11-46_xshootu_sed}
\end{figure*}

\begin{figure*}
\begin{centering}
\includegraphics[trim={1.5cm 2.3cm 1.5cm 2.1cm},clip,width=0.41\linewidth]{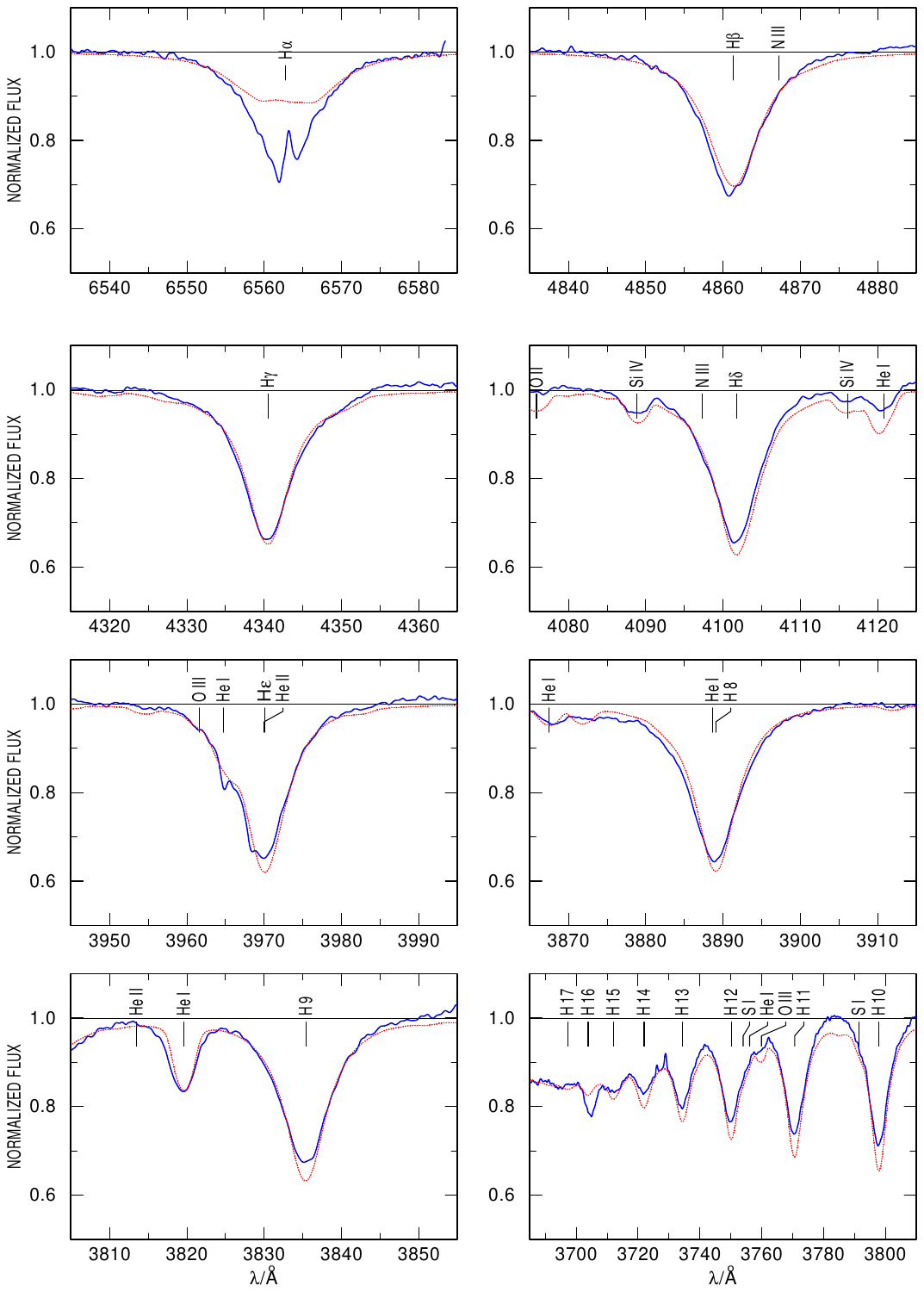}~
\includegraphics[trim={1.5cm 2.3cm 1.5cm 2.1cm},clip,width=0.41\linewidth]{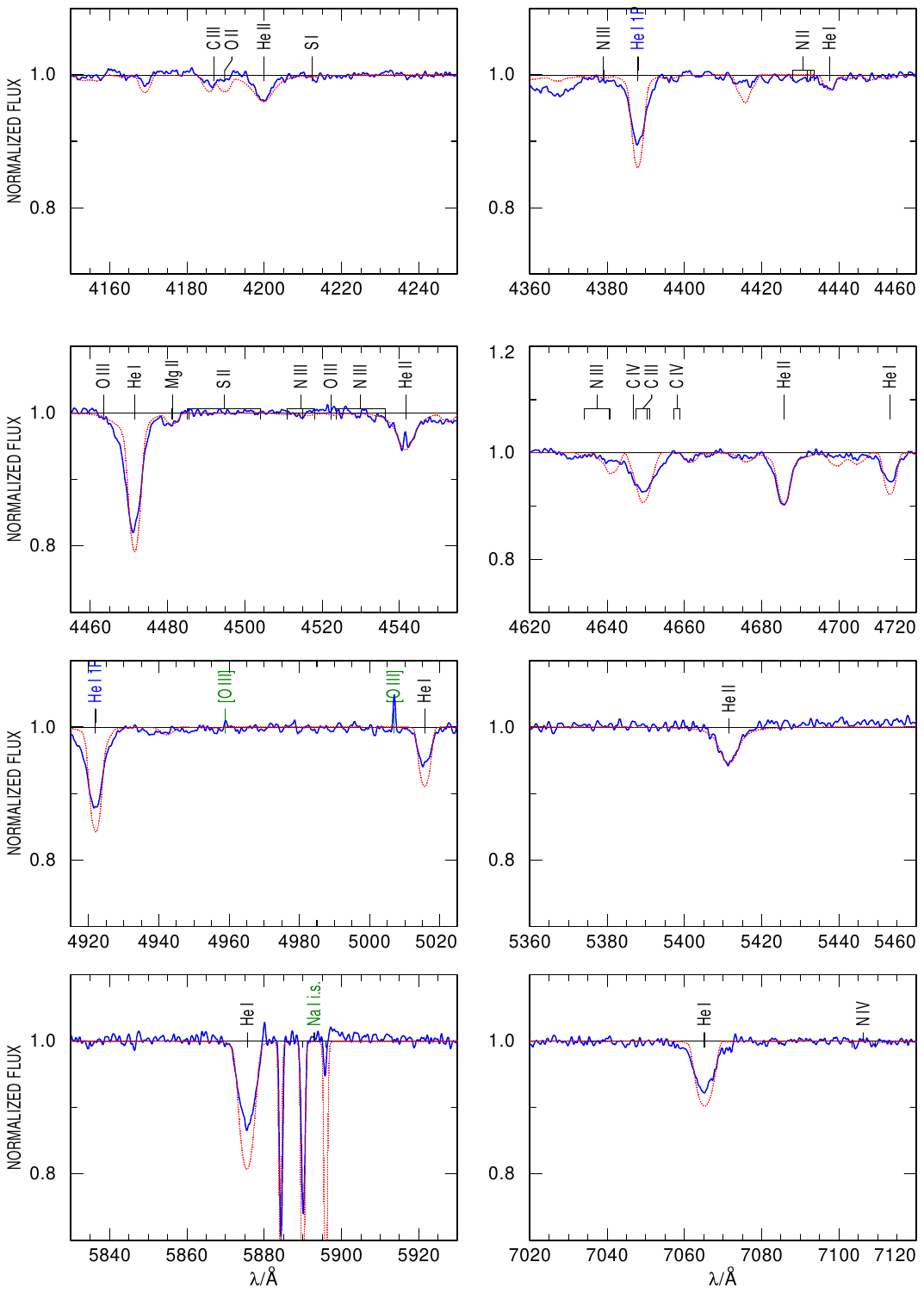}\\
\par\end{centering}
\caption{
PoWR model for N11 046.
({\it left}) Balmer lines in the observed spectrum in comparison with the final model of the star. The \logg\ is determined from the wings of the H lines. \ha\ is considered to determine $\dot{M}$. The \ha\ line is affected by an emission line, most likely of nebular origin.
({\it right}) The most important \hei\ and \heii\ lines, as well as some metals, in comparison with the final model of the star. The \tstar\ of the star is determined by modeling the \hei-\heii\ ratios. Not photospheric lines are indicated with green labels.
\heiiwr\ is crucial to determine $\dot{M}$.
The observed spectra are shown by a blue line and the PoWR model by a red dashed line. The most important lines are identified.
}
\label{fig:n11-46_xshootu_balmer}
\end{figure*}

The PoWR models for the analyzed stars in N11\,B is shown in Fig.~\ref{fig:master_n11-60_sed} to Fig.~\ref{fig:n11-13_xshootu_balmer}.
The SED and photometric magnitudes are shown in the first panel of each Figure.
UV and optical spectra (blue line) are shown normalized to the continuum model.
Balmer lines and the most important \hei\ and \heii\ lines, as well as some metals, are shown separately in comparison with the model.
The models for the three additional objects are shown in Fig.~\ref{fig:sk-69d50_xshootu_sed} to Fig.~\ref{fig:n11-46_xshootu_balmer}.

\section{Check for binarity}
\label{models}

\begin{figure*}
\begin{centering}
\includegraphics[trim={1.0cm 2.3cm 1.0cm 2.1cm},clip,width=0.43\linewidth]{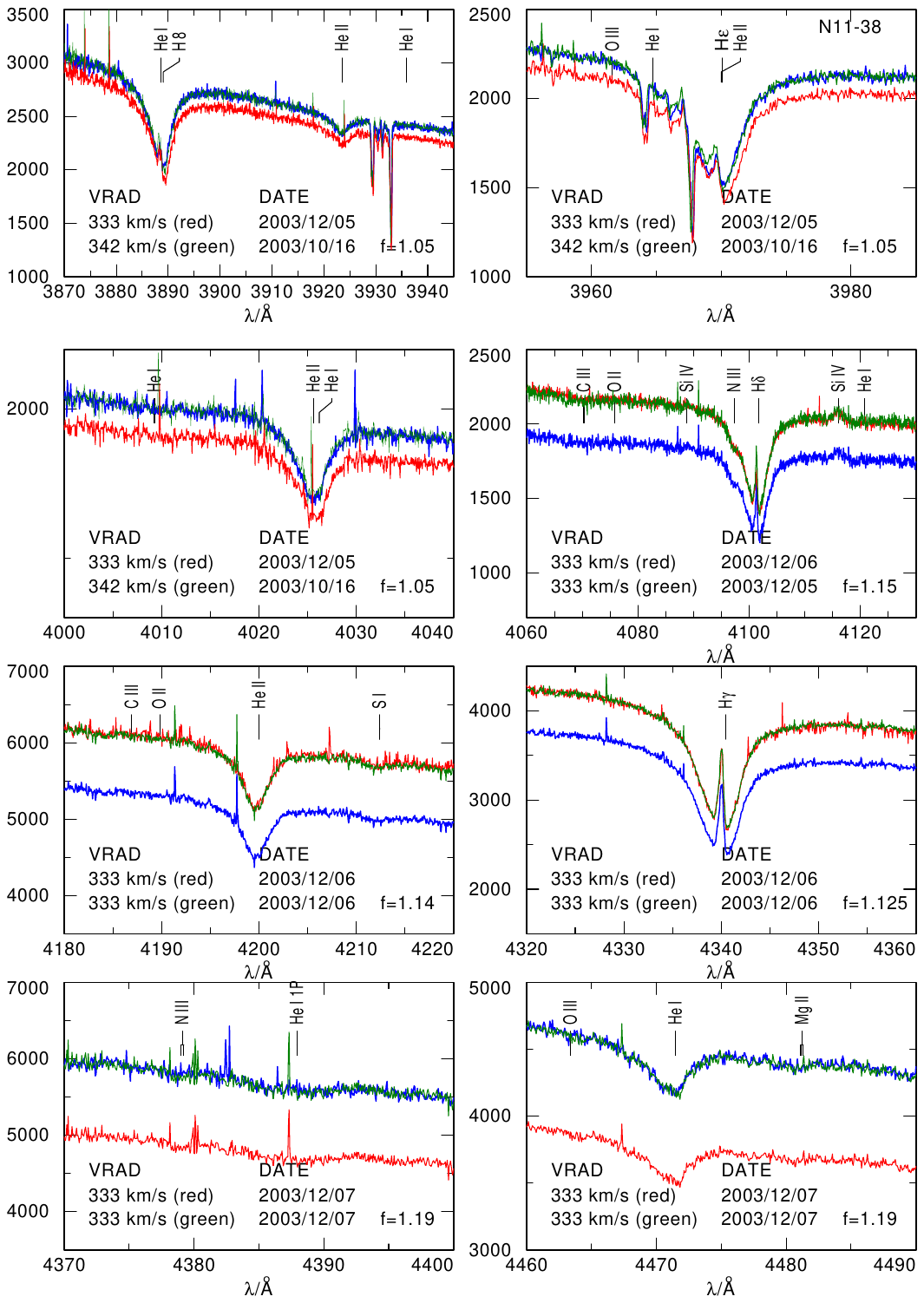}~
\includegraphics[trim={1.0cm 2.3cm 1.0cm 2.1cm},clip,width=0.43\linewidth]{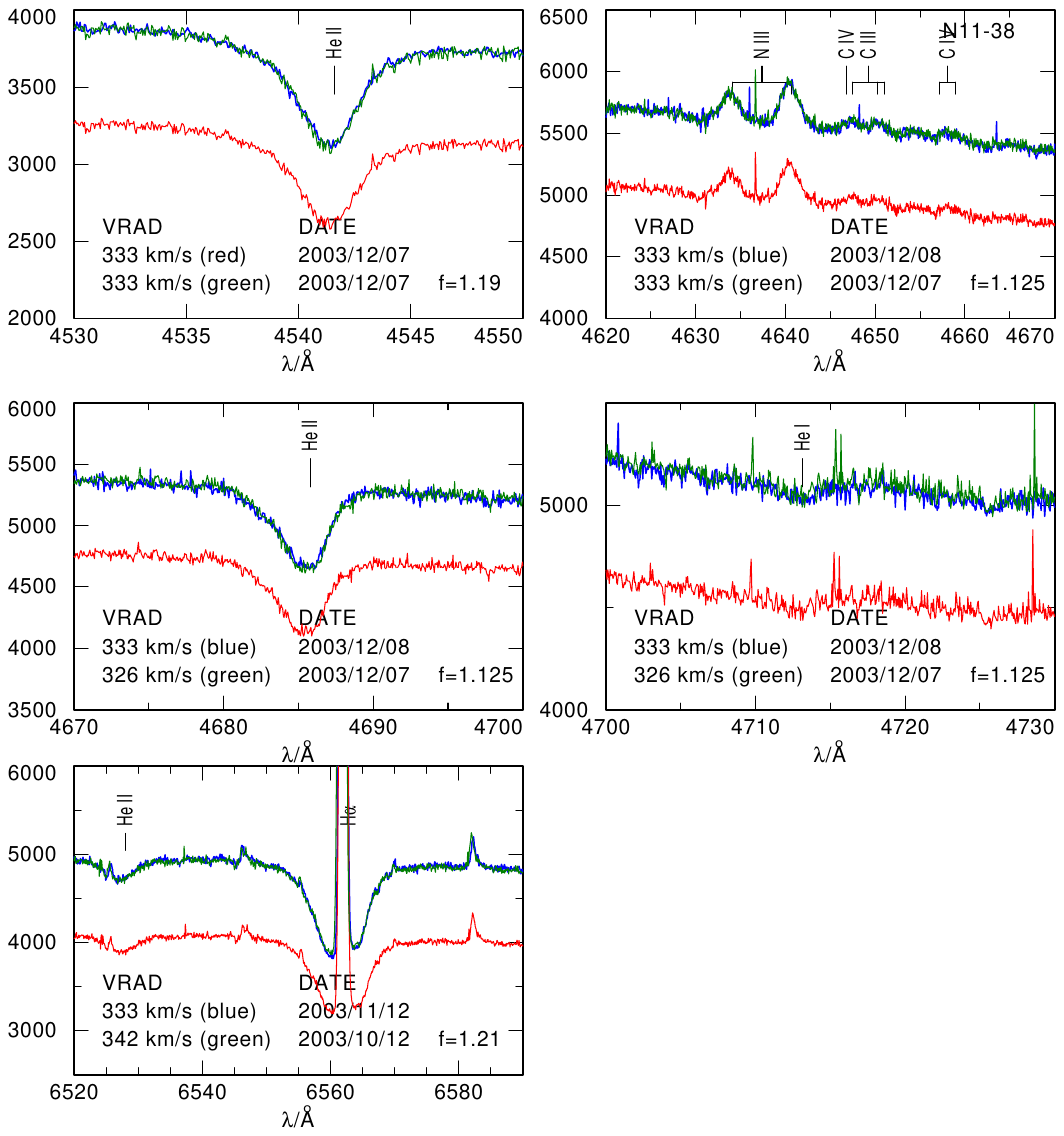}\\
\par\end{centering}
\caption{
PGMW\,3100 \giraffe\ multi-epoch spectra to check for binariry. Only \ha\ displays a variable profile, probably because of wind variability.
Some of the lines have been shifted by the difference in radial velocity, $<9$~\kms\ (green line), to compare their line profiles.
}
\label{fig:n11-38_giraffe}
\end{figure*}

\begin{figure*}
\begin{centering}
\includegraphics[trim={1.0cm 2.3cm 1.0cm 2.1cm},clip,width=0.43\linewidth]{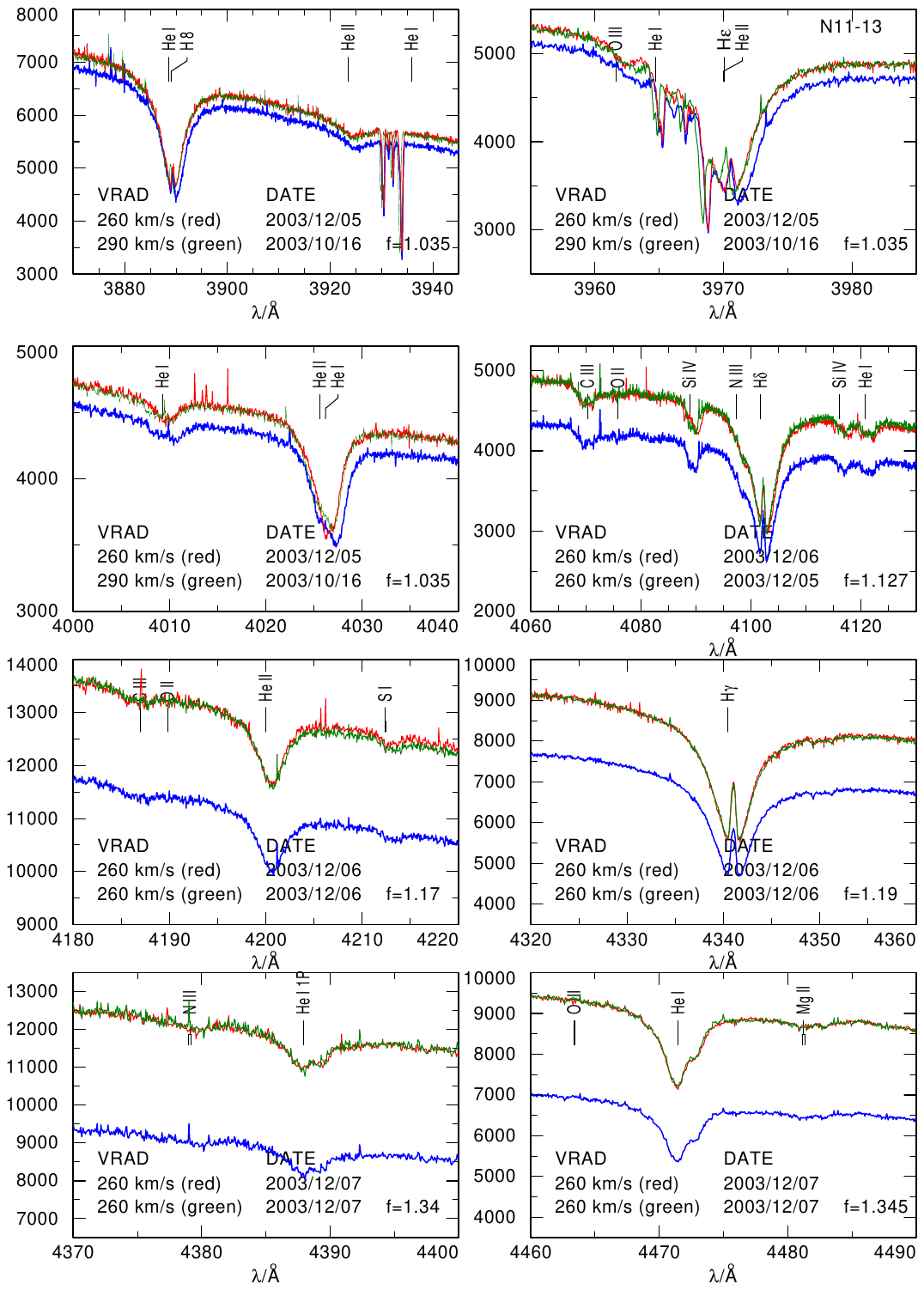}~
\includegraphics[trim={1.0cm 2.3cm 1.0cm 2.1cm},clip,width=0.43\linewidth]{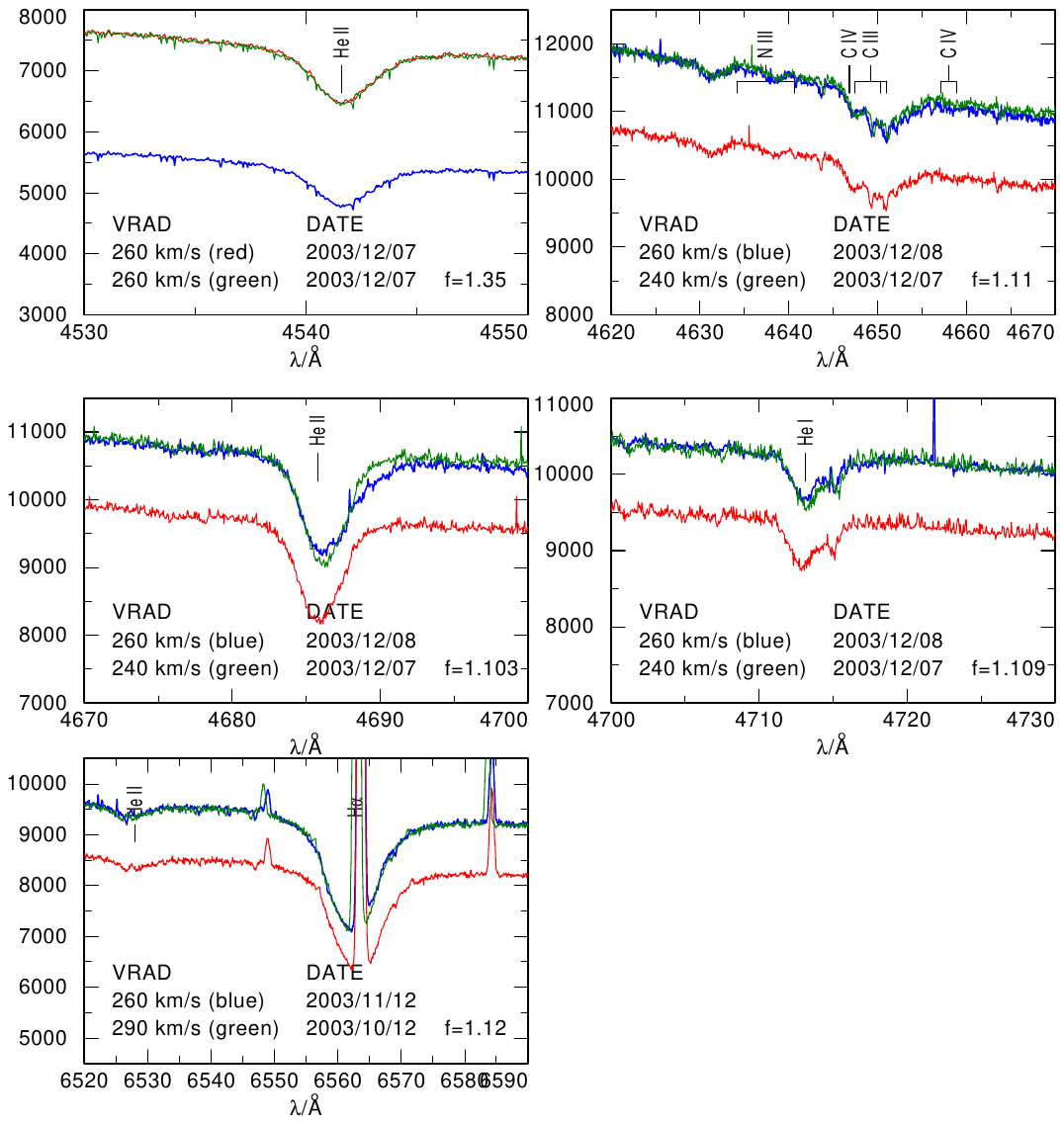}\\
\par\end{centering}
\caption{
PGMW\,3223 \giraffe\ multi-epoch spectra to check for binariry: radial velocity or line profile difference.
Some of the lines have been shifted by the difference in radial velocity, $20-30$~\kms\ (green line), to compare their line profiles.
}
\label{fig:n11-13_giraffe}
\end{figure*}

\begin{figure}
\begin{centering}
\includegraphics[trim={1.0cm 2.3cm 1.0cm 2.1cm},clip,width=0.9\linewidth]{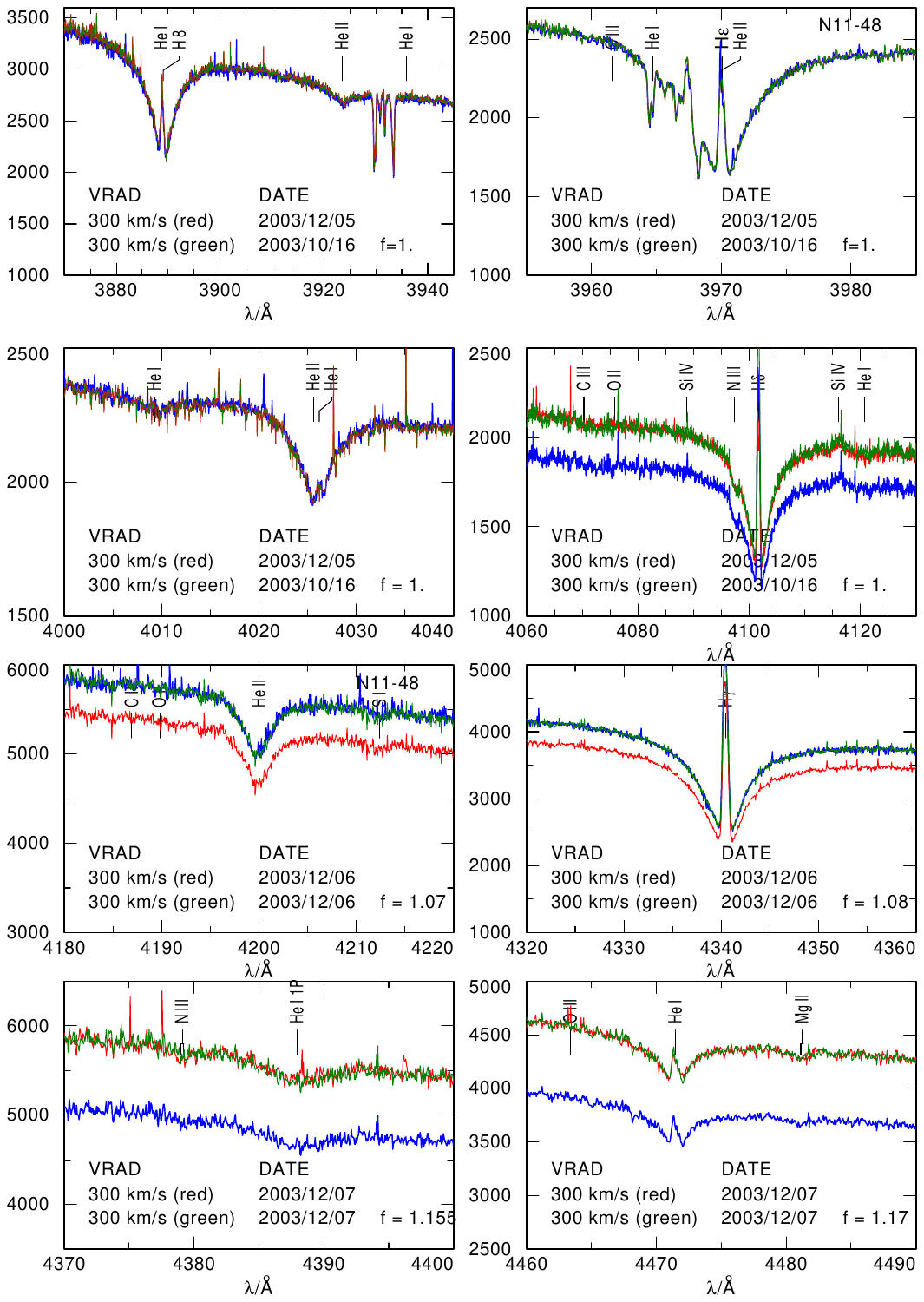}~
\par\end{centering}
\caption{
PGMW\,3204 \giraffe\ multi-epoch spectra to check for binariry: radial velocity or line profile difference. No evidence for binary component in the Balmer nor He lines.
}
\label{fig:n11-48_giraffe1}
\end{figure}

\begin{figure}
\begin{centering}
\includegraphics[trim={1.0cm 2.3cm 1.0cm 2.1cm},clip,width=0.9\linewidth]{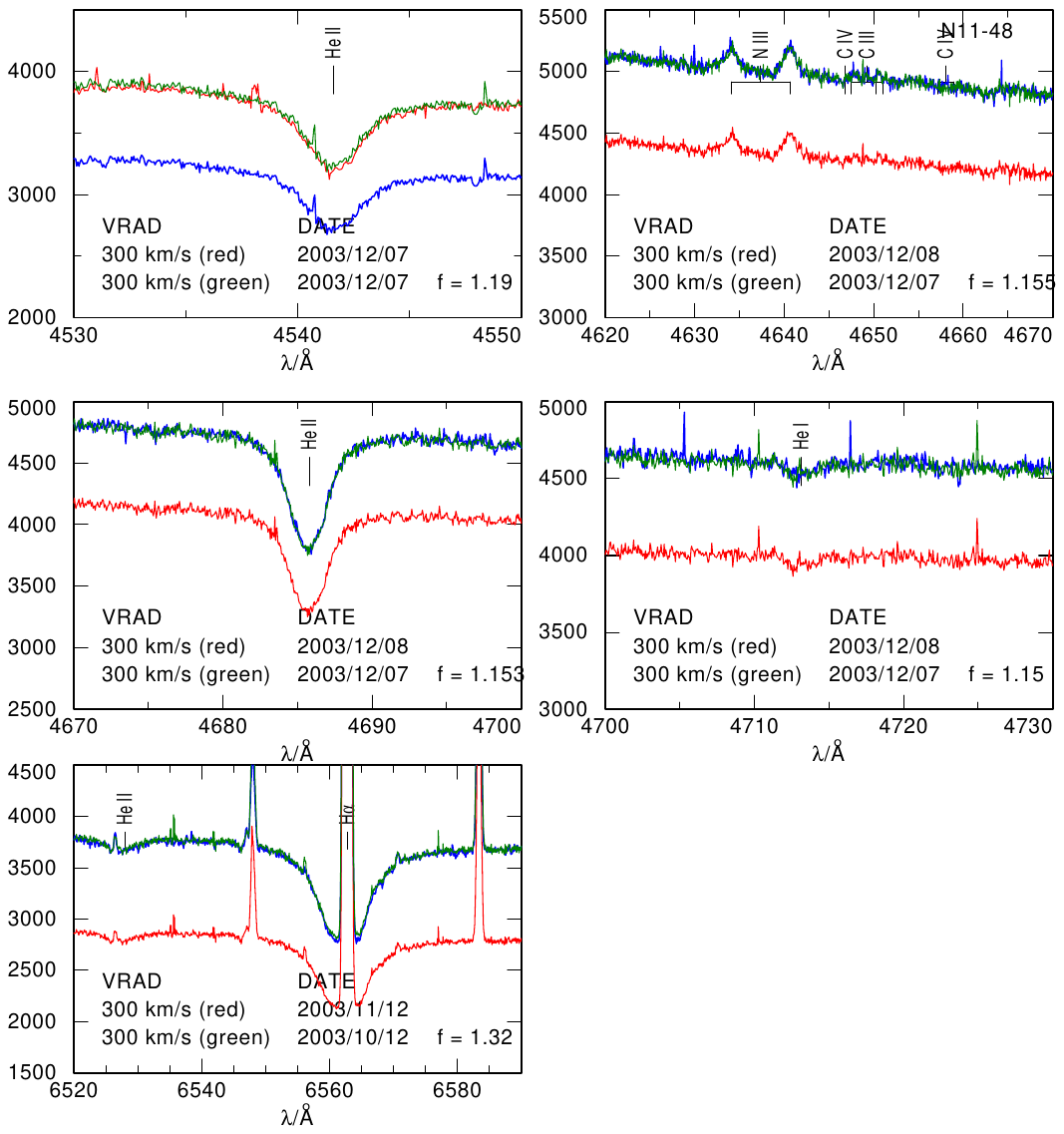}~
\par\end{centering}
\caption{
PGMW\,3204 \giraffe\ multi-epoch spectra to check for binariry: radial velocity or line profile difference. No evidence for binary component in the Balmer nor He lines.
}
\label{fig:n11-48_giraffe2}
\end{figure}

\begin{figure}
\begin{centering}
\includegraphics[trim={1.5cm 5.3cm 1.5cm 5.3cm},clip,width=0.9\linewidth]{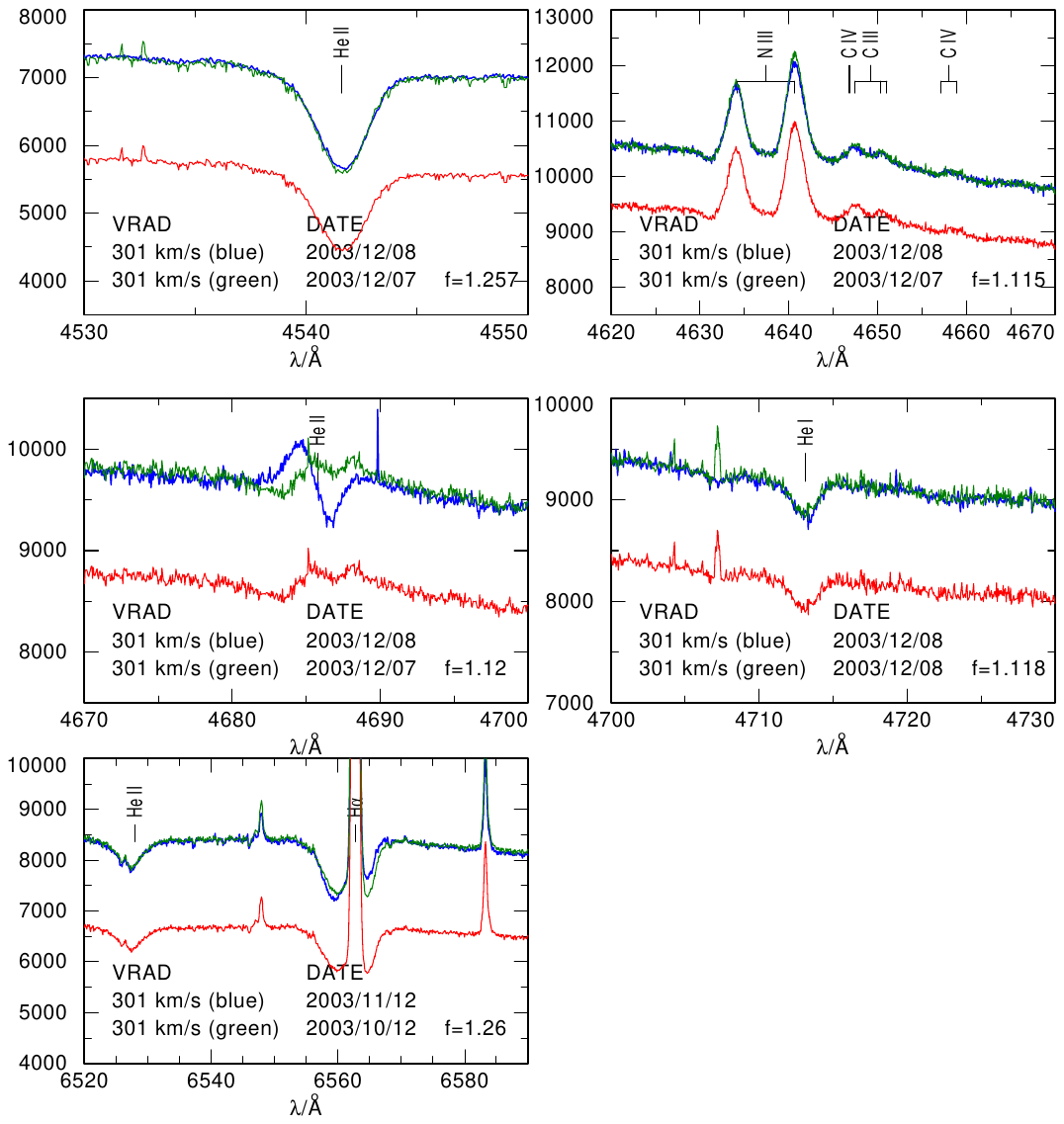}\\
\par\end{centering}
\caption{
PGMW\,3053 \giraffe\ multi-epoch spectra to check for binariry: radial velocity or line profile difference. No evidence for binary component in the Balmer nor He lines. Only \ha\ displays a variable profile, probably because of wind variability.
}
\label{fig:n11-18_giraffe}
\end{figure}

\begin{figure}
\begin{centering}
\includegraphics[trim={1.5cm 5.3cm 1.5cm 5.3cm},clip,width=0.9\linewidth]{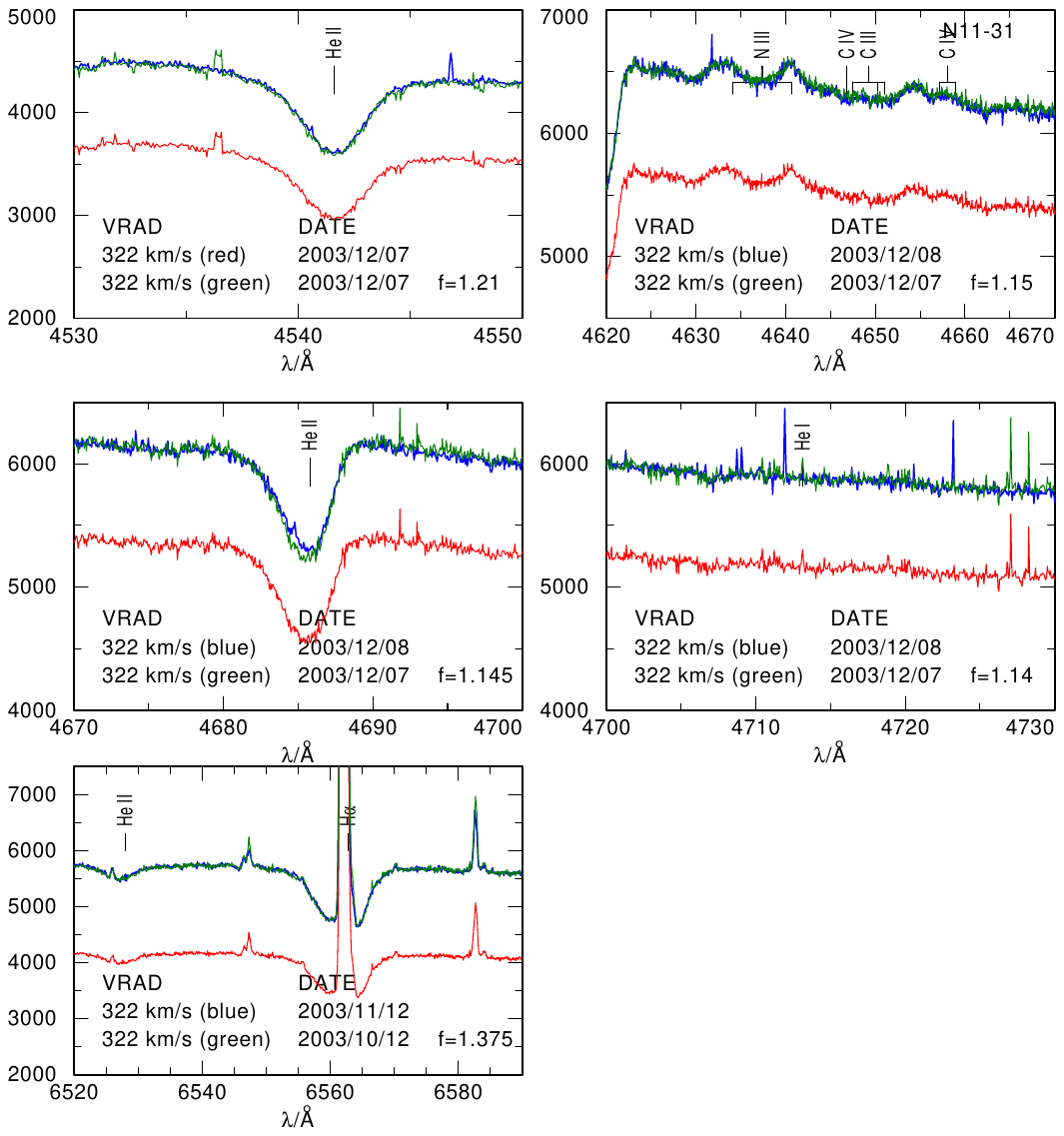}\\
\par\end{centering}
\caption{
PGMW\,3061 \giraffe\ multi-epoch spectra to check for binariry. Only \heiiwr\ displays a variable profile, probably because of wind variability.
}
\label{fig:n11-31_giraffe}
\end{figure}

The search for binarity features in the analyzed stars in N11\,B is shown in Fig.~\ref{fig:n11-38_giraffe} to Fig.~\ref{fig:n11-31_giraffe}.

\section{Additional O-type stars of diffrerent sub-type in LMC}
\label{additional}

The physical parameters obtained with PoWR for three additional O-type stars of different sub-types in LMC are reported in Table~\ref{tab:results4}.
Their chemical abundances and the stellar parameters derived with BONNSAI are listed in Table~\ref{tab:results5} and Table~\ref{tab:results6}, respectively.

\begin{table*}
\begin{center}
\caption[]{Physical parameters obtained with PoWR for O-type stars of different sub-type in LMC.}
\scriptsize\setlength{\tabcolsep}{0.8\tabcolsep}
\begin{tabular}{lcccccccccccccc}
\hline\hline
\rule{0cm}{2.2ex} ID         & \tstar\      &\logg\        & \loglstar\   & $\log\dot{M}$         & \vinf\          & \rstar\ & \mstar\ & \logQH\    & \logQHeI\  & \multicolumn{2}{c}{\logQHeII} & \loglmec\ & \lxlbol\ & \Dmom\ \s \\
   &           &           &          &           &     &    &           &   &   & -- & XR &   &   \\
 & [kK]         &[cm\,s$^{-2}$]& [L$_\odot$]  &[M$_{\odot}$~yr$^{-1}$]& [km\,s$^{-1}$]  & [\Rsun] & [\msol] & [ph\,s$^{-1}$] & [ph\,s$^{-1}$] &  [ph\,s$^{-1}$] &  [ph\,s$^{-1}$] & [\lsun] & & \\
(1)  & (2)          & (3)          & (4)         & (5)          & (6)   & (7)  & (8)         & (9) & (10) & (11) & (12) & (13) & (14) & (15) \\
\hline
Sk $-66^{\circ}$ 171 & 30.5$\pm$0.5 & 3.3$\pm$0.05 & 5.7$\pm$0.1 & $-6.0\pm0.1$ & 1750$\pm50$ & 25.4 & 43 & 49.1 & 47.3 & 38.5 & 41.6 & 2.4 & $-6.4$ & $-2.5$\\
Sk $-69^{\circ}$ 50  & 33.0$\pm$0.5 & 3.4$\pm$0.05 & 5.4$\pm$0.1 & $-6.3\pm0.1$ & 1750$\pm50$ & 15.4 & 19 & 49.0 & 47.7 & 38.6 & 41.1 & 2.8 & $-6.5$ & $-2.0$\\
N11-046              & 31.5$\pm$0.5 & 4.1$\pm$0.05 & 5.0$\pm$0.1 & $-7.8\pm0.1$ &  250$\pm50$ & 10.7 & 46 & 48.1 & 45.9 & 39.4 & 40.3 & 1$\times10^{-4}$ & $-7.6$ & $-4.8$\\ 
\hline
\end{tabular}
\label{tab:results4}
\end{center}
(1) ID \citep[][]{Parker1992};
(2) effective temperature (\tstar);
(3) surface gravity (\logg);
(4) bolometric luminosity (\lstar);
(5) mass-loss rate ($\log\dot{M}$);
(6) terminal wind velocity (\vinf);
(7) stellar radius (\rstar) via the Stefan-Boltzmann law;
ionizing photons of (9) H (\QH);
(10) \hei\ (\QHeI); (11) and \heii\ (\QHeII);
(12) mechanical luminosity;
(13) X-ray to bolometric luminosity ratio (\lxlbol);
(14) modified wind momentum (\Dmom).
\end{table*}

\begin{table}
\begin{center}
\caption[]{Chemical abundances obtained with PoWR for O-type stars of different sub-type in LMC.}
\small\setlength{\tabcolsep}{0.6\tabcolsep}
\begin{tabular}{lcccccc}
\hline\hline
\rule{0cm}{2.2ex} ID   & \multicolumn{2}{c}{$X_{\rm C}$}       & \multicolumn{2}{c}{$X_{\rm N}$}     & \multicolumn{2}{c}{$X_{\rm O}$}         \s    \\ % & $X_{\rm P}$   
(1)       & mass  & number  & mass & number & mass & number \\
     & ($\times10^{-4}$)& ($\times10^{-5}$) & ($\times10^{-5}$)& ($\times10^{-5}$) & ($\times10^{-3}$) &  ($\times10^{-4}$)       \\ % & ($\times10^{-6}$) \\
(1)       & (2)   & (3)    & (4)   & (5)   & (6) & (7) \\
\hline
Sk-66 171 & 14.1  & 14.6   & 54.6  &   4.9 & 1.9 & 1.5          \\ % & 2.9 \\
Sk-69 50  & 9.4   &  9.8   & 15.6  &  13.9 & 2.9 & 2.3           \\ % & 0.58  \\
N11 046   & 2.3   &  2.4   &  7.8  &   0.7 & 2.6 & 1.8          \\ % & 2.9 \\ 
\hline
\end{tabular}
%\vspace{0.4cm}
\label{tab:results5}
\end{center}
(1) ID;
mass fractions by mass and number of (2, 3) carbon ($X_{\rm C}$);
(4, 5) nitrogen ($X_{\rm N}$), and (6, 7) oxygen ($X_{\rm O}$).
\end{table}

\begin{table}
\begin{center}
\caption[]{Parameters derived with BONNSAI \citep[][]{Schneider2014} for O-type stars of different sub-type in LMC.}
\small\begin{tabular}{lccccc}
\hline\hline
\rule{0cm}{2.2ex} ID   & \mini\       & \mev\       & Age [Myr]   & \vini\  \s   \\
PGMW & [\msol]      & [\msol]      & [Myr]       & [\kms]     \\
(1)  & (2)          & (3)          & (4)         & (5)        \\
\hline
Sk-66 171 & 34$\pm4$ & 32$\pm4$ & 3.7$\pm0.4$ & 120$\pm62$ \\
Sk-69 50  & 43$\pm4$ & 41$\pm4$ & 3.3$\pm0.2$ & 110$\pm77$ \\
N11-046   & 20$\pm1$ & 19$\pm1$ & 5.9$\pm0.5$ & 130$\pm38$ \\ 
\hline
\end{tabular}
\label{tab:results6}
\end{center}
(1) PGMW\,$\#$ ID \citep[][]{Parker1992};
(2) initial mass (\mini); (3) actual mass (\mev); (4) age of the star; (5) initial rotational velocity (\vini).
\end{table}
\end{appendix}

%%%%%%%%%%%%%%%%%%%%%%%%%%%%%%%%%%%%%%%%%
\end{document}